*Université Claude Bernard Lyon 1*

Mémoire présenté en vue de l'obtention de l'
# Habilitation à Diriger des Recherches

# Vers une approche intégrative de l'étude des interactions cerveau-environnement chez le primate humain et non-humain

## David THURA, PhD

**Centre de Recherche en Neurosciences de Lyon – Équipe Impact**
Université Claude Bernard Lyon 1 – Inserm U1028 – CNRS UMR5292

### Composition du jury

DR. **THOMAS BORAUD**, DR CNRS (HDR), RAPPORTEUR
DR. **THOMAS BROCHIER**, DR CNRS (HDR), RAPPORTEUR
DR. **THOMAS MICHELET**, ENSEIGNANT-CHERCHEUR (HDR), RAPPORTEUR

DR. **JEAN-RENE DUHAMEL**, DR CNRS (HDR), EXAMINATEUR
DRE. **BJØRG E. KILAVIK**, DR CNRS (HDR), EXAMINATRICE
PR. **YVES ROSSETTI**, PU (HDR), EXAMINATEUR

# Table des matières

*1er neurone enregistré dans le laboratoire - 13 octobre 2022*

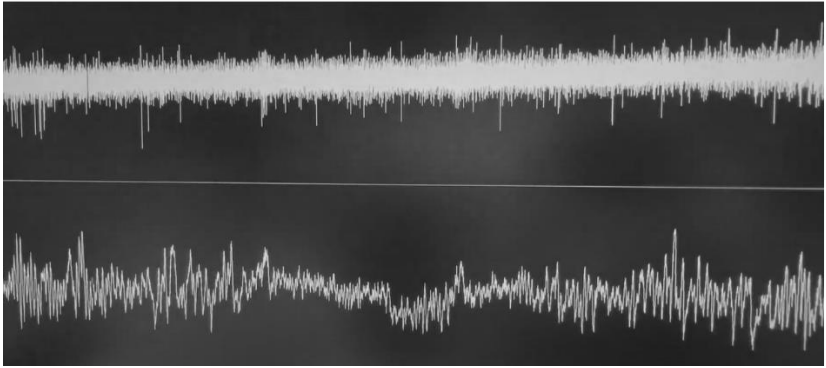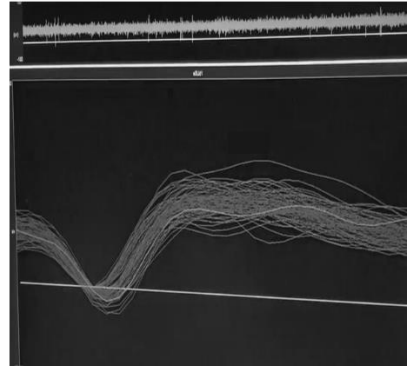

# Résumé

Le dénominateur commun à tout organisme vivant est d'être soumis à un environnement avec lequel il doit interagir de façon à subvenir à ses besoins. Souvent considéré comme un assemblage de modules permettant le traitement de l'information, le cerveau est en réalité un système de contrôle dynamique du comportement, conçu pour permettre aux animaux d'implémenter ces interactions dans des contextes variés et changeants, la plupart du temps de façon imprévisible.

L'étude en laboratoire de cet organe si complexe est un défi. Aussi, pour faciliter son exploration, les chercheurs ont souvent tendance à décomposer les fonctions cérébrales étudiées en différents processus indépendants les uns des autres. Dans le but de faciliter également l'interprétation des données obtenues, les tâches comportementales utilisées sont souvent simplifiées et donc éloignées de la réalité écologique à laquelle la plupart des animaux sont confrontés. Enfin, les analyses utilisées ne permettent pas toujours de décrire un comportement dans sa globalité et sa complexité, en cherchant toujours à isoler et expliquer des paramètres spécifiques que le cerveau est supposé représenter à travers l'activité de ses neurones, parfois à l'échelle des cellules uniques.

Or, cette vision fragmentée, modulaire et réductrice de la méthodologie employée pour étudier la neurophysiologie des interactions cerveau-environnement s'avère limitée dans sa capacité à mettre en évidence ce pourquoi le cerveau a été conçu et ce pourquoi il a évolué.

En retraçant mon parcours scientifique débuté il y a 20 ans, je mets en évidence dans ce mémoire la nécessité de considérer l'organisation du cerveau, certes globalement hiérarchique, mais aussi hautement distribuée et mixte dans la réponse neuronale de ses différentes aires (en me focalisant sur le réseau cortex sensorimoteur-ganglions de la base). Je souligne également l'importance d'adopter des paradigmes comportementaux reflétant autant que faire se peut les caractéristiques des scénarios rencontrés dans la vie réelle des animaux. Et enfin je mentionne l'importance de « désintégrer » la façon dont l'analyse des données est traditionnellement effectuée, et de tenir compte de la nature dynamique du comportement, en privilégiant par exemple l'étude de variables comportementales et d'activités neuronales dites « latentes ».

Je conclue ce mémoire en exposant la vision de mon laboratoire idéal dans une perspective de 5 à 10 années à partir d'aujourd'hui. Ce laboratoire disposerait de deux dispositifs expérimentaux, un premier, classique, pour la réalisation de tests en milieu contraint et donc non naturel, dont les données seraient comparées à celles issues d'un second dispositif dit « naturalistique » dans lequel les animaux pourraient potentiellement exprimer l'ensemble de leur répertoire comportemental. Les tâches testées seraient caractérisées par des principes écologiques forts, tels que dans celles simulant les propriétés de la recherche de nourriture (*foraging*), et les données permettraient de tester des hypothèses basées sur l'addition progressive de composants écologiques dans ces tâches, tout ceci afin de garder un contrôle sur l'interprétabilité de ces données complexes par nature.

# Partie I – Parcours académique

## 1. Formation

| | |
|---|---|
| **1998** | Baccalauréat général, option scientifique, Besançon, France |
| **2000** | Diplôme d'Études Universitaires Générales, Sciences de la vie, Université de Franche-Comté, Besançon, France |
| **2001** | Licence, Biologie, Université de Franche-Comté, Besançon, France |
| **2002** | Maitrise, Biologie cellulaire et physiologie, Université Claude Bernard Lyon I, Lyon, France |
| **2003** | Diplôme d'Études Approfondies, Neurosciences, Université Claude Bernard Lyon I, Lyon, France |
| **2007** | Doctorat, Neurosciences, Université de la Méditerranée (Aix-Marseille II), Marseille, France – Supervision : Dr. Driss Boussaoud |
| **2008-2018** | Post-doctorat, Université de Montréal, Département de neurosciences, Montréal, Canada – Laboratoire du Dr. Paul Cisek |
| **2016** | Certificat universitaire, Journalisme, Université de Montréal, Faculté de l'éducation permanente, Montréal, Canada |

En **juillet 2017**, j'ai reçu la bourse du concours ATIP/Avenir pour étudier le rôle des ganglions de la base dans la régulation de la prise de décision et de l'exécution de l'action. Suite à cette nomination, j'ai participé aux **campagnes de recrutement 2018** des chercheurs CNRS et Inserm. J'ai été recruté par l'Inserm en **août 2018**. J'ai débuté mon programme de recherche indépendant en tant que Chargé de Recherche de Classe Normale Inserm au sein de l'équipe Impact du Centre de Recherche en Neurosciences de Lyon le **1er septembre 2018** et j'ai obtenu ma titularisation le **8 juillet 2019**.

## 2. Publications dans des revues internationales à comité de lecture

### a) Publications associées aux travaux de thèse

Thura D, Hadj-Bouziane F, Meunier M, Boussaoud D. Hand position modulates saccadic activity in the Frontal Eye Field. *Behavioural Brain Research*. 186:148-153, 2008.

Thura D, Boussaoud D, Meunier M. Hand position influences saccadic reaction times in monkey and man. *Journal of Neurophysiology*. 99:2194-2202, 2008.

Thura D, Hadj-Bouziane F, Meunier M, Boussaoud D. Hand modulation of visual, preparatory, and saccadic activity in the monkey frontal eye field. *Cerebral Cortex*. 21(4): 853-864, 2011.

### b) Publications ultérieures

Thura D, Beauregard-Racine J, Fradet C-W, Cisek P. Decision-making by urgency-gating: theory and experimental support. *Journal of Neurophysiology*. 108(11): 2912-30, 2012.

Thura D, Cisek P. Deliberation and commitment in the premotor and primary motor cortex during dynamic decision-making. *Neuron*. 81(6): 1401-1416, 2014.

Thura D, Cos I, Trung J, Cisek P. Context-dependent urgency influences speed-accuracy trade-offs in decision-making and movement execution. *Journal of Neuroscience*. 34(39): 16442-16454, 2014.

Carland M, Thura D, Cisek P. The urgency-gating model can explain the effects of early evidence. *Psychonomic Bulletin & Review*. 22(6):1830-1838, 2015.

Thura D, Cisek P. Modulation of premotor and primary motor cortical activity during volitional adjustments of speed-accuracy trade-offs. *Journal of Neuroscience*. 36(3):938-956, 2016.

Carland M, Marcos E, Thura D, Cisek P. Evidence against sensory integration models of perceptual decisions. *Journal of Neurophysiology*. 115(2):915-930, 2016.

Thura D. How to discriminate conclusively among different models of decision-making? *Journal of Neurophysiology*. 115(5): 2251-2254, 2016.

Thura D, Guberman G, Cisek P. Trial-to-trial adjustments of speed-accuracy tradeoffs in the premotor and primary motor cortex. *Journal of Neurophysiology*. 117(2): 665-683, 2017.

Thura D, Cisek P. The basal ganglia do not select reach targets but control the urgency of commitment. *Neuron.* 95(5):1160-1170, 2017.

Revol P, Colette S, Boulot Z, Foncelle A, Imai A, Thura D, Jacquin S, Osiurak F, Rossetty Y. Thirst for intention? Grasping a glass is a thirst-controlled action. *Front. Psychol*. 10:1248, 2019.

Derosiere G, Thura D, Cisek P, Duque J. Motor cortex disruption delays motor processes but not deliberation about action choices. *Journal of Neurophysiology.* 122(4): 1566-1577, 2019.

Carland M, Thura D, Cisek P. The urge to decide and act: implications for brain function and dysfunction. *The Neuroscientist.* 25(5): 491-511, 2019.

Thura D. Decision urgency invigorates movement in humans. *Behavioural Brain Research.* 382:112477, 2020.

Thura D, Cisek P. Microstimulation of dorsal premotor and primary motor cortex delays the volitional commitment to an action choice. *Journal of Neurophysiology.* 123(3):927-935, 2020.

Reynaud AJ, Saleri Lunazzi C, Thura D. Humans sacrifice decision-making for movement execution when a demanding motor control is required. *Journal of Neurophysiology*. 124(2):497-509, 2020.

Derosiere G, Thura D, Cisek P. Duque J. Trading accuracy for speed over the course of a decision. *Journal of Neurophysiology.* 126(2):361-372, 2021.

Thura D. Reducing behavioral dimensions to study brain-environment interactions. Invited commentary on "Précis of Vigor: Neuroeconomics of movement control", by Reza Shadmehr and Alaa A. Ahmed. *Behavioral and Brain Sciences*. Volume 44, 2021, e135.

Saleri Lunazzi C, Reynaud AJ, Thura D. Dissociating the impact of movement time and energy costs on decision-making and action planning in humans. *Frontiers in Human Neuroscience*. 15:715212, 2021.

Derosiere G, Thura D, Cisek P. Duque J. Hasty sensorimotor decisions rely on an overlap of broad and selective changes in motor activity. *PLOS Biology*. 20(4): e3001598, 2022.

Cisek P, Thura D. Models of decision-making over time. *The Oxford Research Encyclopedia of Neuroscience*. 7 Sep. 2022.

Thura D, Cabana J-F, Féghaly A, Cisek P. Integrated neural dynamics of sensorimotor decisions and actions. *PLOS Biology*. 20(12): e3001861, 2022.

Saleri Lunazzi C, Thura D*, Reynaud A*. Impact of decision and action outcomes on subsequent decision and action behaviors. *European Journal of Neuroscience*. 57(7), 1098-113, 2023. *: même contribution

Leroy É, Koun É, Thura D. Integrated control of non-motor and motor efforts during perceptual decision-making and action execution: a pilot study. *Sci Rep* 13, 9354, 2023.

Kaduk K, Henry T, Guitton J, Meunier M, Thura D. Hadj-Bouziane F. Atomoxetine and Reward equally improve task engagement and perceptual decision but differently affect movement execution. *Neuropharmacology*. 2023 Sep 27; 241:109736, 2023.

### c) Publication en cours d'expertise

Saleri C, Thura D. Evidence for interacting but decoupled controls of decisions and movements in non-human primates. *bioRxiv*. doi: https://doi.org/10.1101/2024.01.29.577721

## 3. Financements de recherche obtenus et distinctions scientifiques

| | |
|---|---|
| **2008-2010** | Bourse postdoctorale de la fondation FYSSEN |
| **2011-2013** | Bourse postdoctorale du Groupe de Recherche sur le Système Nerveux Central (Université de Montréal, Montréal, Canada) |
| **2012** | Travel award – 22ème conférence annuelle de la *Society for Neural Control of Movement* |
| **2018** | Bourse ATIP/Avenir du CNRS et de l'Inserm **– 180 k€** |
| **2019** | Bourse des jeunes chercheurs recrutés à l'Inserm **– 30 k€** |
| **2021** | Extension de la bourse ATIP/Avenir du CNRS et de l'Inserm **– 60 k€** |
| **2023-2027** | Contrat ANR en tant que partenaire (coordinateur : Dr. David Robbe) – **350 k€** |

En 2020 l'équipe Impact du CRNL a obtenu un financement de **204 k€** de l'Université Claude Bernard Lyon 1 pour l'achat d'équipements visant à réaliser un projet dont je serai le responsable scientifique, à savoir l'étude comportementale et neurophysiologique des primates humains et non-humains en condition « naturalistique[1] ». Ce projet sera décrit dans la partie II.4b de ce mémoire.

## 4. Enseignements réalisés

**2014-2015**   Enseignant assistant du cours NSC-2002 (discussion d'un article scientifique), Niveau M2, Université de Montréal – 2h

**2015-2016**   Enseignant assistant du cours NSC-2006 (Introduction à Matlab, équations différentielles), Niveau M2, Université de Montréal – 2h

**2016-2017**   Enseignant du cours #11 du programme NSC-2002 (Planification des mouvements, aires corticales associées), Niveau M2, Université de Montréal – 2h

Enseignant du cours #7 du programme NSC-2004 (Temps de réaction), Niveau M2, Université de Montréal – 2h

Enseignant assistant du cours NSC-2002 (discussion d'un article scientifique), Niveau M2, Université de Montréal – 2h

**2017-2018**   Enseignant du cours #7 du programme NSC-2004 (Temps de réaction), Niveau M2, Université de Montréal – 2h

Enseignant assistant du cours NSC-2002 (discussion d'un article scientifique), Niveau M2, Université de Montréal – 2h

**2018-2019**   Master 2 Neurosciences fondamentales et cliniques – UE *Neural basis of cognition* – Sujet : *Electrophysiology approaches to decision-making* – Université Claude Bernard Lyon 1 – 2h

Enseignant assistant du cours NSC-2004 (Temps de réaction), Niveau M2, Université de Montréal – 2h

**2019-2020**
**2020-2021**   Master 2 Neurosciences fondamentales et cliniques – UE *Neural basis of cognition* – Sujet : *Electrophysiology approaches to decision-making* – Université Claude Bernard Lyon 1 – 2h

Master 1 Neurosciences – UE Neurosciences cognitives et imagerie cérébrale – Sujet : *Neural bases of action selection and execution* – Université Claude Bernard Lyon 1 – 2h

**2021-2022**   Master 2 Neurosciences fondamentales et cliniques – UE *Neural basis of cognition* – Sujet : *Behavioral and electrophysiological approaches to decision-making* – Université Claude Bernard Lyon 1 – 2h

**2022-2023**   Master 2 Neurosciences fondamentales et cliniques – UE *Computational neuroscience* – Sujet : *Models of decision-making in changing conditions* – Université Claude Bernard Lyon 1 – 2h

Master 2 Neurosciences fondamentales et cliniques – UE *Neural basis of cognition* – Sujet : *Behavioral and electrophysiological approaches to decision-making* – Université Claude Bernard Lyon 1 – 2h

**2023-2024**   Enseignant pour la formation règlementaire Inserm pour les concepteurs de protocoles expérimentaux utilisant le primate non-humain. Sujet : Neuro-anatomie – 2h

---

[1] De l'anglais *naturalist* ou *naturalistic neuroscience*, une approche visant à comprendre le fonctionnement du cerveau « dans la nature ». Naît de la nécessité d'aller au-delà des situations expérimentales simplistes et non écologiques souvent utilisées en laboratoire pour faciliter les études de neurosciences.

## 5. Responsabilités scientifiques et administratives

- Membre de la **structure en charge du bien-être animal** (SBEA) de la plateforme Primage et de l'animalerie de l'équipe Impact du CRNL.

- Membre du **comité d'éthique lyonnais pour les neurosciences expérimentales** (CELYNE, CNREEA #42).

## 6. Encadrement et supervision académique

### a) En tant que chercheur principal

| | |
|---|---|
| **2024-2025** | Supervision du Dre **Clara Saleri,** stage postdoctoral |
| **2024** | Supervision de **Mathilde Pauchard**, étudiante en M1, Université Claude Bernard Lyon 1 |
| **2024** | Supervision de **Nathan Lambert**, étudiant en M2, Université d'Angers – Mémoire de stage : Rôle des synchronisations entre le cortex prémoteur dorsal et le striatum dorsal pendant la coordination entre la prise de décision et l'action |
| **2023-2024** | Supervision du Dre **Claire Poullias,** stage postdoctoral |
| **2022** | Co-supervision (avec le Dre Hadj-Bouziane) de **Chloé Rivière**, étudiante en DUT, Université Claude Bernard Lyon 1 |
| **2021-2022** | Supervision d'**Élise Leroy**, étudiante en M2, Université Lyon 2 – Mémoire de stage : Prise de décision et action : ressources partagées ? |
| **2020-2024** | Supervision de **Clara Saleri**, étudiante en thèse, Université Claude Bernard Lyon 1 – Dérogation d'encadrement sans HDR obtenue le 4 septembre 2020. Mémoire de thèse : Principes comportementaux et neuronaux de la coordination entre la prise de décision et de l'action chez le primate humain et non-humain. Thèse soutenue publiquement le 13 juin 2024 au Neurocampus du CRNL |
| **2020-2021** | Supervision de **Clara Saleri**, étudiante en M2, Université Claude Bernard Lyon 1 – Mémoire de stage : *Impact of temporal and energy costs of movements on decision making strategy* |
| **2019-2022** | Supervision du Dre **Amélie Reynaud,** stage postdoctoral |
| **2019** | Co-supervision (avec le Dre Fadila Hadj-Bouziane) de **Martin Leclerc**, étudiant en M1, Université de Rouen ; et de **Raphaëlle Schlienger**, étudiante à l'INSA de Lyon |

### b) En tant que chercheur postdoctoral

- Supervision de 4 étudiants de niveaux M1/M2 pendant des stages de 3 mois dans le laboratoire du Dr. Paul Cisek

- Aide à la supervision de 2 étudiants (**Alexandre Pastor-Bernier** et **Ayuno Nakahashi**) au doctorat dans le laboratoire du Dr. Paul Cisek

# Partie II – Mémoire d'HDR : Vers une approche intégrative de l'étude des interactions cerveau-environnement chez le primate humain et non-humain

## 1. Préambule et problématiques générales

En initiant la rédaction de ce mémoire, j'ai envisagé plusieurs angles sous lesquels je pourrais développer une ligne conductrice retraçant mon parcours et mes réflexions à mi-carrière, motiver mes futurs objectifs et décrire la façon dont je transmettrais ma vision de la recherche à mes prochain(e)s étudiant(e)s. Celui qui s'est très vite imposé peut-être résumé avec une question posée par John Kalaska lors de sa présentation de clôture du 23ème congrès de la *Society for the Neural Control of Movement* en 2013, question qui m'avait alors particulièrement interpellé : Malgré des décennies de recherches, pourquoi avons-nous à ce jour échoué à décrire avec précision et exhaustivité la correspondance entre un comportement supposément simple, le déplacement du bras d'un point de l'espace vers un autre, et l'activité neuronale dans des régions cérébrales dont nous connaissons le rôle crucial dans le contrôle moteur depuis bientôt un siècle et demi (Ferrier, 1875; Penfield and Boldrey, 1937; Evarts, 1968; Gross, 2007) ?

Dans ce mémoire, je propose de mettre l'emphase sur trois raisons pouvant expliquer selon moi cette difficulté :

a) Des cadres théoriques prévoyant la décomposition artificielle du comportement en étapes successives et indépendantes les unes des autres.
b) Des protocoles expérimentaux non écologiques et contraignants, ne permettant pas de tester ce pourquoi le cerveau a été élaboré et les raisons pour lesquelles il a évolué.
c) Des analyses comportementales et neuronales ne tenant pas compte de la nature intégrée et dynamique du comportement.

### a) La théorie sérielle de l'organisation du comportement et du fonctionnement cérébral

La plupart des études contemporaines en neurosciences se réfèrent à cadres théoriques issus de travaux en neuropsychologie datant des 19ème et 20ème siècles dans lesquels les chercheurs ont voulu expliquer le comportement humain, éminemment complexe par nature, en le décomposant en différents processus indépendants les uns des autres (James, 1890). Cette approche, réminiscence du concept de dualisme cher à René Descartes, a notamment donné naissance à la **théorie sérielle de l'organisation du comportement** (on parle également de modèle perception-cognition-action, de computationnalisme ou encore de métaphore de l'ordinateur, Figure 1) (Fodor, 1983; Pylyshyn, 1984). Dans ce cadre, le cerveau est un **dispositif de traitement de l'information** dans lequel la perception engendre des représentations internes du monde servant d'entrées aux systèmes cognitifs. Ces modules cognitifs permettent la manipulation des représentations pour la construction de connaissances complexes, l'élaboration de raisonnements et la prise de décision. Enfin, les centres moteurs mettent en œuvre les plans d'action élaborés par les processus cognitifs. Dès lors, ce modèle établit une taxonomie plus ou moins élaborée dans laquelle les chercheurs peuvent identifier les grandes fonctions qu'ils étudient (le système perceptif, la cognition, le contrôle moteur), et/ou leurs sous-catégories associées (proprioception, reconnaissance d'objets, attention, mémoire, prise de décision, modèles internes, planification de trajectoires, etc.).

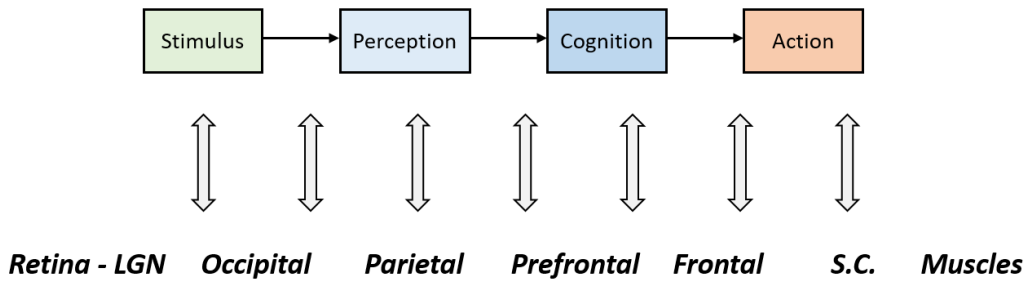

**Figure 1**: Illustration des cadres théoriques de l'organisation sérielle et modulaire du comportement (haut) et du fonctionnement cérébral (bas, pour la modalité visuelle). LGN : Corps géniculé latéral du thalamus (*lateral geniculate nucleus*) ; S.C. : Colliculus supérieur (*superior colliculus*).

Cette vision ségrégée de l'organisation du comportement trouve sa correspondance au niveau de l'organisation du fonctionnement cérébral lorsque **l'on associe chaque fonction (dont celles mentionnées ci-dessus) au recrutement de régions spécifiques du cerveau** (Figure 1). Un exemple de ce modèle modulaire concerne la prise de décision, très souvent associée à l'activité du cortex préfrontal, notamment orbitofrontal (OFC) and ventromédial (vmPFC) (Padoa-Schioppa, 2011). Selon cette proposition, tous les facteurs pertinents pour le choix d'une offre sont intégrés dans l'OFC et le vmPFC, ils sont ensuite comparés, indépendamment des processus sensorimoteurs, et la meilleure option est sélectionnée. L'action appropriée pour exprimer ce choix est enfin planifiée et exécutée avec la mise en œuvre des régions cérébrales motrices, les cortex sensorimoteurs et moteurs du lobe frontal et les ganglions de la base notamment.

Cependant, le comportement naturel des animaux n'obéit pas à un fonctionnement sériel de type entrée-sortie, avec des processus cognitifs entre ces deux étapes (Hurley, 2001). Celui-ci est en effet plutôt de type **interactif**, favorisant **un système de contrôle dynamique fonctionnant en boucle fermée** dans lequel les mouvements façonnent les actions potentielles actuelles et futures (ou *affordances*), attribuant diverses importances aux stimuli en fonction de ces opportunités, et **influençant ainsi directement les décisions** (Cisek, 1999, 2007; Buzsáki et al., 2014; Yin, 2014; Pezzulo and Cisek, 2016; Yoo et al., 2021). De même, les données neurophysiologiques sous-tendant le comportement dirigé vers un but ne sont pas compatibles avec la plupart des hypothèses basées sur une organisation sérielle et modulaire du comportement et du fonctionnement cérébral. Par exemple, des études sur les mécanismes neuronaux de

- 11 -

la prise de décision ont montré depuis plusieurs décennies que **les corrélats des processus de décision sont distribués dans tout le cerveau**, notamment dans les régions corticales et sous-corticales fortement impliquées **dans le contrôle du mouvement** (Cisek and Kalaska, 2010). De façon réciproque, deux études électrophysiologiques récentes, menées pour l'une chez la souris, pour l'autre chez le singe, ont révélé **un niveau de mixité des réponses neuronales à l'échelle du cerveau entier** et dans le cortex préfrontal, respectivement, jusqu'alors insoupçonné. En effet, ces travaux montrent que dans des tâches de prises de décision, ce sont les mouvements, dont ceux non directement liés à la tâche, qui expliquent le plus la variance du taux de décharge des neurones enregistrés, y compris dans les aires traditionnellement considérées comme sensorielles ou cognitives (Musall et al., 2019; Tremblay et al., 2023a). Cette "contamination" des activités cérébrales non motrices par des mouvements met en évidence l'importance du contrôle moteur pour le fonctionnement global du cerveau, soutenant ainsi l'idée selon laquelle le cerveau existe avant tout pour contrôler le comportement (Fine and Hayden, 2022).

b) Des protocoles expérimentaux non écologiques et contraignants

Pour simplifier l'étude du comportement et du fonctionnement cérébral comme le prévoient les modèles théoriques sériels mentionnés ci-dessus, l'approche traditionnelle en neurosciences des systèmes est l'utilisation de **protocoles expérimentaux contraints** qui limitent considérablement la gamme des comportements qui peuvent être exprimés par les sujets testés. Chez le singe par exemple, les sujets réalisent des tâches en étant assis dans une chaise « de contention », qui, comme son nom l'indique, n'autorise que les mouvements de l'effecteur étudié. Le but est d'isoler les processus pertinents à l'étude en question, assurer une réplicabilité de ces processus au cours d'une session expérimentale et entre différentes sessions, le tout en se **prémunissant des facteurs confondants** rendant difficile l'interprétation des résultats. Une autre caractéristique de ces protocoles est l'obligation pour les sujets d'exécuter **plusieurs dizaines, voire centaines, d'essais par condition expérimentale**. Du point de vue de l'expérimentateur, ce nombre important de répétitions est nécessaire afin d'atteindre une puissance statistique acceptable pour valider l'interprétation des résultats. Ce design artificiel est même à la base d'hypothèses très influentes sur l'optimisation du comportement des sujets humains et non-humains, telles que la notion selon laquelle le taux de succès[2], plus que la précision du comportement, serait le paramètre que ces sujets cherchent à optimiser (Balci et al., 2011). Cependant, dans la vie quotidienne, nous ne sommes jamais confrontés à des situations nécessitant l'exécution successive de plusieurs centaines de décisions exprimées par des mouvements. Les tâches de laboratoire sont donc pour la plupart **détachées de l'expérience naturelle des sujets testés**, supprimant ainsi l'essentiel de la variabilité et de la complexité qui façonnent intrinsèquement le comportement dans des contextes écologiquement valables (Gomez-Marin and Ghazanfar, 2019; Maselli et al., 2023).

---

[2] Le taux de succès est défini par la probabilité de succès du comportement, moins les efforts demandés pour effectuer ce comportement, le tout divisé par le temps nécessaire à la réalisation du comportement. On parle de taux de succès, voire de taux de capture pour les expériences menées chez l'humain, alors que l'on parlera plutôt de taux de récompense pour les expériences menées chez le singe.

Dans le domaine de la prise de décision par exemple, une tâche comportementale largement utilisée et à partir de laquelle des résultats comportementaux et neurophysiologiques très influents ont émergé (Roitman and Shadlen, 2002; Gold and Shadlen, 2007) est la discrimination de direction de points en mouvements (*random dot motion discrimination task*) (Britten et al., 1992). Les analyses comportementales sur ce stimulus très peu écologique concernent souvent les temps de réaction et la précision des décisions (Palmer et al., 2005). Chez l'humain, l'expression du choix se fait souvent par un mouvement simple comme l'appui sur un bouton poussoir (Servant et al., 2021). Pourtant, la prise en compte de la cinématique des mouvements des yeux, des doigts, de la souris d'ordinateur ou du corps entier (Grießbach et al., 2022) pendant la prise de décision permet d'accéder à de riches indices comportementaux dit cognitifs, tels que l'incertitude décisionnelle, les changements d'avis, la confiance et la motivation, soit autant de facteurs indissociables du comportement naturel des animaux. Ces choix expérimentaux enrichissant la gamme des réponses pouvant être exprimées par les sujets ont également permis d'initier **la mise en évidence des interactions entre les processus moteurs et non-moteurs** au niveau comportemental (Burk et al., 2014; Gallivan et al., 2018).

c) Des analyses réductrices

Même lorsqu'une tâche est éthologiquement compatible, la façon dont l'analyse du comportement est effectuée peut impacter notre capacité à comprendre un phénomène particulier. Par exemple, pour les mouvements d'atteinte réalisés par les primates, la plupart des modèles computationnels supposent que le **système moteur planifie et contrôle explicitement les caractéristiques spécifiques de ces mouvements** sur lesquelles nous semblons être capables d'imposer un contrôle volontaire, telles que la direction, la position finale, la trajectoire spatiale, la durée, la vitesse et la force d'exécution (Bullock and Grossberg, 1988; Kalaska et al., 1997). Or, comme nous le détaillons ci-dessous (p. 15-16), les tentatives de mise en évidence d'un mécanisme neurophysiologique effectif basé sur une correspondance entre l'activité neuronale motrice et ces variables spécifiques a aujourd'hui globalement échoué (Kalaska, 2019). Dès lors, on peut se demander si le cerveau a été conçu pour « encoder » chacune de ces variables comportementales liées à l'exécution d'un mouvement d'atteinte. En effet, la plupart des mesures énumérées ci-dessus varient fortement les unes avec les autres. Par exemple, l'augmentation de la vitesse d'un mouvement pour une amplitude donnée réduit généralement sa durée et augmente son coût énergétique. Pourquoi le cerveau prendrait-il la peine de calculer autant de quantités qui sont toutes si étroitement liées ? Ne bénéficierait-il pas, en termes de compromis précision-dépense énergétique, de calculer des variables moins nombreuses **mais intégrées qui capturent, sinon la totalité, la plus grande variabilité** représentée par l'ensemble de ces paramètres ?

Un exemple illustrant le bénéfice potentiel de cette approche intégrative de l'analyse du comportement est celui de l'étude de **la vigueur** de ce comportement. La notion de vigueur a gagné en popularité ces dernières années en partie parce que l'on pense qu'elle reflète la façon dont le cerveau évalue l'utilité[3] du mouvement (Choi et al., 2014; Berret and Jean, 2016). La vigueur est une notion plutôt intuitive a priori, mais elle est communément associée dans la littérature à des variables très différentes. Elle fait en effet parfois référence à la vitesse et/ou à la durée du mouvement (Muhammed et al., 2020), à la vitesse ou à la durée échelonnée en fonction de l'amplitude (Choi et al., 2014; Berret et al., 2018), à la cinématique du mouvement et au temps de réaction (Milstein and Dorris, 2007; Summerside et al., 2018), ou au temps de réaction uniquement (Griffiths and Beierholm, 2017). Je pense depuis quelques années que la neuropsychologie ainsi que les neurosciences comportementales, y compris mes propres travaux, souffrent d'une tentation de vouloir **« désintégrer » le comportement afin que les sous-processus artificiellement définis par cette division paraissent plus faciles à analyser**. Je crois qu'une telle approche en réalité contribue à confondre la terminologie, encourager un comportement non-écologique et limiter finalement la mise en évidence d'éléments pertinents pouvant expliquer la raison pour laquelle les systèmes neuronaux ont été conçus et ce pourquoi ils ont évolué (Thura, 2021). Comme d'autres (Shadmehr and Ahmed, 2020), je pense que la vigueur du mouvement nous informe sur la façon dont nous valorisons nos actions. Cependant, en tant qu'outil d'évaluation, il manque à ce jour une définition claire de ce qu'est la vigueur ainsi qu'une façon de la quantifier. Je pense que cela est dû au fait que la vigueur est l'un **des paramètres de faible dimension qui reflète une approximation de la durée et/ou la vitesse à laquelle le cerveau interagit avec son environnement** dans des circonstances particulières. Par conséquent, la vigueur ne peut pas être quantifiée avec précision, mais seulement caractérisée de manière approximative sur la base de variables de décision et/ou de mouvement sélectionnées arbitrairement.

Concernant l'analyse de l'activité neuronale à l'origine du comportement, la question de son niveau d'intégration se pose également. Par exemple, lorsque les études neurophysiologiques sur animal vigile ont débuté dans les années 1960, le champ était dominé par des modèles dits « représentationnels » de la fonction cérébrale, qui supposaient que l'activité de **neurones uniques exprimait explicitement des types d'informations spécifiques**, telles que des propriétés particulières d'un stimulus sensoriel ou les paramètres particuliers d'un mouvement (Figure 2A-B).

Dans le domaine du contrôle sensorimoteur, ce fondement conceptuel a motivé de nombreuses études utilisant une grande variété de tâches pour tenter d'identifier les paramètres de sortie motrice et les cadres de coordination exprimés par les neurones dans différentes aires motrices corticales, notamment le cortex moteur primaire (M1), le cortex prémoteur dorsal (PMd), le cortex prémoteur ventral (PMv), l'aire motrice motrice supplémentaire (SMA), l'aire 5 du cortex pariétal et le cortex intrapariétal médian (MIP). En accord avec les prédictions d'une hypothèse représentationnelle des mécanismes corticaux du contrôle moteur volontaire, ces études ont révélé des différences importantes dans les propriétés de réponse d'un seul neurone ainsi que dans l'importance et le *timing* des corrélations avec différents paramètres de mouvements à la fois au sein et entre les aires corticales, reflétant vraisemblablement les différents rôles joués par chaque population neuronale dans le contrôle moteur (Georgopoulos et al., 1982; Kalaska, 2009).

---

[3] L'évaluation subjective du rapport entre ce que rapporte et ce que coûte un comportement.

Cependant, **ces travaux n'ont pas fourni à ce jour de consensus** quant à l'identité du ou des paramètres contrôlés, ou des transformations de coordonnées codées dans une aire motrice corticale. Les raisons de cet échec incluent des **corrélations non stationnaires** entre l'activité d'un neurone unique et les paramètres d'exécution motrice à différents moments de la tâche, notamment avant et pendant le mouvement, des **propriétés qui se chevauchent** entre les neurones dans différentes zones corticales et des **corrélations partielles** de l'activité d'un neurone unique avec plusieurs paramètres d'exécution motrice, ceci s'expliquant en partie par le fait que différents paramètres moteurs sont couplés par les lois du mouvement et la biomécanique des membres (Kalaska, 2019).

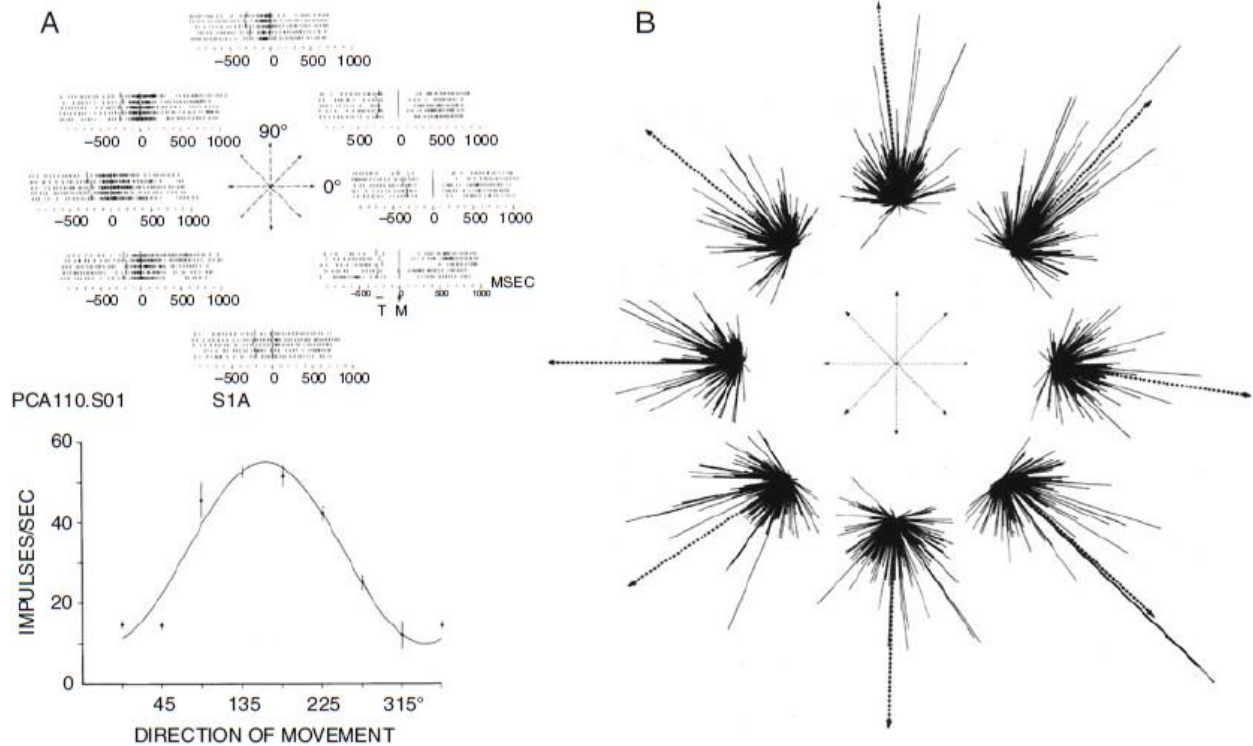

**Figure 2**: "Représentation" de la direction du mouvement dans le cortex moteur primaire. ***A***. Haut : tracés de type « rasters » de l'activité d'un neurone du cortex moteur primaire (M1) lors de 5 mouvements dans 8 directions différentes dans un plan (2D), alignés sur le début du mouvement. Bas : Courbe de préférence directionnelle de l'activité de ce neurone, centrée sur sa direction préférée. ***B***. Représentation vectorielle de la répartition de l'activité directionnelle dans une population neuronale de M1 pour 8 directions de mouvement (adaptés de Georgopoulos et al., 1982, 1983).

Un exemple très intéressant qui illustre cette difficulté à établir une correspondance entre activité unitaire et paramètre spécifique des mouvements d'atteinte vient d'une étude de Mark Churchland et collaborateurs publiée en 2006. Dans cette étude, les auteurs ont entraîné les singes à réaliser des mouvements de type

« centre vers l'extérieur » (« *center-out* » *task*) à un signal donné (*go signal*). La direction, la distance et la vitesse d'exécution (lente ou rapide) du mouvement étaient indiquées par une consigne visuelle fournie entre 400 et 800 ms avant le signal de départ. Des enregistrements neuronaux ont été réalisés dans la partie caudale du cortex prémoteur dorsal (PMd) et la partie rostrale du cortex moteur primaire (M1). En plus de mettre en évidence l'effet classique de la direction du mouvement sur la réponse des neurones dans les deux régions, les auteurs ont observé que la quasi-totalité d'entre eux étaient également sensibles à la vitesse d'exécution du mouvement d'atteinte. De façon remarquable cependant, **la direction préférée des cellules variait en fonction des différentes conditions de vitesse testées** (Churchland et al., 2006). Les auteurs de cette étude et d'autres ont interprété ce manque de stabilité en rappelant que le rôle du système moteur est **de produire un mouvement, pas de le décrire**, et que les modèles d'activité neuronale qui mettent en œuvre l'action ne sont pas nécessairement contraints de représenter les variables du mouvement dans un cadre de référence particulier (Fetz, 1992). Avec le grand nombre de cellules qui existent dans le cortex, **le système moteur est incroyablement redondant**. Pour produire n'importe quel *pattern* d'activité descendant vers les muscles, on pourrait utiliser un nombre presque infini de motifs d'activation neuronale. Cela signifie que les cellules individuelles n'ont pas besoin de coder explicitement les variables pertinentes pour la réussite de la tâche (qu'il s'agisse de la direction, de la vitesse, etc.) tant que **la population dans son ensemble** spécifie ces variables de manière appropriée (Cisek, 2006a).

L'intérêt pour ce type de **codage populationnel** connait une explosion de popularité depuis une petite quinzaine d'années. On parle même de la **doctrine populationnelle**, comme le principe unique selon lequel la population de neurones, et non le neurone unique, est l'unité de codage fondamentale utilisée par le cerveau (Saxena and Cunningham, 2019). Les théories populationnelles ne sont pas nouvelles, des chercheurs tentent de comprendre comment les neurones fonctionnent en assemblées depuis les années 1940 (Hebb, 1949). Cependant, avec le développement et l'accessibilité de nouvelles technologies permettant d'enregistrer de très nombreux neurones simultanément, l'attrait pour l'étude des populations neuronales à grande échelle connait un nouvel essor (Ebitz and Hayden, 2021). Au-delà des moyens technologiques repoussant les limites de notre capacité à étudier une multitude de neurones chez l'animal vigile en comportement, une explosion de **nouveaux concepts et d'analyses** est venue définir la conception moderne de l'approche neurophysiologique de l'étude du cerveau, que cela concerne des processus dits cognitifs, tels que la prise de décision, ou le contrôle moteur (Mante et al., 2013; Shenoy et al., 2013; Jazayeri and Afraz, 2017; Humphries, 2020; Churchland and Shenoy, 2024).

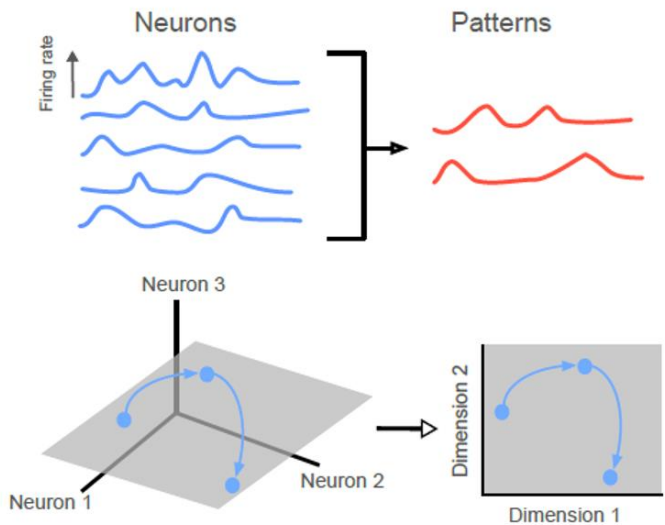

**Figure 3**: Principe de réduction de dimensions de l'activité neuronale unitaire. *Haut* : Pour l'activité de cinq neurones au cours du temps (bleu), l'application d'une réduction dimensionnelle résume cette activité en motifs d'activité les plus courants et donc les plus partagés par ces neurones – ici deux (rouge). *Bas* : Imaginons l'activité conjointe de trois neurones évoluant dans un espace tridimensionnel, une dimension par neurone (à gauche). Représenter l'évolution de leurs taux de décharge au cours du temps (flèches) crée une trajectoire d'activité. L'activité des neurones varie de telle sorte qu'elle reste toujours dans le même plan (souvent appelé *manifold*, gris). Ce qui signifie que seules deux dimensions sont nécessaires pour capturer la variation d'activité de ces trois neurones (à droite) (adapté de Humphries, 2020).

En particulier, pour un neurophysiologiste des populations, l'analyse canonique n'est plus l'histogramme péri-stimulus, mais **la trajectoire neurale dans un espace d'état**. Dans ce cadre, l'activité de *n* neurones enregistrés comprend un « espace d'état » à *n* dimensions dans lequel l'activité de chaque neurone forme un axe (une dimension) de cet espace. L'activité instantanée de l'ensemble de la population enregistrée occupe un point précis dans cet espace d'état à un instant donné. L'activité générant un mouvement trace une trajectoire dans cet espace à *n* dimensions au cours du temps. Un des principes de ce système dynamique est que **l'activité de sous-ensembles de neurones** qui contribuent au déroulement de la trajectoire neuronale **co-varie** de différentes manières via des signaux d'entrée partagés et en fonction des interactions synaptiques entre les neurones du circuit (Vyas et al., 2020). Au vu de cette complexité, vouloir décrire le fonctionnement du circuit en quantifiant le taux de décharge de chaque neurone à chaque instant est simplement impossible. Pour ce faire, une approche plus efficace et maintenant largement utilisée est la « **réduction de dimensionnalité** » (Cunningham and Yu, 2014). La réduction de dimension permet de mettre en évidence des modèles de co-variation de réponses dynamiques (variables latentes) qui sont partagées par de nombreux neurones de la population (Figure 3). Cela réduit l'ensemble de l'espace d'activité neuronale à *n* dimensions en un espace de variables latentes **de basse dimension beaucoup plus intégré** et traitable. Cet espace **représente la majorité de la variance totale de l'activité neuronale** qui façonne la trajectoire de l'activité de la population au cours du temps. De fait, plutôt que d'essayer de déterminer si un paramètre de mouvement est « codé » par un seul neurone, la réduction de dimension analyse **la structure de covariance statistique de l'activité de la population** pour identifier les corrélats multi-neurones de différents processus computationnels.

## 2. Approches utilisées et lignes conductrices

Depuis le début de ma formation en neurosciences et de mes premiers travaux en tant qu'étudiant en doctorat (en 2004), je m'intéresse aux **mécanismes neuronaux qui sous-tendent les fonctions cérébrales permettant l'exploration de l'espace qui nous entoure et les interactions avec notre environnement**. J'ai étudié ces fonctions dans le système oculomoteur et dans le système moteur du bras, à l'aide de techniques comportementales chez l'humain sain, de techniques comportementales et électrophysiologiques chez le singe macaque rhésus (enregistrements mono- et multi-électrodes, analyse des activités neuronales unitaires et des potentiels de champ locaux, micro-stimulation intra- et sous-corticale), le tout combiné à une approche de modélisation informatique visant à formaliser les mécanismes étudiés et proposer de nouvelles hypothèses testables expérimentalement sur le fonctionnement cérébral en lien avec ce comportement interactif.

Au cours de ces vingt dernières années, les cadres théoriques dans lesquels j'ai entrepris mes travaux et les approches méthodologiques choisies pour la réalisation de ces travaux se sont progressivement rapprochés des principes édictés dans la section II-1 ci-dessus. Et comme je le détaillerai dans la partie « Projets de recherche » de ce mémoire (section II-4), j'ai pour ambition de poursuivre sur cette trajectoire et m'orienter toujours davantage vers **une approche intégrative de l'étude des interactions entre le cerveau et l'environnement** qui nous entoure.

Ci-dessous, je décris brièvement quelques-uns de ces principes adoptés systématiquement dans le cadre de mes recherches.

### a) Étude des interactions entre fonctions

Sans questionner directement la question de l'organisation sérielle et modulaire du comportement et du fonctionnement cérébral mentionnée ci-dessus, mes travaux passés m'ont permis d'être très vite sensibilisé à la notion de neuro-organisation certes globalement hiérarchique du cerveau mais également **au principe de sélectivité mixte** des réponses neuronales, ainsi qu'à l'**organisation anatomo-fonctionnelle très distribuée** des aires cérébrales. En effet, mon projet de thèse consistait à démontrer la présence de signaux liés à la position de la main dans une structure corticale connue alors pour contrôler uniquement les mouvements oculaires, le champ oculaire frontal (voir section II-3a). Peu de temps avant le début de ma thèse, mon superviseur, le Dr. Driss Boussaoud, avait démontré la présence de signaux liés à la position des yeux dans l'orbite dans différentes régions contrôlant la préparation et l'exécution des mouvements d'atteinte (signaux suspectés de contribuer à la coordination œil-main) (Boussaoud et al., 1998). Par la suite, ma formation postdoctorale dans le laboratoire du Dr. Paul Cisek m'a donné l'opportunité de tester l'hypothèse selon laquelle le cortex sensorimoteur frontal, celui-là même responsable de la planification et de l'exécution des mouvements d'atteinte, participe aussi **à la sélection** de ces mouvements d'atteinte (voir section II-3b). Ces travaux m'ont permis de démontrer au niveau comportemental le **lien étroit et bidirectionnel existant entre sélection et exécution de l'action** (voir section II-3c). En tant que chercheur indépendant, je poursuis l'étude de ces interactions entre prise de décision et action aux niveaux comportemental et neuronal, conformément au cadre théorique selon lequel le cerveau est **un système de contrôle dynamique fonctionnant en boucle fermée** dans lequel les décisions influencent les actions et réciproquement (voir section II-3d)**.**

b) Corrélation et causalité entre comportement et activité neuronale à l'échelle unitaire

Les mécanismes auxquels je m'intéresse depuis mon initiation aux neurosciences en 2004 (coordination œil-main, sélection et exécution de l'action) sont des **phénomènes rapides**, de l'ordre de la dizaine de millisecondes, **et dynamiques**. Ils sont sous le contrôle de **sous-populations de neurones spécifiques,** nichées dans des régions cérébrales à la fois corticales et sous-corticales, formant des réseaux anatomo-fonctionnels souvent caractérisés par des activités récurrentes, prenant la forme de réverbérations. La technique de mesure de l'activité neuronale employée dans mes travaux, l'enregistrement des potentiels d'action unitaires extracellulaires et des potentiels de champs locaux à l'aide de microélectrodes, **possède les résolutions à la fois temporelle et spatiale nécessaires à l'étude de ces mécanismes**. Les informations qu'elle permet d'obtenir ne seraient pas accessibles par une approche en imagerie fonctionnelle, par exemple l'imagerie par résonance magnétique fonctionnelle, qui mesure l'activité cérébrale à l'échelle du voxel, volume contenant des millions de neurones, avec une résolution temporelle de l'ordre de la seconde.

J'ai débuté en doctorat en réalisant des enregistrements à l'aide d'une microélectrode unique insérée et retirée du cerveau à chaque séance, permettant de recueillir l'activité d'un à 3 neurones (dans le meilleur des cas) simultanément. J'ai poursuivi en post-doctorat avec des enregistrements (toujours aigus) réalisés à l'aide de 3 ou 4 électrodes descendues conjointement dans nos structures d'intérêt, permettant de collecter l'activité d'un à 10 neurones (très rarement) simultanément. Actuellement, nous enregistrons jusqu'à 20 neurones simultanément dans 3 régions différentes avec un système incorporant 32 microélectrodes implantées de façon chronique sur l'animal et pouvant être manipulées indépendamment les unes des autres (Dotson et al., 2017). Ces dernières acquisitions permettent d'effectuer **des analyses populationnelles de la dynamique neuronale**, conformément au principe énoncé ci-dessus selon lequel la population, et non le neurone isolé, est l'échelle d'analyse pertinente pour le type de mécanisme étudié dans mes projets. Nous enregistrons également les potentiels de champs locaux, même si leur analyse n'a pas représenté une part importante de mes travaux jusqu'à présent (mais voir la section II-4).

Enfin, l'approche corrélative adoptée de façon systématique dans mes études et qui consiste à mettre en relation l'activité neuronale et le comportement a parfois (pas assez souvent à mon goût) été associée à une approche permettant de se rapprocher **d'une démonstration d'un lien de causalité entre activité neuronale générée par le cerveau et comportement résultant de cette activité**. Ceci a été notamment entrepris pendant mon post-doctorat par le biais de la micro-stimulation intra- et sous-corticale, et je compte poursuivre cette approche dans mes futurs projets.

- 19 -

c) Comparaison Humains-Singes

L'axe principal de mon travail consiste à étudier et décrire les bases neurales de différents types de comportements typiques de la vie quotidienne des humains en examinant le fonctionnement de structures cérébrales particulières **chez le singe macaque rhésus vigile** à l'aide d'enregistrements électrophysiologiques ou de perturbations de l'activité dans ces structures. Evidemment, le but ultime de tout neuroscientifique est de **comprendre comment le cerveau humain fonctionne**. Or, même si le singe macaque est à ce jour le modèle animal le plus proche phylogénétiquement de l'humain pour la recherche en neurosciences et que plusieurs études aient mis en évidence **des similarités comportementales** entre les deux espèces (Hanks and Summerfield, 2017; Evans and Hawkins, 2019), en particulier lorsque les paradigmes comportementaux sont adaptés aux capacités cognitives et motrices de l'espèce testée (Blanchard and Hayden, 2015), elles présentent aussi **des spécificités dont il faut tenir compte**. Par exemple, la vitesse des saccades oculaires des macaques rhésus est environ deux fois plus rapide que celle des humains (Straube et al., 1997). Les singes ont une biomécanique oculaire différente de celle des humains, mais une fois ces différences prises en compte, il subsiste des différences persistantes dans la **vigueur des mouvements** (Shadmehr et al., 2010). De même, les singes présentent un taux de **dévaluation temporelle des récompenses** plus élevé que celui des humains (Jimura et al. 2009).

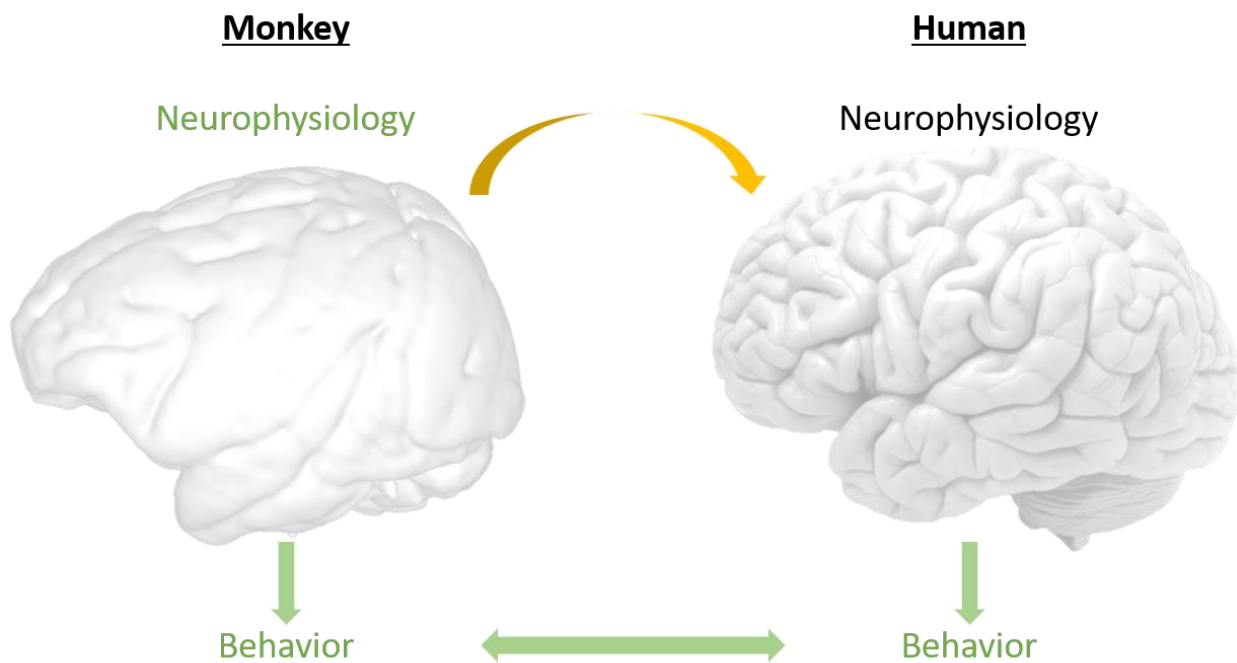

**Figure 4**: Approche basée sur la comparaison du comportement des humains et des singes pour inférer la neurophysiologie des humains sur la base des données électrophysiologiques obtenues chez le singe (flèche orange).

Dès lors, afin de pouvoir inférer la neurophysiologie humaine en se basant sur les propriétés neuronales observées chez le singe, une étape initiale essentielle consiste à déterminer si **ces deux espèces**, ayant divergé au cours de leur évolution respective il y a 25 millions d'années, **sont capables des mêmes stratégies comportementales dans le contexte des tâches testées en laboratoire**, et si tel n'est pas le cas, **identifier les raisons pouvant expliquer ces différences**. L'approche expérimentale adoptée depuis le début de ma carrière a donc été de **tester systématiquement les deux espèces**, primates humains et non-humains, sur **les mêmes tâches comportementales** et dans **des conditions expérimentales similaires**. En ce faisant, la comparaison du comportement entre les deux espèces permet d'établir avec davantage de confiance un pont entre les mécanismes neurophysiologiques observés chez le singe et le fonctionnement cérébral humain inféré de ces observations (Figure 4).

### 3. Vingt années de recherches passées

#### a) Travaux de doctorat (2004-2007)

Ma thèse portait sur la coordination œil-main et les champs oculaires frontaux (FEF), une région majeure du réseau oculomoteur chez le primate (Figure 5) (Bruce and Goldberg, 1984), et qui est aujourd'hui considérée comme contribuant également à diverses fonctions non motrices telles que la prise de décision (Kim and Shadlen, 1999). **Le but de mon travail de doctorat était d'évaluer le rôle du FEF dans l'exploration oculaire de l'espace péri-personnel, avec l'hypothèse spécifique que le FEF reçoit des signaux visuels et proprioceptifs de la main pour assurer une exploration oculaire adaptée de l'espace entourant le corps.** En effet, s'il est communément admis que lors des comportements engageant à la fois les yeux et la main (regarder puis prendre une tasse de café par exemple), les yeux influencent fortement à la fois le moment de l'initiation et la précision des mouvements d'atteinte, la composante réciproque, c'est-à-dire l'influence de la position du corps (la main en particulier) par rapport à la cible d'une saccade sur le comportement oculomoteur était sous-étudiée.

Dans le laboratoire dirigé par Driss Boussaoud à Lyon (Institut de Sciences Cognitives) puis à Marseille (Institut des Neurosciences Cognitives de la Méditerranée), nous avons d'abord mené une expérience psychophysique chez l'humain et le singe dans laquelle nous avons montré chez les deux espèces que la position de la main affecte fortement les temps de réaction (ou latences) des saccades exécutées vers des cibles visuelles situées à différentes distances de celle-ci (Figure 6A). Plus précisément, nous avons observé que lorsque le mouvement de la main précède l'apparition de la future cible de la saccade, le déclenchement des saccades est facilité (i.e. les latences sont raccourcies) lorsque la main et la cible des saccades sont éloignées l'une de l'autre.

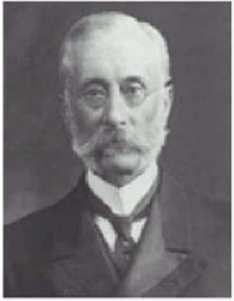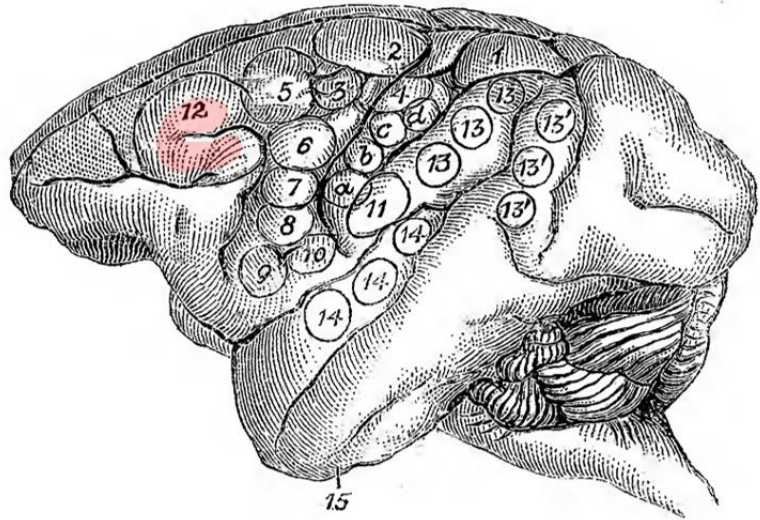

**Figure 5** : Portrait de David Ferrier et localisation originale des champs oculomoteurs frontaux (région numérotée 12, en rouge), représentée sur une vue latérale d'un cerveau de singe macaque, d'après les stimulations électriques réalisées à la fin du 19ème siècle (Ferrier, 1875).

De façon intéressante, lorsqu'un délai est introduit entre le positionnement de la main et l'apparition de la cible, le résultat décrit ci-dessus tend à diminuer, voire à s'inverser, suggérant des phénomènes attentionnels de type « inhibition de retour » à l'origine de ces observations.

Les résultats complets de cette étude ont été publiés en 2008 dans *Journal of Neurophysiology* (Thura et al., 2008a).

Ensuite, nous avons exploré l'activité du champ oculomoteur frontal (FEF) chez le singe entraîné à effectuer des mouvements oculaires saccadiques vers des cibles visuelles tandis que la position de sa main (visible et invisible) variait d'un essai à l'autre. Nous avons observé que les signaux de position de la main, issus de la vision et de la proprioception, modulaient l'activité d'une large population de neurones visuels et saccadiques du FEF, que la main soit visible ou non, et particulièrement lorsque la cible de la saccade était située dans la région de l'espace péri-personnel de l'animal (c'est-à-dire à distance d'atteinte) (Figure 6B).

Ces résultats sont décrits en détails dans deux articles publiés en 2008 dans *Behavioral Brain Research* (Thura et al., 2008b) et en 2011 dans *Cerebral Cortex* (Thura et al., 2011).

Ensemble, ces études comportementales et neurophysiologiques menées à la fois chez l'humain et le singe ont démontré que chez le primate, **les signaux de position des membres supérieurs influencent l'exécution des saccades programmées dans le système oculomoteur** et qu'au sein de ce réseau, **les champs oculaires frontaux contribuent au processus neuronal de coordination œil-main dans l'espace péri-personnel en intégrant les signaux positionnels du bras** (Thura, 2007).

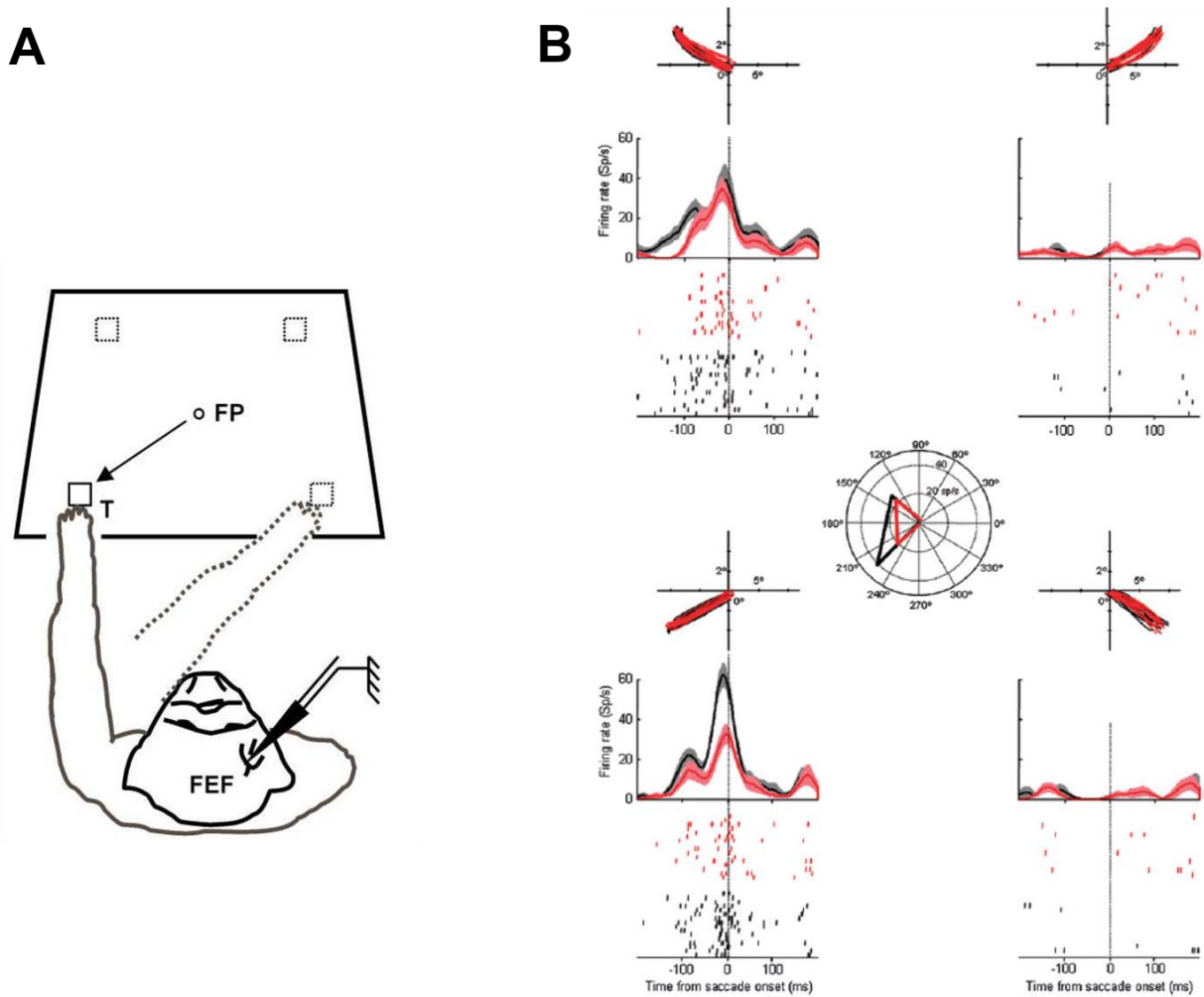

**Figure 6** : *A* : Configurations des positions cible-main pour l'étude comportementale et neurophysiologique de l'influence de la position de la main sur le comportement oculomoteur et sur l'activité du champ oculomoteur fronral (FEF). FP : *Fixation point* ; T : *Target*. *B* : Exemple d'un neurone saccadique de FEF modulé par la position de la main. L'activité neuronale est représentée sous forme de « rasters » où chaque ligne représente un essai et chaque point représente un potentiel d'action. Une fonction de densité de potentiels d'action (SDF, sp/s) est calculée toutes les 30ms en faisant convoluer le train de potentiels d'action avec une gaussienne d'écart type égal à 20 ms. L'activité est alignée sur l'initiation de la saccade. Les 4 panneaux représentent la réponse du neurone pour chacune des 4 cibles. Pour chaque cible, la main était placée soit en bas à gauche (noir) soit en bas à droite (rouge) de l'écran. Les traces oculaires individuelles sont représentées par des trajectoires de saccade du point de fixation vers chacune des 4 cibles (au-dessus de chaque graphique). Le tracé polaire au centre représente l'activité moyenne (prise de 50 ms avant à 30 ms après le début de la saccade) pour chaque direction de saccade et pour chaque position de main. Adapté de Thura et al., 2011.

b) Travaux de post-doctorat (2008-2013)

Ma formation postdoctorale dans le laboratoire de Paul Cisek à l'Université de Montréal m'a donné l'opportunité de démarrer un nouveau programme de recherche tout en élargissant mes compétences techniques et théoriques en enregistrements multi-électrodes et en modélisation computationnelle. **Mon projet était d'étudier les mécanismes computationnels et neurophysiologiques de la prise de décision en condition « dynamique », lorsque l'information influençant le choix varie pendant la** **période de délibération**. En abordant cette question, nous avons remis en question deux modèles théoriques jusqu'alors largement acceptés et défendus dans la communauté :

- Les décisions sont généralement considérées comme des représentations abstraites d'informations, déterminées dans un centre exécutif unique situé dans les zones préfrontales du cortex cérébral (Padoa-Schioppa, 2011). Cependant, l'aspect dynamique de nombreuses décisions de la vie quotidienne nous a conduit à proposer que ces décisions soient plutôt de nature motrice et donc directement déterminées dans les aires cérébrales impliquées dans la préparation et l'exécution des mouvements, le système sensorimoteur.
- La plupart des modèles de prise de décision proposent que le cerveau prend une décision en accumulant au fil du temps les informations pertinentes en faveur des différentes options possibles et concurrentes jusqu'à ce qu'un seuil critique soit atteint (Gold and Shadlen, 2007; Ratcliff et al., 2016). Cependant, des données du laboratoire de Paul Cisek suggéraient que ces modèles de type « accumulation d'information » n'expliquent pas le comportement des sujets lorsque cette information varie au cours des essais (Cisek et al., 2009). Un nouveau modèle, le modèle d'urgence (*urgency-gating model*), a donc été proposé, étant plus performant dans cette situation écologique. Dans ce modèle, les informations sensorielles fournies au sujet pour guider son choix sont intégrées au cours du temps **mais sur une durée très brève**, **puis elles sont combinées à un signal « d'urgence »** qui pousse l'activité neuronale vers le seuil de choix pendant délibération (Figure 7). Pour contribuer à un débat animé sur ce sujet (Thura and Cisek, 2016a), j'ai publié en 2016 un article de type "NeuroForum" dans *Journal of Neurophysiology* sur la manière de tester de manière convaincante ces différents modèles de prise de décision (Thura, 2016).

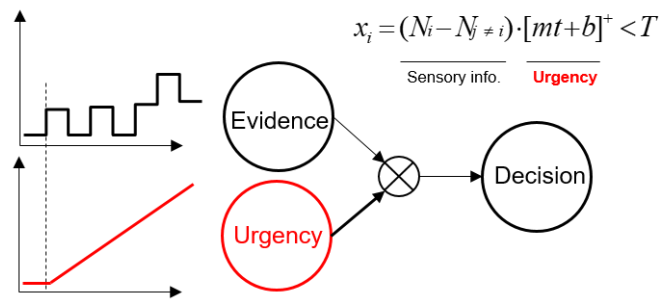

**Figure 7** : Le modèle d'urgence (*urgency-gating model*), dans lequel l'information sensorielle (*Evidence*) est multipliée par un signal croissant lié à l'urgence de répondre (*Urgency*). Le produit de ces deux signaux provoque une augmentation d'activité jusqu'à ce qu'un seuil fixe de décision soit franchi (*Decision*).

Pour répondre à ces questions de manière empirique, nous avons d'abord mené une étude comportementale chez l'humain dans laquelle nous avons trouvé des preuves solides en faveur du modèle d'urgence. Les sujets devaient exécuter une version modifiée de la fameuse tâche consistant à déterminer la direction nette décrite par un ensemble de points évoluant entre autres par des mouvements aléatoires (*random-dot motion discrimination task*) (Britten et al., 1992). Dans notre paradigme, la direction décrite par les points évoluait par paliers, toutes les 200ms, lors de chaque essai et les sujets devaient prédire quelle serait la direction finale décrite par l'ensemble de ces points avec des déplacements non aléatoires à la fin de l'essai. Contrairement aux prédictions des modèles « d'accumulation » de la prise de décision, nous avons observé que les sujets employaient une stratégie consistant à évaluer rapidement chaque nouvel état de l'information visuelle fournie **sans l'accumuler dans le temps**, et à combiner cette information « instantanée » avec un signal d'urgence permettant au système d'atteindre le seuil de décision.

Les arguments expérimentaux et théoriques de cette étude ont été publiés dans *Journal of Neurophysiology* en 2012 (Thura et al., 2012).

Ensuite, nous avons mené une étude électrophysiologique chez deux singes macaques entraînés à effectuer une tâche de prise décision (la tâche des jetons, Figure 8A) conceptuellement similaire à celle décrite ci-dessus et au cours de laquelle les animaux devaient choisir l'une des deux options proposées sur la base d'indices visuels (les jetons) évoluant au cours de chaque essai, puis effectuer des mouvements d'atteinte pour exprimer leurs choix.

Nous avons d'abord montré que le comportement des singes était comparable à celui des humains engagés dans des tâches similaires, notamment l'observation selon laquelle la **quantité d'information utilisée par les animaux pour prendre une décision diminuait avec la durée de la délibération**, résultat compatible avec un mécanisme lors duquel le critère d'exigence du sujet diminue avec le temps qui passe pendant la prise de décision, et **suggérant la présence d'un signal d'urgence** poussant les sujets à choisir dans des situations d'incertitude.

Nous avons ensuite enregistré l'activité de neurones isolés dans le cortex prémoteur dorsal (PMd) et moteur primaire (M1) de ces mêmes animaux, et nous avons montré que les deux régions étaient fortement impliquées à la fois dans la formation de la décision (l'activité de ces neurones reflète l'évolution des indices visuels présentés et utilisés par le singe lors de la délibération, Figure 8B) et dans l'engagement dans le choix (*commitment*), les neurones "signalant" la fin du processus de délibération en atteignant un seuil d'activité relativement fixe, quelles que soient la durée et la difficulté de la décision (Figure 8C). Nous avons également constaté que chacune **des deux régions combinait les informations sensorielles avec un signal lié au temps qui s'écoule pendant la délibération**, que nous interprétons comme étant de l'**urgence**, comme prédit par le modèle du même nom.

Ces données font l'objet d'un article publié en 2014 dans *Neuron* (Thura and Cisek, 2014).

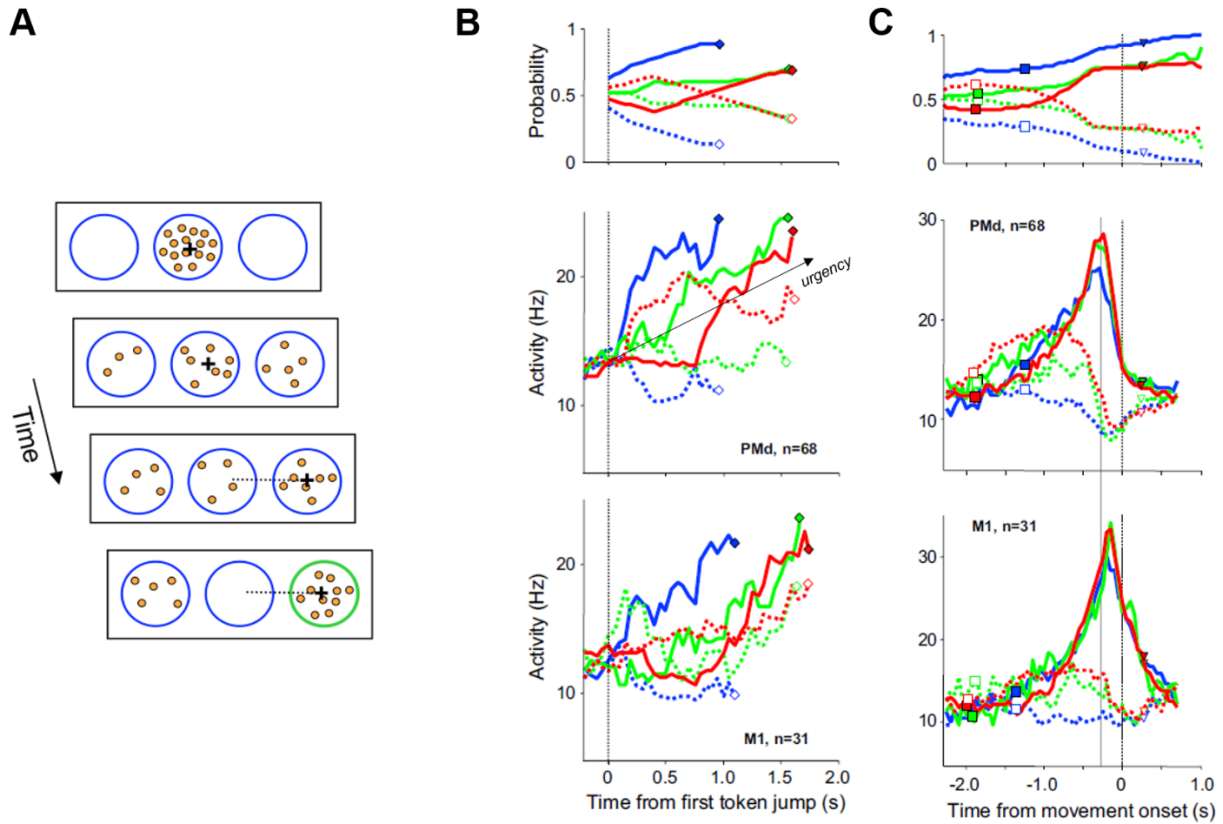

**Figure 8** : *A* : La tâche des jetons. Dans cette tâche, les singes regardent 15 jetons sautant un à un toutes les 200 ms d'un cercle central à l'un des deux cercles latéraux et doivent deviner quel cercle recevra la majorité des jetons avant qu'ils n'aient tous sautés. Ils expriment leurs choix en déplaçant avec leur bras un levier dans le cercle choisi. La décision peut être prise à tout moment, et lorsque l'un des deux cercles est atteint, les sauts du jeton s'accélèrent, permettant au singe de gagner du temps en devinant rapidement. *B* : Le panel du haut montre l'évolution de la probabilité de succès du cercle préféré des cellules, alignée sur le début de la délibération, lors d'essais faciles (tous les jetons ont tendance à sauter dans la cible correcte, bleus), ambigus (les 5-6 premiers jetons se répartissent entre les deux cibles, verts) et trompeurs (les 2-3 premiers jetons sautent dans la mauvaise cible, rouges), dans lesquels le singe a correctement choisi le cercle préféré (lignes pleines) ou le cercle opposé au cercle préféré des cellules (lignes pointillées). Le panel du milieu montre l'activité moyenne de 68 neurones de PMd alignée sur le premier saut de jeton (donc le début de la délibération) dans les trois types d'essais testés. La flèche noire illustre l'effet de l'urgence décisionnelle qui amplifie l'activité neuronale au cours du temps. À noter que ces cellules sont celles possédant une préférence spatiale au moment de l'initiation du mouvement. Le panel du bas montre l'activité moyenne de 31 cellules de M1 possédant également une direction préférée, toujours alignée sur le début de la délibération et dans les trois types d'essais testés, faciles, ambigus et trompeurs. *C* : Mêmes probabilités de succès (haut) et mêmes activités que dans le panel B mais alignées sur le début du mouvement exprimant le choix du singe. Adapté de Thura and Cisek, 2014.

Plus récemment, nous avons testé le rôle causal de ces cellules dans le processus de décision en utilisant la micro-stimulation électrique intra-corticale (ICMS) pendant que les singes réalisaient la tâche des jetons. Nous avons constaté **qu'une brève micro-stimulation** (dont l'intensité est inférieure au seuil de déclenchement des mouvements) **dans PMd ou M1 retarde le moment du choix** (*commitment*), avant l'exécution de l'action. Les mêmes stimulations dans une tâche contrôle nous ont permis de constater que la perturbation électrique n'affectait pas l'initiation de l'action elle-même. Le résultat très intéressant de cette étude est que la micro-stimulation avait un effet significatif uniquement lorsqu'elle était administrée **peu de temps avant que le choix de l'animal ne soit arrêté**. Ces résultats sont compatibles avec la proposition selon laquelle PMd et M1 participent (de façon causale) au processus de prise de décision. Nous interprétons l'effet retardateur de la micro-stimulation sur la prise de décision comme une conséquence de la perturbation de **la compétition dynamique entre les populations de cellules « votant » pour chacune des options** (Cisek, 2006b). En particulier, si la stimulation perturbe l'activité neuronale impliquée dans le processus de délibération **en ajoutant du bruit**, elle aura alors tendance à augmenter davantage l'activité du groupe de neurones « perdant » que l'activité du groupe « gagnant », car l'activité de ce dernier groupe, de fait plus élevée, est vraisemblablement aussi plus proche de la saturation. Cela aura pour conséquence de **réduire le contraste qui se développe entre les activités des deux groupes de cellules**, retardant ainsi la résolution de la compétition et augmentant la durée de la décision. Ceci ne peut avoir lieu que si les cellules sont déjà suffisamment influencées par l'information sensorielle, ce qui expliquerait que seules les stimulations délivrées peu de temps avant le choix aient eu une influence significative.

Ces données font l'objet d'une publication dans *Journal of Neurophysiology* parue en 2020 (Thura and Cisek, 2020).

L'ensemble de ces résultats comportementaux et neurophysiologiques suggère que les décisions entre les actions sont déterminées dans les zones prémotrices et motrices du cortex cérébral selon un mécanisme d'urgence, en contradiction avec les hypothèses communément admises à l'époque stipulant que lors de la prise de décision, une accumulation d'informations a lieu dans un "centre exécutif" unique situé dans le cortex préfrontal, et le résultat de ce calcul est transmis dans les centres moteurs chargés d'exécuter l'action sélectionnée au niveau préfrontal. Des enregistrements préliminaires dans le cortex préfrontal dorsolatéral (dlPFC) dans la tâche des jetons soutiennent cette hypothèse et montrent que le dlPFC fournit des informations sensorielles pertinentes pour la prise de décision, mais n'est pas impliqué dans la détermination du choix lui-même (Thura and Cisek, 2017).

### c) Travaux en tant que chercheur assistant (2013-2018)

De 2013 à 2018, j'ai travaillé dans le même laboratoire dirigé par Paul Cisek à l'Université de Montréal mais en tant qu'associé de recherche (le statut de post-doctorant étant limité à 5 années au Canada). Cette continuité m'a permis de poursuivre mes travaux sur les bases neurales de prise de décision entre actions et le rôle de l'urgence en testant l'hypothèse selon laquelle, lors d'une prise de décision dynamique, les singes **ajustent leur stratégie** (traduction imprécise de *policy*) **de compromis vitesse-précision** (*speed-accuracy tradeoff*, *SAT*) **en adaptant leur niveau d'urgence en fonction du contexte dans lequel la tâche est effectuée**.

Au cours de ces cinq années, nous avons mis en évidence des effets à la fois au niveau comportemental et au niveau neurophysiologique appuyant cette hypothèse. En termes de comportement, lorsque les animaux effectuent la tâche des jetons dans des blocs d'essais où ils ont la possibilité (sans instruction ni contrainte) **d'augmenter leur taux de récompense en décidant plus rapidement, la durée de la délibération est effectivement raccourcie** (Figure 9A). Bien que cela entraîne une légère diminution de la probabilité de choisir la bonne cible (échange de la précision contre du temps), ce taux de récompense est effectivement considérablement augmenté dans ces blocs « rapides ». De façon intéressante, ces données comportementales sont très bien prédites par le modèle d'urgence mentionné ci-dessus, **prédictions basées sur l'ajustement du niveau d'urgence décisionnelle dans les deux conditions**, à la fois en termes de niveau de base de l'urgence (plus élevé dans la condition favorisant la vitesse des choix) et au niveau de sa pente (plus faible dans cette même condition « rapide ») (Figure 9B).

C'est à cette époque nous avons découvert ce qui allait constituer dès lors mon thème de recherche principal pour les années suivantes (jusqu'à aujourd'hui). **Nous avons en effet observé que le même signal d'urgence influençant la prise de décision semble également moduler les propriétés des mouvements d'atteinte effectués par les animaux pour exprimer leur choix.** Ces mouvements sont en effet plus rapides après de longues délibérations, lorsque les informations sensorielles sont ambiguës et qu'un niveau d'urgence élevé est donc requis pour prendre la décision, par rapport aux décisions rapides basées sur des informations sensorielles fortes, sans nécessité d'urgence supplémentaire. Les mouvements d'atteinte sont également plus rapides dans la condition de SAT favorisant les choix rapides et risqués (Figure 9C). Il est particulièrement intéressant de noter que la vitesse des saccades oculaires effectuées au cours de la délibération était également modulée de la même manière (Figure 9D), ce qui suggère que le système oculomoteur et le système moteur du bras **reçoivent un signal d'urgence commun qui sert de source globale de vigueur pour la prise de décision et l'exécution des mouvements**. C'est ce que nous avons appelé **l'hypothèse de régulation unique de la décision et de l'action par l'urgence** (Figure 9E).

Conformément aux observations comportementales décrites ci-dessus, nous avons démontré qu'au cours du processus de délibération dans la tâche des jetons, l'activité de base et le gain d'une population importante de **cellules de PMd et de M1 liées à la décision étaient plus élevés lorsque les singes adoptaient une stratégie hâtive de compromis vitesse-précision** (Figure 9F). D'autres cellules présentaient une activité qui augmentait ou diminuait avec le temps jusqu'au moment où le choix était entériné. Concernant les neurones liés au mouvement, que ce soit dans PMd ou dans M1, ils étaient souvent plus actifs après des décisions plus longues, comme s'ils exprimaient l'influence du même signal d'urgence contrôlant le gain d'une activité liée à la décision.

Les données comportementales étayant ces découvertes importantes sont décrites en détails dans un article publié dans *Journal of Neuroscience* en 2014 (Thura et al., 2014), et leurs corrélats neuronaux font l'objet d'un article publié en 2016 dans le même journal (Thura and Cisek, 2016b).

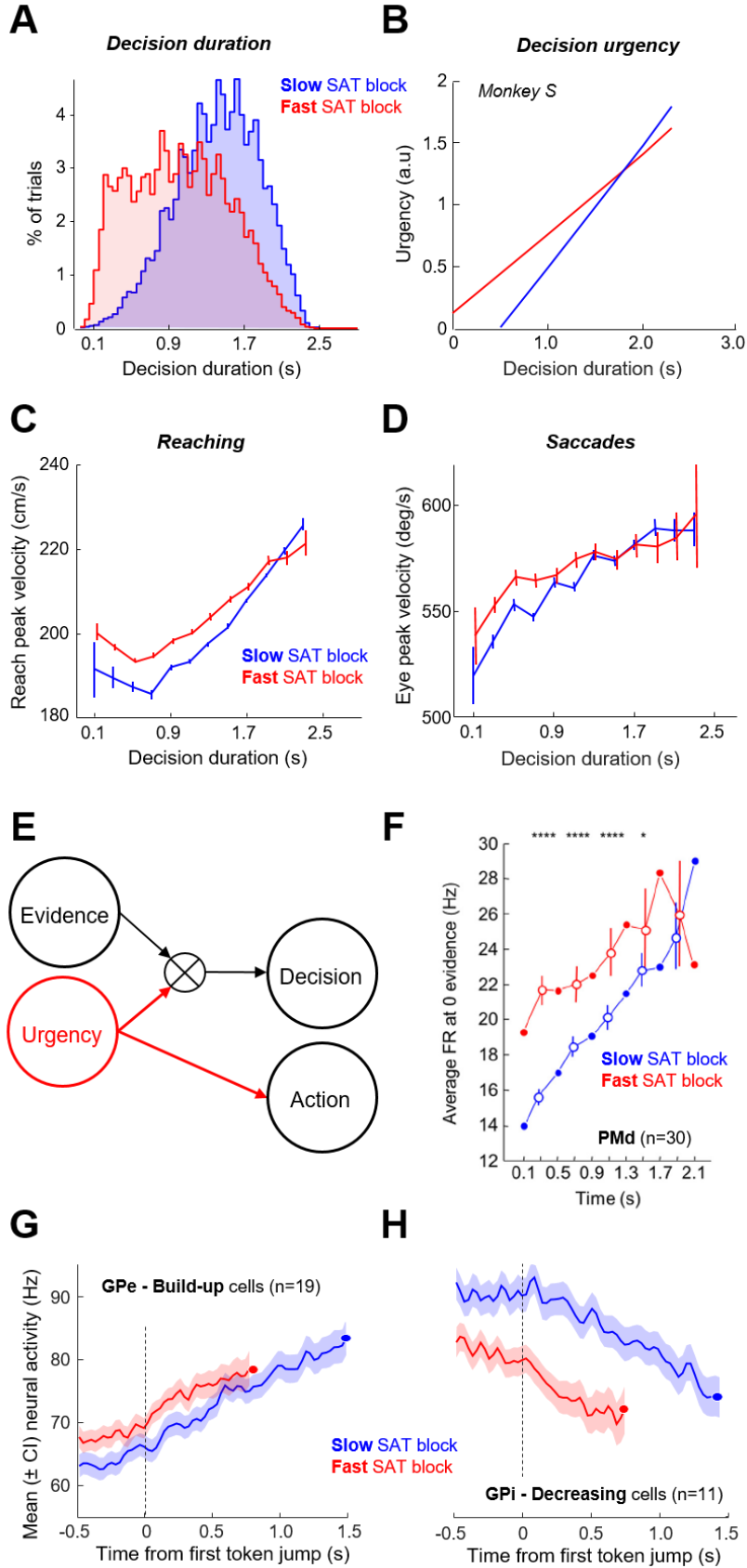

**Figure 9** : Observations comportementales et corrélats neuronaux de l'ajustement du compromis vitesse-précision (SAT) pendant la prise de décision. ***A*** : Distribution des durées de décisions d'un singe exécutant la tâche des jetons dans les blocs d'essais favorisant la vitesse (rouge) ou la précision (bleu) des choix. ***B*** : Signaux d'urgence dans les deux même blocs d'essais (rapide en rouge, précision en bleu) estimés sur la base du comportement du singe par le modèle d'urgence. ***C*** : Pic de vitesse des mouvements d'atteinte exécutés par un singe dans la tâche des jetons pour exprimer ses choix en fonction de la durée des décisions (axe des abscisses) et de la condition de SAT dans laquelle la tâche est effectuée, favorisant la vitesse (rouge) ou la précision (bleu) des choix. ***D*** : Pic de vitesse des saccades oculaires exécutées par le même singe dans la tâche des jetons en fonction de la durée des décisions (axe des abscisses) et des mêmes conditions de SAT. ***E*** : Modèle illustrant l'hypothèse de régulation unique de la durée des décisions et de la vitesse des mouvements par l'urgence. ***F*** : Activité neuronale moyennée de 30 cellules de PMd impliquées dans la prise de décision, alignée sur le début de la délibération, en fonction de la durée de la délibération (axe des abscisses) et de la condition de SAT dans laquelle la tâche est effectuée (vitesse, rouge ou précision, bleu). Pour se prémunir de l'effet de l'information sensorielle sur ces réponses, seules l'activité observée quand l'information sensorielle est égale pour les deux cibles potentielles (autant de jetons présents dans une cible que dans l'autre) est considérée. ***G*** : Réponse moyenne d'une population de 19 cellules du globus pallidus externe alignée sur le début de la délibération en fonction de la condition de SAT dans laquelle la tâche est effectuée. ***H*** : Identique à G pour une population de 11 cellules du globus pallidus interne. Modifiés de Thura et al., 2014 (A-D); Thura and Cisek, 2016b (F), 2017 (G-H).

Plus récemment, mes travaux se sont focalisés sur **l'identification de la source du signal d'urgence dans le cerveau**. Alors que de nombreuses aires corticales et sous-corticales pouvaient être envisagées, notre choix s'est porté sur **les ganglions de la base** (ou noyaux gris centraux), un ensemble de noyaux sous-corticaux dont le rôle dans la prise de décision, l'apprentissage par renforcement et le contrôle moteur est suspecté depuis longtemps (Redgrave et al., 1999; Schultz, 2006; Turner and Desmurget, 2010), même si une hypothèse unifiée sur leur(s) rôle(s) n'a toujours pas été proposée à ce jour. Nous avons enregistré l'activité de neurones isolés dans deux noyaux des ganglions de la base, le globus pallidus externe (GPe) et le globus pallidus interne (GPi), tandis que des singes effectuaient la tâche des jetons dans les deux conditions de compromis vitesse-précision, l'une encourageant la vitesse et la prise de risque lors des choix et l'autre le conservatisme et la précision.

Nos données montrent que contrairement aux régions corticales pré-motrices et motrices, **les neurones pallidaux (GPe et GPi) ne contribuent pas à la sélection de la cible.** En revanche, leur activité varient de façon monotonique (en augmentant ou diminuant) pendant la délibération, **reflétant l'urgence croissante de s'engager dans un choix.** De façon intéressante, cette activité variant au cours du choix était également **modulée en fonction de la condition de SAT adoptée par les animaux.** Dans les blocs d'essais durant lesquels les singes favorisaient la vitesse et la prise de risque pendant la délibération, l'activité des neurones pallidaux qui augmentait pendant le choix était amplifiée par rapports aux blocs d'essais marqués par des décisions plus lentes et plus prudentes (Figure 9G). L'effet inverse était observé pour les cellules qui diminuaient leur activité pendant la prise de décision (Figure 9H). Une fois la cible sélectionnée dans les régions corticales, **l'activité du globus pallidus confirmerait l'engagement et envigorerait[4] le mouvement effectué par les animaux** pour exprimer leur choix. Ensemble, ces données sont compatibles avec l'hypothèse selon laquelle ces cellules des ganglions de la base **fournissent un**

---

[4] Traduction délicate de *invigorate*, utilisé ici pour signifier la détermination du niveau de vigueur du mouvement.

**signal permettant de contrôler la règle de compromis vitesse-précision adoptée par les animaux en variant le niveau d'urgence décisionnelle pendant la délibération**.

Ces résultats importants ont fait l'objet d'une publication dans la revue *Neuron* en 2017 (Thura and Cisek, 2017).

Enfin, nous avons étudié les corrélats neuronaux **des ajustements locaux du SAT au cours de décisions successives**. En effet, outre des ajustements globaux en fonction d'un contexte général (un bloc d'essais au cours duquel rapidité et risque s'avèrent payants par exemple), la stratégie optimale d'un sujet engagé dans une série de décisions peut évoluer sur une plus **petite échelle de temps, d'une décision à l'autre**, sur la base du résultat de ces décisions dans un essai donné. Un exemple d'un tel ajustement local est connu sous le nom de **ralentissement post-erreur** (*post-error slowing*) (Danielmeier and Ullsperger, 2011). Nous avons constaté, toujours dans la tâche des jetons, que les singes adoptaient cette stratégie et que l'activité des neurones de PMd et de M1 était compatible avec ce mécanisme. En particulier, dans le cortex prémoteur dorsal, 23 % des cellules présentaient **une activité de base significativement plus faible après des essais lors desquels le singe avait commis une erreur de décision**, et pour environ 30 % d'entre elles, cette baisse d'activité persistait pendant la totalité de la période de délibération. De façon intéressante, ces cellules contribuaient au processus décisionnel en combinant les informations sensorielles liées aux sauts de jetons avec l'urgence croissante de s'engager dans un choix. Nous avons également observé que **l'activité de 22 % des cellules PMd augmentait après des erreurs décisionnelles**. Contrairement aux neurones précédemment cités, ces neurones augmentant leur activité après une erreur n'étaient pas impliqués dans la prise de décision, suggérant une rôle fonctionnel différent pour ces deux populations de neurones modulées de façon opposée en fonction de la performance décisionnelle dans l'essai précédent (décision et possiblement inhibition, respectivement). Des résultats similaires ont été observés pour les cellules de M1. Ces données suggèrent que **PMd et M1 appartiennent à un réseau d'aires cérébrales** (voir aussi Purcell and Kiani, 2016) **impliquées dans les ajustements locaux de compromis vitesse-précision,** établis d'un essai à l'autre à l'aide de l'histoire récente des gains de récompenses.

Ce travail est paru en 2017 dans *Journal of Neurophysiology* (Thura et al., 2017).

### d) Travaux en tant que chercheur statutaire (2018-présent)

Les travaux entrepris dans l'équipe Impact du Centre de Recherche en Neurosciences de Lyon en tant que chercheur statutaire indépendant à l'Inserm l'ont été avec pour objectif de mettre en œuvre le projet m'ayant permis **d'obtenir la bourse ATIP/Avenir du CRNS et de l'Inserm en 2017**. Ce projet constitue la suite directe des études menées à Montréal (décrites en sections II-3b et II-3c) et dont je rappelle brièvement le contexte ci-dessous.

Au cours de leur comportement naturel, les animaux (y compris les humains) décident et agissent pour optimiser ce qui leur tient le plus à cœur : **le taux de récompense** (Balci et al., 2011). Pour optimiser ce taux, les sujets ont naturellement tendance **à optimiser le compromis vitesse-précision** de leur comportement. Le programme ATIP/Avenir a été défini pour poursuivre l'étude des principes comportementaux et des mécanismes neuronaux du compromis vitesse-précision **des décisions et des actions**, dans le contexte d'un comportement interactif, **avec l'hypothèse selon laquelle décisions et actions sont des processus interdépendants**. À la fin de ma formation postdoctorale, j'ai proposé que **les ganglions de la base sont potentiellement impliqués dans le contrôle unifié de l'urgence décisionnelle et de la vigueur du mouvement**. Le programme découlait naturellement de ces travaux pour lesquels de nombreuses questions nécessitaient d'être approfondies.

i. <u>Expériences comportementales chez l'humain</u>

Mon travail en tant que postdoctorant indique que le niveau d'urgence au moment de la décision influence les caractéristiques des mouvements d'atteinte effectués **par les singes** pour exprimer leurs décisions. Bien que ces résultats impliquent un ajustement unifié et intégré du compromis vitesse-précision (SAT) lors du comportement dirigé vers un but, la tâche des jetons ne permettait pas une compréhension complète de ce phénomène. En particulier, elle n'a pas été élaborée pour faire **varier indépendamment les caractéristiques des décisions et des actions**. De plus, il fallait tester si cette observation importante était **vraie chez des sujets humains** qui ne subissent évidemment pas la même intensité d'entraînement que les singes. Enfin, **la réciprocité de cet effet devait être également testée** : les contraintes de mouvement affectent-elles également la façon dont les sujets prennent leurs décisions ?

Effet de l'urgence décisionnelle sur les caractéristiques des mouvements d'atteinte des humains

L'objectif des premiers travaux comportementaux du laboratoire était ainsi de tester l'hypothèse selon laquelle la régulation commune de la décision et de l'action par l'urgence, précédemment observée chez des singes « experts », **n'est pas spécifique à cette espèce, ni une conséquence du surentraînement des animaux**. Si tel n'est pas le cas, les sujets humains sains naïfs adopteront une stratégie de décision basée sur l'urgence dès qu'ils effectueront leur première session expérimentale et cette stratégie décisionnelle influencera la manière dont ils expriment leurs choix via des mouvements d'atteinte. De par son caractère supposément global, l'hypothèse prédit également que l'urgence de la décision impactera également les saccades oculaires.

Pour ce faire, j'ai mené une expérience dans laquelle 17 participant(e)s ont effectué une version modifiée de la tâche des jetons décrite plus haut (Figure 10A), tâche de prise de décision perceptive dans laquelle les informations sensorielles guidant le choix évoluent continuellement au cours d'un essai, et où les paramètres temporels peuvent être manipulés afin **d'encourager les ajustements du compromis vitesse-précision** (*speed-accuracy trade-off, SAT*) des décisions, permettant de tester l'effet de ce contexte de SAT sur **les prises de décision et les caractéristiques des mouvements**. Cette version de la tâche des jetons diffère de celle précédemment utilisée car les cibles des mouvements d'atteinte sont dissociées des cercles contenant les jetons, et les propriétés de ces cibles sont contrôlées de manière indépendante, permettant l'obtention de données cinématiques plus homogènes, et permettant de tester les effets des contraintes motrices sur la prise de décision (sujet de l'expérience suivante). Les participants ont effectué **au moins deux séances** au cours desquelles ils devaient réaliser un nombre donné d'essais corrects, les motivant indirectement à maximiser leur taux de succès (nombre d'essais corrects par unité de temps).

Les résultats reproduisent d'abord l'observation faite chez les singes selon laquelle une stratégie de décision basée sur l'urgence, dépendante du contexte, est adoptée par des sujets humains confrontés à des conditions changeantes. Les décisions rapides sont en effet basées sur une forte quantité d'information sensorielle (sans urgence), alors que les décisions longues sont prises avec peu d'informations pertinentes (ce qui est typiquement le cas lorsque les choix sont difficiles et ambigus), principalement sous l'effet de l'urgence qui croit au cours d'un essai (Figure 10B-C). Nous avons également observé plusieurs autres phénomènes importants : (1) la stratégie de décision au sein et entre les conditions de SAT **est affinée par l'expérience**, coïncidant avec une augmentation du taux de récompense avec l'entraînement ; (2) La **vigueur des mouvements des bras et des yeux augmente** systématiquement à **mesure que la durée de la décision augmente,** quelle que soit la condition de SAT (Figure 10D) ; (3) La stratégie de décision **dépendante de la condition de SAT** influence également la cinématique des mouvements d'atteinte, mais pas des saccades, des sujets humains, avec des **mouvements globalement plus vigoureux dans les blocs d'essais correspondant à la condition marquée par un haut niveau d'urgence décisionnelle** (Figure 10D-E). **De façon importante, ces effets ont été observés dès la première séance expérimentale effectuée par les sujets**.

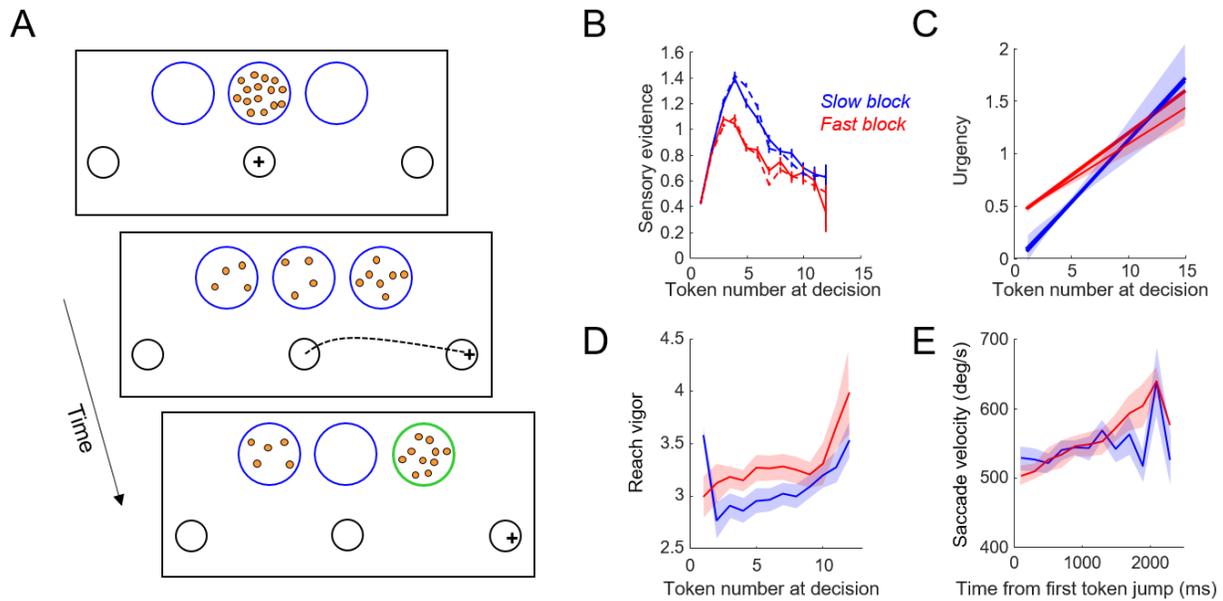

**Figure 10** : Impact de l'urgence décisionnelle sur la prise de décision et sur la vigueur des mouvements des humains sains. ***A*** : Décours temporel d'un essai dans la tâche des jetons. Les cercles bleus illustrent les stimuli de décision. Les jetons sautent successivement du cercle central vers l'un des deux cercles latéraux. Les cercles noirs illustrent les cibles des mouvements d'atteinte. Les sujets déplacent un levier (croix) d'un cercle central de « départ » vers l'une des deux cibles latérales, selon leur choix. ***B*** : Quantité d'information sensorielle (basée sur la différence du nombre de jetons présents dans les deux cibles) moyenne (± erreur standard, SE) utilisée par les sujets au moment de prendre leurs décisions, en fonction de la durée des décisions (axe des abscisses) et de la condition de SAT (bleu, précision ; rouge, vitesse). Les lignes solides montrent les données, les lignes pointillées les prédictions du modèle d'urgence. ***C*** : Signaux d'urgence moyens (± intervalles de confiance) dans les deux mêmes blocs d'essais (rapide en rouge, précision en bleu) estimés par le modèle d'urgence sur la base du comportement des sujets. ***D*** : Vigueur (vitesse normalisée en fonction de l'amplitude) moyenne (± SE) des mouvements d'atteinte des sujets en fonction de la durée des décisions (axe des abscisses) et de la condition de SAT. ***E*** : Vitesse moyenne (± SE) des saccades oculaires des sujets en fonction de la durée des décisions (axe des abscisses) et de la condition de SAT. Adapté de Thura, 2020.

Ces premiers résultats du laboratoire indiquent qu'un pan de l'hypothèse de la « régulation partagée » **existe chez les sujets humains**, et n'est pas **une conséquence du surentraînement** (des singes). J'ai proposé à l'époque que **l'urgence pourrait constituer le lien mécanistique** permettant d'établir un **contrôle intégré** de la durée des décisions et de la vigueur de l'action, afin d'optimiser le taux de récompense.

Cette étude a été publiée dans *Behavioural Brain Research* en 2020.

### Effet de la contrainte motrice sur la prise de décision chez l'humain

L'impact de la prise de décision sur l'action plaide fortement en faveur d'un mécanisme dans lequel un signal de régulation commun, l'urgence, module à la fois la durée des choix et la vitesse des mouvements. Cependant, afin de valider cette hypothèse de façon exhaustive, il était nécessaire de tester sa composante **réciproque**, à savoir **l'impact de la vigueur du mouvement sur la durée et la précision des décisions** (Figure 11A). En effet, si un signal unique régule à la fois les choix et les mouvements, nous devrions

- 34 -

observer que si un contexte particulier encourage la baisse de la vitesse des mouvements d'atteinte (sous l'effet d'une contrainte quelconque), l'urgence décisionnelle suivra cette baisse de vigueur motrice et diminuera également, conduisant à des durées de décision augmentées avant l'exécution de ces mouvements.

Pour tester cette hypothèse, vingt sujets humains (parmi eux, 17 étaient les sujets de la première étude décrite ci-dessus) ont effectué la même tâche de décision probabiliste que celle décrite sur la figure 10A, dans laquelle des choix perceptuels étaient exprimés en exécutant des mouvements d'atteinte vers des cibles dont la taille et la distance par rapport à une position de départ variaient dans des blocs d'essais distincts (petites vs grandes et proches vs éloignées, Figure 11B). En supposant que ces variations de taille et de distance allaient entrainer des variations de vigueur motrice (ce qui fut le cas, Figure 11C), ce *design* devait nous permettre d'**évaluer l'impact de cette vigueur motrice sur la stratégie de décision des sujets humains**.

Les résultats de cette analyse indiquent une **influence importante du contexte moteur sur la stratégie de prise de décision** de la plupart des sujets. Mais, **contrairement aux prédictions de l'hypothèse de la « régulation partagée »**, nous avons observé que **les mouvements lents exécutés dans les blocs moteurs les plus exigeants en termes de précision** (les conditions de petites cibles) **étaient souvent précédées de décisions plus rapides** (Figure 11D), prises avec une quantité d'information moindre (Figure 11E) et donc globalement légèrement moins précises (Figure 11F) par rapport aux blocs d'essais dans lesquels de grandes cibles permettaient d'exprimer des choix avec des mouvements rapides.

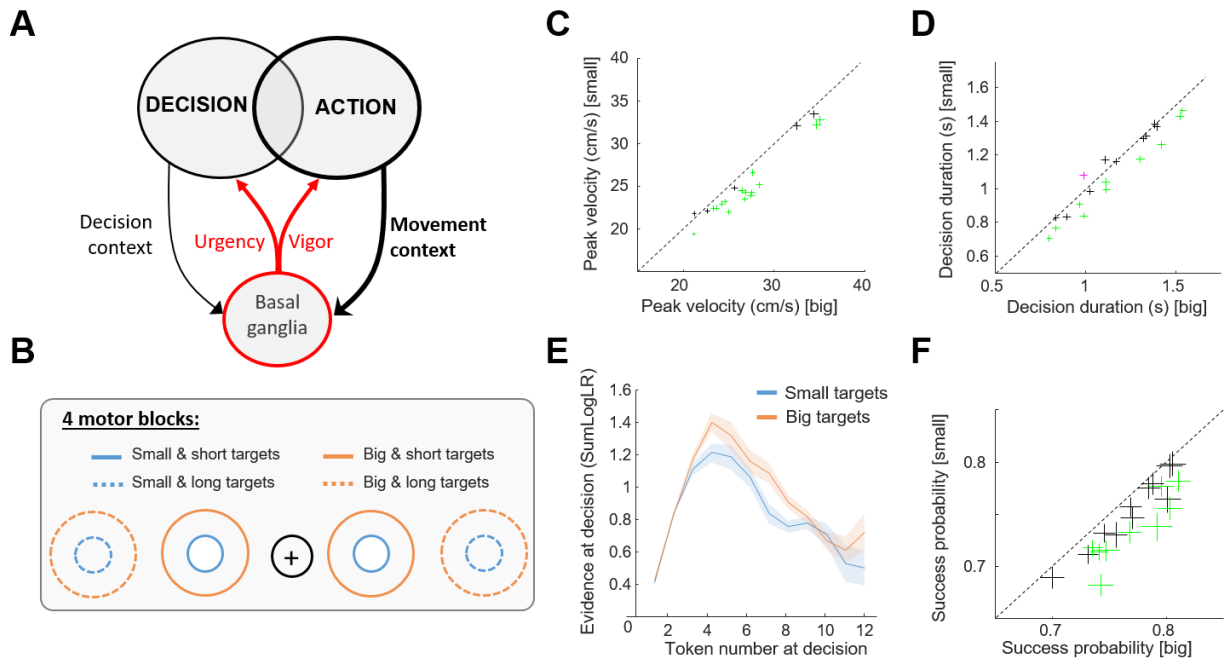

**Figure 11** : Influence de la vigueur motrice sur la prise de décision. ***A*** : Effet du contexte moteur sur la prise de décision (flèche noire épaisse) dans le cadre de l'hypothèse de régulation partagée de la décision et de l'action. ***B*** : Conditions motrices testées. Les sujets humains devaient exprimer leurs choix (cercles de décisions et jetons non illustrés ici) en déplaçant le levier d'un cercle central (noir) vers des cibles latérales dont la taille et la distance par rapport au cercle de départ variaient (cercles larges ou petits, éloignés ou pas du cercle central) dans des blocs d'essais dédiés. ***C*** : Pic de vitesse moyen (± SE) des mouvements d'atteinte exécutés par les 20 sujets humains pour exprimer leurs choix dans la condition « cibles larges » (abscisses) et « cibles petites » (ordonnées). Les croix colorées illustrent les sujets pour lesquels la différence de vitesse entre les deux conditions est significative. ***D*** : Durée moyenne (± SE) des décisions des 20 sujets humains dans la condition « cibles larges » (abscisses) et « cibles petites » (ordonnées). Mêmes conventions que pour C. ***E*** : Quantité d'information moyenne utilisée par les sujets pour prendre leurs décisions en fonction de la durée de ces choix et de la condition motrice testée (petites ou larges cibles). ***F*** : Probabilité de succès moyenne (± SE) des décisions des 20 sujets humains dans la condition « cibles larges » (abscisses) et « cibles petites » (ordonnées). Mêmes conventions que pour C. Adapté de Reynaud et al., 2020.

Ces résultats suggèrent que la prise de décision et le contrôle moteur **ne sont finalement pas régulés par un signal d'envigoration unique** et partagé déterminant à la fois l'urgence de décision et la vigueur de l'action, mais plus probablement par **des signaux d'urgence décisionnelle et de vigueur motrice indépendants**, mais **en interaction les uns avec les autres**.

Cette étude a été publiée dans *Journal of Neurophysiology* en 2020.

Avec ces deux études, j'ai répondu aux questions soulevées dans le volet comportemental de mon projet ATIP/Avenir 2017. Dans le paragraphe suivant, je décris la suite logique de ces expérimentations menées sur des sujets humains.

Effet de la durée des mouvements et de leur cout énergétique sur la prise de décision des humains

Dans l'étude décrite ci-dessus, les sujets prennent des décisions plus rapidement dans les blocs d'essais où les mouvements de réponse sont plus contraignants, donc plus longs (c'est-à-dire moins vigoureux). On peut interpréter cette observation en s'appuyant sur deux **principes d'optimisation économique**, qui ne s'excluent pas mutuellement.

- Il est d'un côté possible que les sujets aient raccourci leurs décisions dans le contexte moteur nécessitant le plus de contrôle et d'effort (celui des petites cibles) **pour prioriser et investir la quantité d'énergie nécessaire à l'exécution de ces mouvements plus exigeants**. Comme le titre de l'article l'indique, les sujets auraient en quelque sorte « **sacrifié** » leurs décisions aux bénéfices de leurs actions.

- Une autre possibilité est que les sujets aient cherché **à limiter la dévaluation temporelle** d'un essai correct **en raccourcissant les décisions** précédant l'exécution des mouvements les plus précis et donc **les plus longs**.

Le protocole de cette étude ne permet pas de trancher entre ces deux interprétations. En effet, la manipulation de la vigueur des mouvements module à la fois la durée de ces mouvements et la dépense énergétique associée. Il était donc important de réaliser **une seconde étude permettant de manipuler spécifiquement le coût temporel et le coût énergétique des mouvements dans des blocs d'essais distincts,** ceci afin de tester leur impact respectif sur la prise de décision des sujets.

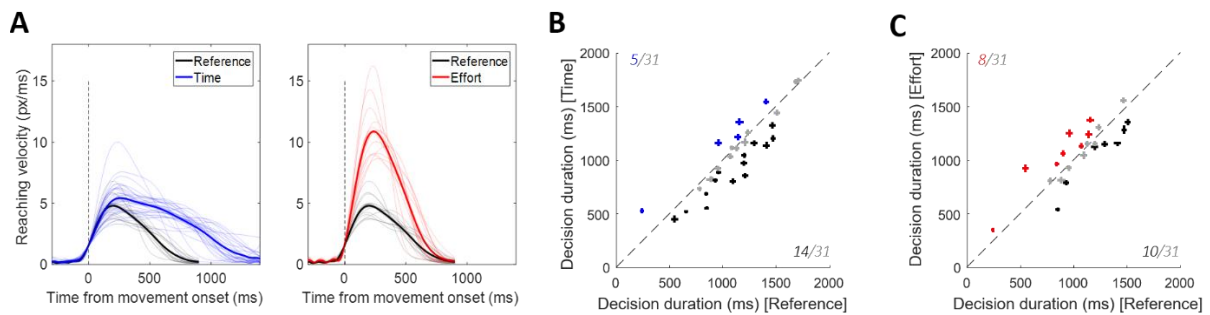

**Figure 12** : Effet de la durée et du cout énergétique des mouvements d'atteinte sur la durée des prises de décision des sujets humains. ***A*** : Comparaison des profils de vitesse des mouvements d'atteinte des sujets dans les conditions « Référence » et « Temps » (gauche) et dans les conditions « Référence » et « Effort » (droite). Les courbes fines illustrent les données individuelles et les courbes épaisses la moyenne pour tous les sujets. Les mouvements sont deux fois plus longs dans la condition « Temps » que dans la condition « Référence », pour un coût énergétique équivalent. Les mouvements sont deux fois plus coûteux énergétiquement dans la condition « Effort » par rapport à la condition « Référence », pour une durée globalement similaire. ***B*** : Comparaison des durées de décision moyennes (± SE) des sujets dans les conditions « Référence » (abscisses) et « Temps » (ordonnées). Les croix colorées illustrent les cas pour lesquels la différence est significative. ***C*** : Similaire à B mais pour la condition « Effort » comparée à la condition « Référence ». Adapté de Saleri Lunazzi et al., 2021.

Dans cette nouvelle étude, la même tâche décisionnelle était donc réalisée (la tâche des jetons) par 31 sujets humains sains, mais ces décisions pouvaient être exprimées dans trois conditions motrices distinctes : une condition **« Référence »** dans laquelle les sujets exécutaient les mouvements exprimant leurs choix à **leur vitesse spontanée**, une condition « **Temps** » dans laquelle les sujets devaient exécuter des mouvements **deux fois plus longs** qu'en condition « Référence » (pour une dépense énergétique comparable) et une condition « **Effort** » dans laquelle les sujets devaient effectuer des mouvements deux fois plus coûteux énergétiquement qu'en condition « Référence » (pour une même durée) (Figure 12A).

Le résultat principal de cette étude est qu'**environ la moitié des sujets ont réduit leurs durées de décision dans la condition « Temps »**, caractérisée par l'exécution de mouvements très longs (Figure 12B), par rapport à la condition « Référence ». En revanche, l'impact du coût énergétique des actions sur les durées de décision était très mitigé parmi la population testée (32% des sujets ont raccourci leurs décisions, 26% les ont allongées, toujours par rapport à la condition de référence) (Figure 12C). La précision des décisions était globalement similaire dans toutes les conditions motrices.

Ces résultats indiquent que les ajustements décisionnels les plus marqués, c'est-à-dire des raccourcissements de la durée des décisions, ont lieu dans le contexte moteur imposant un coût temporel important aux mouvements, suggérant que **le temps est le coût considéré comme le plus pénalisant par les sujets ayant effectué cette tâche**. Nous avons interprété ce résultat en proposant que les sujets compensent la durée importante de leur mouvement en raccourcissant celui de leur décision, illustrant **le caractère très intégré du contrôle du comportement**, et ceci afin de limiter la dévaluation temporelle de la récompense et la baisse du taux de succès.

> Les résultats de cette étude ont été publiés dans *Frontiers in Human Neuroscience* en 2021.

Comme souvent, de nouvelles questions ont débouché de cette étude. Notamment :

1- Est-ce que le contrôle intégré des décisions et des actions, permettant la compensation temporelle de l'un ou l'autre des processus, ne peut se mettre en place que dans des **blocs de plusieurs dizaines d'essais**, pour possiblement optimiser le taux de récompense à **l'échelle globale d'une session** ? Ou peut-il également s'observer **plus localement**, **d'un essai à l'autre**, sur la base de la performance comportementale dans un essai donné, pour possiblement aussi limiter la dévaluation temporelle **d'un succès immédiat** ?

2- Est-il possible que la compensation principalement temporelle que nous avons observée dans l'étude décrite ci-dessus soit **le résultat de la structure et du design de la tâche**, permettant plus **facilement les « échanges » temporels** entre processus décisionnels et moteurs, que les **échanges d'effort** ?

Dans les paragraphes suivants, je décris brièvement une analyse et une étude entreprise pour répondre à ces deux questions.

### Impact de la performance comportementale sur la prise de décision et l'action dans l'essai suivant

Pour répondre à la première question, nous avons exploité les jeux de données des deux études décrites ci-dessus (Reynaud et al., 2020; Saleri Lunazzi et al., 2021) avec pour objectif d'analyser les ajustements comportementaux **des sujets d'un essai à l'autre**, ceci afin d'approfondir davantage notre compréhension des mécanismes régissant les interactions entre la décision et l'action.

Pour cette analyse, nous avons classé les essais en fonction de la performance décisionnelle et motrice des sujets dans ces essais (décisions correctes, erreurs de décision, mouvements correctement exécutés, mouvements trop rapides, trop lents), et nous avons quantifié l'ajustement de la durée des décisions et de la vitesse des mouvements entre ces essais et l'essai suivant. Par exemple, on sait qu'après une erreur de décision, le choix fait dans l'essai suivant est plus long (pour une même difficulté). C'est le ralentissement post-erreur évoqué plus haut. Mais nous nous sommes demandés si un tel ralentissement post-erreur des décisions entrainait aussi un ajustement des mouvements exécutés pour exprimer ces choix plus lents. De la même manière, si un mouvement est trop long dans un essai donné, le sujet va l'exécuter plus rapidement dans l'essai suivant. On voulait savoir si cette augmentation de la vitesse des mouvements d'un essai à l'autre était aussi accompagnée d'une modification locale de la prise de décision.

Les résultats de cette analyse indiquent qu'après une mauvaise décision, les sujets décident plus lentement et avec plus de précision dans l'essai suivant, conformément au phénomène de ralentissement post-erreur, mais uniquement lorsque leurs mouvements ne sont pas contraints. Fait intéressant, ils raccourcissent également **la durée des mouvements** exprimant ces choix plus longs en augmentant leurs vitesses d'exécution. À l'inverse, nous avons constaté que les erreurs de mouvements contraints influençaient non seulement la vitesse et l'amplitude du mouvement suivant (comme attendu) **mais aussi la durée de la décision**. En effet, si le mouvement devait être ralenti dans l'essai suivant (suite à un mouvement trop rapide), la décision qui précédait ce mouvement plus lent était raccourcie, et vice versa.

Ensemble, ces résultats indiquent que **le contrôle intégré des choix et des mouvements peut intervenir localement, d'un essai à l'autre**, sur la base de la performance comportementale des sujets. La direction de ces ajustement locaux suggèrent qu'ils permettraient la limite de la dévaluation temporelle des succès immédiats. Les résultats confirment également que les humains **cherchent à déterminer une durée comportementale globale**, décision et action, plutôt que d'optimiser chacun des compromis vitesse-précision pendant la décision et l'action indépendamment l'un de l'autre.

Les résultats de cette étude ont été publiés dans *European Journal of Neuroscience* en 2023.

Pour les trois dernières études décrites dans ce mémoire (Reynaud et al., 2020; Saleri Lunazzi et al., 2021, 2023), j'ai supervisé le travail d'**Amélie Reynaud**, post-doctorante, et de **Clara Saleri**, alors étudiante en M2 puis doctorante à l'Université Claude Bernard Lyon 1.

### Partage des ressources énergétiques motrices et non-motrices chez l'humain

Les études récentes décrites ci-dessus indiquent que les décisions et les actions partagent des principes importants, dont **l'optimisation de leur durée** lorsque le contexte l'exige. Dans une étude récente dont je rapporte les principaux résultats ci-dessous, nous avons voulu tester l'hypothèse selon laquelle **la gestion** **des ressources énergétiques liées à l'effort est également partagée entre la décision et l'action**, et que la raison pour laquelle nous n'avons pas pu la mettre en évidence de façon robuste auparavant, notamment dans l'étude de 2021 publiée dans *Frontiers in Human Neuroscience* (Saleri Lunazzi et al., 2021), tenait en partie au *design* de la tâche comportementale utilisée, permettant plus facilement les échanges de durées entre décision et mouvement que les échanges d'effort entre ces deux même processus.

Des sujets humains en bonne santé ont effectué une nouvelle tâche de prise de décision perceptuelle dans laquelle ils devaient **choisir entre deux niveaux d'effort à investir dans la prise de décision** (c'est-à-dire deux niveaux de difficulté de perception, facile ou difficile) et exprimer leurs choix en exécutant un **mouvement d'atteinte** d'un cercle central vers l'une des deux cibles latérales présentées aux participants. Le point clé de cette expérience est que **l'exigence de précision des mouvements d'atteinte augmentait progressivement d'un essai à l'autre, et ce avec l'augmentation de la performance décisionnelle des participants**. En d'autres termes, en début de session, les cibles latérales étaient grandes, et les mouvements faciles à effectuer. Puis, plus les sujets enchainaient les bonnes décisions (ce qui leur permettait d'accumuler des points), plus la taille des cibles diminuait, jusqu'à devenir très petite et imposer des mouvements d'une très grande précision (effort moteur important lié au haut niveau de contrôle requis). Une autre caractéristique importante de cette tâche est qu'une **décision difficile correcte rapportait 5 points** (ou retirait 5 points en cas d'erreur) alors qu'une **décision facile correcte ne rapportait qu'un point** (ou retirait 1 point en cas d'erreur), et les sujets devaient accumuler 200 points pour terminer la séance.

L'hypothèse d'un contrôle intégré, non indépendant, des ressources énergétiques entre décision et action prédit qu'en début de séance, les sujets choisiront principalement les décisions difficiles (ils choisiront d'investir leurs ressources énergétiques sur la décision), celles qui leur permet d'amasser plus de points, car à ce stade de l'expérience les mouvements d'atteinte sont faciles et nécessitent donc peu d'effort. Puis, avec l'accumulation de points et **la diminution de la taille des cibles**, les participants choisiront de plus en plus souvent les décisions faciles, même si elles ne rapportent qu'un point, **car les ressources énergétiques seront progressivement transférées au système moteur**, et ce, au détriment du système décisionnel.

Contrairement à cette prédiction, les résultats indiquent **un impact globalement modéré de l'augmentation croissante de la difficulté motrice sur le choix de l'effort décisionnel** à investir dans chaque essai. De même, cette augmentation de l'effort moteur n'**a pas impacté de façon significative la performance décisionnelle** des participants testés. En revanche, nous avons mis en évidence que les **performances motrices étaient fortement dégradées** avec l'augmentation de l'effort moteur requis, ce qui était attendu, mais **également en fonction de la difficulté des décisions**. Ce résultat suggère que lorsque le sujet doit prendre une décision difficile, le besoin de ressources non-motrices pour effectuer cette tâche difficile impacte sa capacité à effectuer un mouvement précis pour exprimer ce choix. De façon assez impressionnante, cet effet a été observé quelle que soit la taille de la cible des mouvements d'atteinte.

Ensemble, les résultats soutiennent l'**hypothèse d'une gestion intégrée des ressources énergétiques liées à l'effort entre décision et action**. Ils suggèrent qu'à l'instar des ressources temporelles, ces **ressources énergétiques peuvent également être partagées entre la décision et l'action**, et que dans le cas particulier de cette étude, les ressources mutualisées sont principalement allouées au processus de décision au détriment des mouvements.

> Cette étude est décrite en détails dans un article publié dans *Scientific Reports* en 2023.

> Pour cette étude (Leroy et al., 2023), j'ai supervisé le travail d'**Elise Leroy**, alors étudiante en M2 à l'Université Lyon 2.

ii. <u>Expérience comportementale chez le singe</u>

L'un des objectifs majeurs de mon travail en tant que chercheur indépendant est de décrire **les bases neurales de la coordination entre la décision et l'action**, en particulier en examinant le fonctionnement de la **boucle sensorimotrice cortex-ganglions de la base chez le singe** à l'aide d'enregistrements électrophysiologiques. Cependant, comme illustré en section II-2c (Figure 4), afin de pouvoir inférer la neurophysiologie de la coordination décision/action chez l'humain en se basant sur les propriétés neuronales observées chez le singe, une étape importante consiste **à s'assurer que les singes macaques sont capables d'utiliser les mêmes stratégies de coordination que les humains**. En effet, bien qu'une étude ait montré chez le singe que le contexte décisionnel dans lequel des décisions étaient prises impactait les décisions et les actions (Thura et al., 2014), le mode de compensation temporel mis en évidence chez l'humain (Reynaud et al., 2020; Saleri Lunazzi et al., 2021, 2023) n'a quant à lui jamais été démontré chez le primate non-humain. Ainsi, il était nécessaire de mener les mêmes études chez les singes que celles décrites plus haut chez les humains, en testant notamment **l'effet du contexte moteur sur la stratégie de prise de décision**.

L'objectif de l'étude décrite dans ce chapitre était donc de tester **la réciproque de l'hypothèse de régulation partagée chez les singes**, tout comme nous l'avons fait chez les humains, en **manipulant le contexte moteur** dans lequel les mouvements exprimant les choix sont exécutés, pour observer les répercussions de ce contexte **sur la stratégie de prise de décision** des animaux. Pour cette étude, nous avons choisi de faire varier uniquement la taille des cibles des mouvements d'atteinte (permettant de faire varier la vigueur des mouvements en fonction du niveau de contrôle moteur requis), ce paramètre étant celui pour lequel nous avons observé le plus d'effets comportementaux chez les humains. De plus, nous avons décidé de ne pas tester le paradigme permettant de dissocier les coûts temporel et énergétique des mouvements chez les singes au vu de sa difficulté à être réalisé par nos sujets humains.

Dans un dispositif expérimental identique à celui utilisé par les sujets humains (Figure 13A), nous avons entraîné **deux singes macaques rhésus** (**Ghana**, singe G et **Bingo**, singe B) sur **la tâche des jetons** (Figure 13B) dans laquelle la taille des cibles à atteindre pour exprimer un choix variait dans deux blocs d'essais dédiés (grandes cibles ou petites cibles). Les animaux ont réalisé **plusieurs dizaines de sessions de cette tâche** ce qui, contrairement aux expériences menées chez les humains, nous a permis d'étudier **la stabilité ou l'évolution de la coordination entre décision et action au cours du temps**.

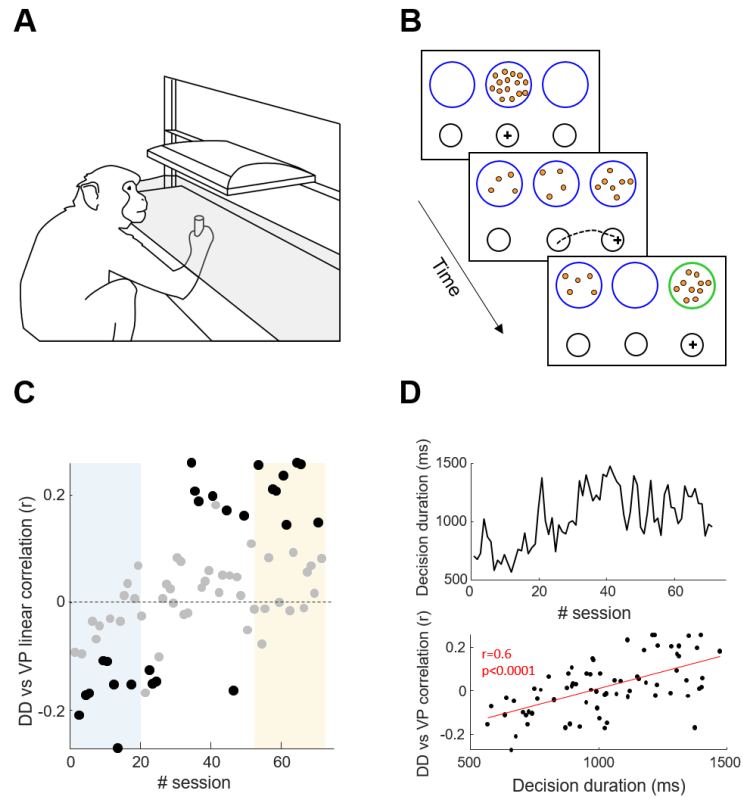

**Figure 13** : Coordination entre décisions et actions chez le singe macaque. *A* : Dispositif expérimental dans lequel les singes ont effectué la tâche des jetons. Les animaux déplacent un levier sur une table numérique en fonction des stimuli qu'ils voient projetés par un moniteur sur un miroir. Ce dispositif donne l'illusion que les stimuli sont projetés directement sur la table, au niveau du bras de l'animal. *B* : La tâche des jetons (voir Figure 10A). *C* : Evolution au cours des sessions effectuées (abscisse) de la corrélation essai par essai entre la durée des décisions et le pic de vitesse des mouvements d'atteinte exécutés par le singe G pour exprimer ses choix. Chaque point représente la corrélation calculée pour une session expérimentale. Une valeur de r négative indique que plus les décisions sont longues, plus les mouvements sont lents. A l'inverse, une corrélation positive témoigne d'une session durant laquelle des décisions longues ont été suivies de mouvements rapides. Les points noirs illustrent les cas pour lesquels la corrélation de Pearson est significative. La zone bleutée illustre les 20 premières sessions effectuées par le singe G, la zone jaune claire les 20 dernières sessions. DD : *Decision duration* ; VP : *Velocity peak*. *D* : Haut. Durée moyenne des décisions par session en fonction des différentes sessions exécutées par le singe G. Bas. Relation entre les corrélations durées de décision-vitesses de mouvement illustrées en C et les durées de décision moyennes illustrées sur le panel du haut. Adapté de Saleri and Thura, 2024.

A l'instar des analyses effectuées chez les humains (Reynaud et al., 2020), nous avons tout d'abord comparé les vitesses de mouvement et les durées de décision des singes entre les deux conditions de contraintes motrices testées, celle nécessitant un haut niveau de contrôle moteur (petites cibles) et celle dans laquelle les mouvements d'atteinte étaient moins contraints (grandes cibles).

En moyennant toutes les sessions effectuées par le **singe G**, lorsque les mouvements étaient correctement exécutés (les décisions pouvaient être correctes ou fausses), leurs **vitesses étaient globalement plus lentes** et leurs **durées plus longues** lorsqu'ils étaient exécutés vers **les petites cibles** que lorsqu'ils visaient les grandes cibles. Ceci était attendu et montrait que nous avons pu faire varier la vigueur des mouvements d'atteinte du singe dans les blocs d'essais caractérisés par la taille de la cible. Il était dès lors intéressant de regarder les conséquences de cet ajustement de la vigueur motrice sur les prises de décision du singe G. Ce faisant, nous avons constaté que **la durée des décisions** du singe G **était globalement plus lente lorsque les mouvements étaient dirigés vers de petites cibles, c'est-à-dire lorsque les mouvements étaient peu vigoureux,** que lorsqu'ils étaient dirigés vers les cibles plus grandes. **Cette observation, compatible avec l'hypothèse de régulation partagée, va à l'encontre de ce qui a été décrit chez l'humain**.

Concernant le singe B, l'analyse ne s'est pas relevée être particulièrement conclusive car la modulation de la vigueur des mouvements d'atteinte par la taille des cibles a été très peu efficace. En effet, la vitesse et la durée des mouvements exécutés par le singe B étaient très similaires que les cibles soient grandes ou petites. De façon peut être liée, nous avons également observé très peu de modulations au niveau décisionnel entre les deux conditions motrices pour cet animal.

Le résultat obtenu chez le singe G, des décisions longues précédant des mouvements lents et longs, pouvait potentiellement suggérer que les singes macaques rhésus, contrairement aux humains, ne sont pas capables d'adopter **un mécanisme de compensation temporelle** pour éventuellement limiter la baisse de leur taux de récompense (défini je le rappelle comme la probabilité de succès du comportement, moins les efforts demandés pour effectuer ce comportement, le tout divisé par le temps nécessaire à la réalisation du comportement). **Ceci pouvait être problématique en vue d'effectuer l'étude neurophysiologique sur ces animaux et proposer sur la base des données obtenues des mécanismes neuronaux présents chez l'humain**.

Cependant, nous avons exploité le fait que les singes réalisent des dizaines de sessions et l'échec de la modulation de la vigueur des mouvements dans les blocs d'essais chez le singe B pour étudier la coordination décision-action sur une autre échelle de temps, celui de l'essai unique, et l'évolution de la coordination à l'échelle de l'essai en fonction de l'expérience des animaux dans la tâche. Cette analyse revient simplement à calculer **une corrélation essai par essai dans chaque session entre la durée des décisions et la vitesse des mouvements** exécutés par les animaux pour exprimer leur choix.

Cette analyse nous a permis de mettre en évidence un résultat très intéressant et nous le pensons crucial pour comprendre les principes régissant la coordination entre la prise de décision et l'action. Pour les deux animaux, nous avons en effet observé que lors des premières sessions effectuées, la corrélation entre durée de décision et vitesse du mouvement était négative, indiquant que lorsque **les décisions des** **animaux étaient longues, les mouvements exécutés étaient lents** (quelle que soit la taille de la cible). Ceci semblait confirmer le résultat mis en évidence en analysant les essais par blocs. De façon très intéressante cependant, nous avons constaté chez le singe G que **cette relation évoluait avec l'expérience de l'animal dans la tâche**, et ce de façon quasi-linéaire au cours de l'entrainement (Figure 13C). Ainsi, les dernières sessions effectuées par le singe G étaient marquées par une corrélation positive entre durée de décision et vitesse de mouvement, indiquant **que les décisions longues étaient exprimées par des mouvements rapides**.

On s'est ensuite demandé **ce qui pouvait expliquer ce changement de stratégie** de coordination décision-action du singe G avec l'expérience dans la tâche. On a alors remarqué qu'au cours de l'entrainement, Ghana a **augmenté la durée moyenne de ses décisions**, surtout entre la première et la quarantième session (Figure 13D, haut). Ceci nous a conduit à interpréter le changement de stratégie, passage d'une co-régulation des décisions et des actions à une compensation temporelle, comme une façon pour le singe G de **limiter la perte de temps au niveau de l'essai due à l'augmentation de la durée de ses décisions**. En accord avec cette hypothèse, nous avons observé **une relation significative entre le niveau de corrélation décision-action décrite plus haut et la durée des décisions du singe G** (Figure 13D, bas). En résumé, plus les décisions du singe G étaient globalement longues (à l'échelle de la session), plus elle a adopté une stratégie de compensation temporelle de ses actions par rapport à ses décisions.

Concernant le singe B, nous avons observé que la corrélation entre durée de décision et vitesse du mouvement était globalement négative, sans réelle évolution au cours de l'entrainement. Ceci pouvait affaiblir le résultat décrit dans le paragraphe précèdent pour le singe G, mais il très intéressant de noter que le singe B avait un comportement général (décision et action) beaucoup plus impulsif que celui du singe G. Ses durées de décisions étaient notamment beaucoup plus courtes (249ms) que celles du singe G (1046ms). Notre interprétation de ce résultat est qu'en adoptant un **comportement global très vigoureux**, ce singe **n'a pas eu besoin, contrairement au singe G, de mettre en place un mécanisme de compensation** entre les deux processus pour limiter la baisse de son taux de récompense.

Dans l'ensemble, cette étude comportementale indique que les singes sont susceptibles d'adopter les mêmes modes de coordination (co-régulation et compensation) entre la décision et l'action que ceux observés chez l'humain. Ces modes varient en fonction du contexte dans lequel les sujets se comportent, et plus particulièrement de leur expérience dans la tâche. Les données issues de cette étude suggèrent aussi que le mode de co-régulation des décisions et des actions (décision lente => action lente ; décision rapide => action rapide) **serait le mécanisme de contrôle des deux processus présents "par défaut"** et que, dans certaines circonstances comme l'augmentation de la durée de l'un des deux processus, un mode de compensation est alors susceptible d'être mis en place.

Cette étude a été déposée sur le serveur bioRxiv en avril 2024 et l'article est actuellement en révision pour être publié dans *Journal of Neurophysiology*.

> Cette étude a été réalisée dans le cadre du doctorat de **Clara Saleri**, thèse défendue le 13 juin 2024 à Lyon.

### iii. Approche électrophysiologique chez le singe

Bien que **l'électrophysiologie chez le singe** soit l'axe majeur de mon travail en tant que chercheur statutaire, c'est l'approche qui a à ce jour produit le moins de résultats depuis mon retour de Montréal en 2018. Ceci s'explique par plusieurs raisons. Tout d'abord, **l'entrainement des animaux sur des tâches sensorimotrices complexes telles que celles utilisées dans le laboratoire est une étape extrêmement longue**, s'étendant typiquement sur plusieurs mois. Ensuite, **les dispositifs d'enregistrement de l'activité neuronale** que j'utilise depuis mon installation à Lyon **nécessite un développement important** et une fabrication sur mesure, afin de maximiser les chances de réussite des chirurgies d'implantation et de précision des localisations des électrodes dans plusieurs régions cérébrales cibles. Ce qui nous mène au défi majeur inhérent à cette approche, **les implantations chirurgicales et le maintien de l'hygiène des implants** sur les animaux sur des périodes de temps pouvant atteindre plusieurs années.

Mes collègues, **Amélie Reynaud**, **Clara Saleri** et **Gislène Gardechaux**, dont je tiens à saluer ici la dévotion et la persévérance, et moi-même, avons dû rencontrer à peu près toutes les difficultés que l'on peut imaginer au cours des cinq dernières années, ce qui a inévitablement entrainé un retard conséquent de ce pan de mes travaux. Comme d'autres, nous avons notamment perdu un singe entrainé peu de temps après la chirurgie d'implantation du système d'enregistrement électrophysiologique, en raison d'une infection probablement contractée au cours d'un soin. Au-delà du drame émotionnel que représente la perte d'un animal que vous avez côtoyé quasiment tous les jours pendant des années, cet accident a réduit à néant plus de deux années de travail… Mais heureusement, l'électrophysiologie chez le singe est aussi parfois **synonyme de joie**, notamment lorsque vient le moment d'enregistrer l'activité du tout premier neurone du laboratoire (page 3). Pour ma collègue Clara Saleri et moi, ce fut lors d'un matin d'octobre 2022 (le 13 pour être précis, soit plus de 4 ans après mon installation à Lyon…). Ci-dessous, je rappelle rapidement le contexte de cette première étude électrophysiologique du laboratoire, qui, étant toujours en cours de réalisation et donc non publiée, sera décrite elle aussi brièvement.

Ce travail correspondait au versant Électrophysiologie du projet ATIP/Avenir pour lequel j'ai intégré le laboratoire de Lyon en 2018, défini afin **d'étudier les bases neurales du contrôle intégré de la décision et l'action chez le singe**. Ce que l'on savait à l'époque est qu'il existait **un recouvrement anatomique important** entre des circuits, **hautement distribués par ailleurs**, impliqués dans la prise de décision et ceux impliqués dans le contrôle sensorimoteur (voir Cisek and Kalaska, 2010 pour une revue). De nombreuses études issues de plusieurs axes de recherche indiquaient en particulier que parmi ces circuits, **la boucle cortex sensorimoteur - ganglions de la base** joue probablement un rôle crucial dans la sélection de l'action (Redgrave et al., 1999; Pasquereau et al., 2007) ainsi que dans le contrôle des mouvements sélectionnés (Turner and Desmurget, 2010). Néanmoins, aucune étude n'avait déterminé le **rôle respectif de chaque structure au sein de cette boucle, ni comment elles permettent la coordination des décisions et des actions, et ni si et comment elles traitent les déterminants cruciaux des comportements dirigés vers un but, tels que la récompense et les coûts moteurs**.

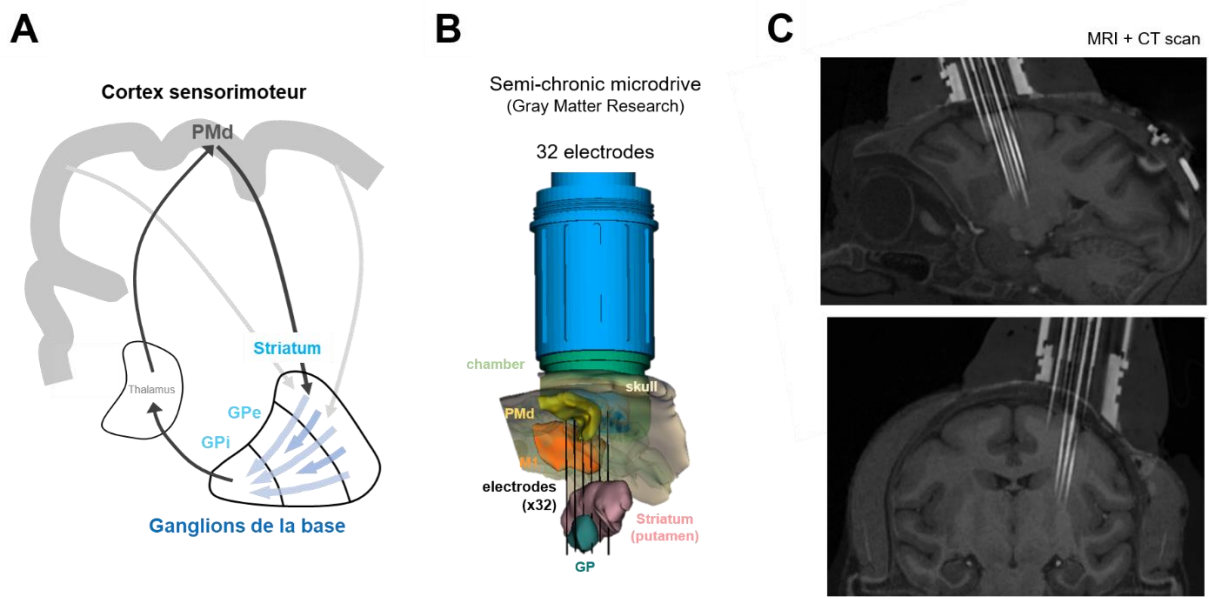

**Figure 14** : Étude de la neurophysiologie de la coordination décision-action dans le réseau cortex sensorimoteur – ganglions de la base du singe macaque. *A* : Schéma illustrant de façon simplifiée la boucle neuro-anatomique formée par le cortex prémoteur dorsal (PMd), les ganglions de la base (le striatum dorsal, ou putamen, globus pallidus externe, GPe et interne, GPi) et le thalamus (Alexander et al., 1986). *B* : Modélisation dans le logiciel 3DSlicer du système d'enregistrement électrophysiologique composé d'une chambre d'enregistrement (vert) et d'un micro-descendeur (bleu) contenant les 32 électrodes (noir), et des structures cérébrales cibles (PMd, striatum dorsal et globus pallidus, GP). *C* : Superposition (dans le même référentiel stéréotaxique) de l'IRM anatomique et du CT-scan du singe G permettant de vérifier la position des électrodes dans le cerveau.

L'objectif de ces travaux était donc d'initier la **description d'un modèle fonctionnel unifié de cette boucle cortex sensorimoteur - ganglions de la base** (Figure 14A) afin de comprendre si et, le cas échéant, comment elle contribue au contrôle intégré et à la coordination de la décision et de l'action en fonction du contexte dans lequel se déroule le comportement. Pour ce faire, l'approche que nous avons utilisée consiste à enregistrer, la plupart du temps **simultanément**, l'activité électrique unitaire de neurones **dans différentes structures clés de cette boucle** chez le singe macaque réalisant une tâche comportementale. Ces structures incluent le **cortex prémoteur dorsal (PMd)**, le **striatum dorsal** ainsi que **les parties externe et interne du globus pallidus (GPe et GPi)**. Nous utilisons un **micro-descendeur semi-chronique** contenant 32 microélectrodes, fabriqué sur mesures par Gray Matter Research (Dotson et al., 2017), et nous planifions la position de ce micro-descendeur à l'aide d'un logiciel, 3DSlicer (Fedorov et al., 2012), permettant de modéliser les implants et les structures cérébrales cibles sur la base de l'IRM anatomique de l'animal (Figure 14B). Ce système d'enregistrement, implanté à demeure sur l'animal, **permet de manipuler tous les jours les électrodes indépendamment les unes des autres**. Cette approche combine la flexibilité des enregistrements aigus, offrant la possibilité d'enregistrer l'activité de nouvelles cellules tous les jours, avec la stabilité, la rapidité et le confort pour l'animal des systèmes chroniques « *plug-and-play* ». Après quelques semaines ou mois d'enregistrements, **la position des électrodes dans le cerveau peut être vérifiée** en faisant passer à l'animal un CT-scan que l'on superpose ensuite à son IRM anatomique (Figure 14C). L'activité neuronale acquise dans chaque structure (activités unitaires et potentiels de champs locaux) est ensuite analysée et comparée entre différentes conditions expérimentales définies par la manipulation des paramètres de récompense et de coûts moteurs dans une nouvelle tâche décisionnelle et motrice, une tâche de *foraging* (voir ci-dessous). À noter que l'on fait également l'acquisition systématique des mouvements oculaires de l'animal testé (travaillant tête fixe) ainsi que de la taille de sa pupille pendant le déroulement de la tâche à l'aide d'un oculomètre vidéo à infra-rouge (EyeLink 1000, SR Research Ltd.).

Le passage de l'utilisation dans le laboratoire de la tâche des jetons à la tâche de *foraging* a été motivé par plusieurs raisons. Idéalement, nous aurions souhaité réaliser les enregistrements électrophysiologiques dans la boucle cortex-ganglions de la base pendant que nos singes exécutaient la tâche des jetons, afin d'étudier les bases neurales du comportement décrit ci-dessus (section II-3-d-ii), notamment les stratégies de coordination décision-action employées par les animaux. Malheureusement, il s'est avéré que notre singe implanté avec un plot de fixation de la tête, Ghana, **n'a jamais pu travailler sur la tâche des jetons en ayant la tête fixée** (les données comportementales décrites en section II-3-d-ii ont été acquises tête libre). La solution que nous avons trouvée pour remédier à ce problème fut de proposer à Ghana une nouvelle tâche comportementale a priori **plus simple à réaliser** pour elle. Nous avons donc entrepris de **transformer cette difficulté en opportunité** pour mettre au point une tâche qui répondait à ma volonté de tester les animaux sur des protocoles **se rapprochant au maximum de leur comportement naturel**. Ce faisant, nous espérions interroger ce pour quoi le cerveau a été conçu et ce pour quoi il a évolué, et augmenter ainsi nos chances de succès dans nos tentatives de mettre en lien l'activité neuronale observée avec le comportement de l'animal dans la tâche.

Nous avons donc développé cette nouvelle tâche basée sur le principe de *foraging*. Le *foraging* est un comportement très naturel que l'on retrouve chez toutes les espèces animales, des insectes aux éléphants en passant par les humains (Calhoun and Hayden, 2015; Trapanese et al., 2019). Dans cette nouvelle tâche, les sujets doivent choisir entre rester dans un « *patch* » qu'ils exploitent, c'est-à-dire que la source de récompense (gouttes de jus de fruit) s'appauvrit de plus en plus pendant la période d'exploitation, ou bien se déplacer vers un autre *patch* pour exploiter une autre source de récompenses disponible (Figure 15). Dans ce paradigme, le choix ne porte donc pas sur la direction du mouvement à exécuter après une exploitation (sa direction est systématiquement connue, contrairement à la tâche des jetons où ces derniers déterminent la cible à atteindre, et donc la direction du mouvement du bras). En revanche, les sujets **doivent décider du meilleur moment de quitter le *patch* en cours d'exploitation pour aller atteindre et exploiter l'autre source de récompense**. Malgré cette différence, cette tâche nous permet d'étudier les interactions décision - action, et plus particulièrement la coordination entre la durée de la décision et la vigueur des mouvements, ainsi que l'influence des coûts moteurs sur cette coordination.

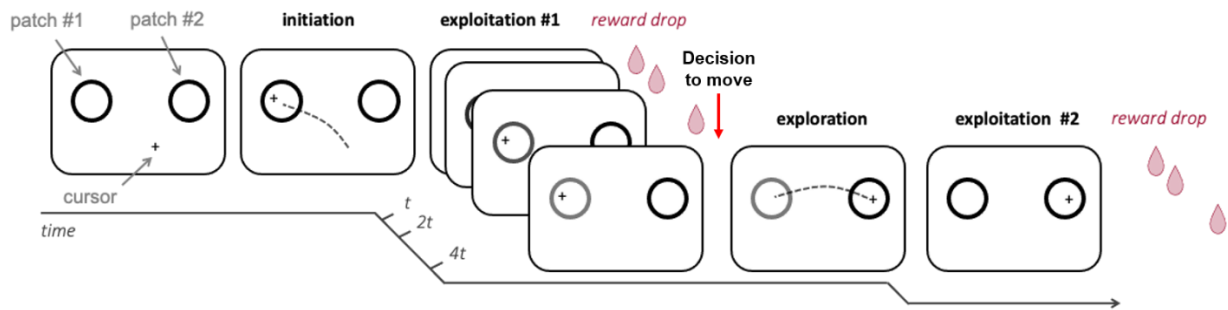

**Figure 15** : La tâche de *foraging*. Les rectangles illustrent un écran sur lequel sont projetés des stimuli visuels. Les stimuli sont constitués de deux cercles correspondant aux sources de récompense, que l'on appelle « *patchs* », et d'une croix marquant la position du levier tenu par le singe pendant la tâche. L'animal prend des décisions successives sur la durée d'exploitation des *patchs* et des « explorations[5] » en déplaçant le levier d'un *patch* à l'autre. Le rythme des gouttes de jus reçues varie d'un essai à l'autre), et les contraintes motrices liées à la taille et à la distance séparant ces *patchs* sont manipulées dans des blocs d'essais dédiés.

Les principes comportementaux sous-tendant l'activité de *foraging* sont bien étudiés dans la littérature, avec l'utilisation de tâches et scénarios variés (e.g. Hayden et al., 2011; Constantino and Daw, 2015; Yoon et al., 2018; Heron et al., 2020), y compris en condition "naturalistique" (Shahidi et al., 2024). De plus, certains de ces principes peuvent être prédits dans **le cadre d'une théorie** très influente, le *Marginal Value Theorem* (Charnov, 1976). En particulier, selon ce théorème, si l'exploration (c'est-à-dire le mouvement) est coûteuse, les individus auront tendance à exploiter plus longtemps dans un *patch* pour maximiser leurs gains avant de décider de se déplacer vers un nouveau *patch*. Le modèle permet ainsi de **prédire le temps d'exploitation optimal en fonction du contexte environnemental** (nombre et qualité des sources de récompenses, coûts liés à ces différentes sources de récompenses). Ainsi, en utilisant ce paradigme de *foraging*, nous pouvons étudier les mécanismes sous-jacents à la coordination entre la décision et l'action, tout en **cherchant à expliquer les principes économiques déterminant ce contrôle intégré**, en **manipulant notamment les contingences de récompenses et, à l'image de nos manipulations effectuées dans la tâche des jetons, les coûts moteurs.** Chaque période d'exploitation est caractérisée par son rythme de récompense qui varie entre 3 niveaux (faible, moyen ou élevé) d'un essai à l'autre. Le singe **ne connaît pas le rythme du *patch* suivant avant de débuter son exploitation**. Pour étudier l'effet du contexte moteur sur la stratégie de prise de décision, la taille et la distance des *patchs* sont modulées par blocs d'essais, encourageant ainsi l'exécution de **mouvements plus ou moins vigoureux.**

Pour évaluer les marqueurs neuronaux du contrôle intégré de la prise de décision et de l'exécution motrice, nous avons réalisé des enregistrements de l'activité électrique extracellulaire unitaire des neurones appartenant aux différentes régions de la boucle cortex sensorimoteur - ganglions de la base évoquées ci-dessus, en vue d'effectuer des **analyses basées sur ces réponses unitaires** ainsi que des **analyses de la dynamique neuronale populationnelle** de type « *state-space* » (Cunningham and Yu, 2014). Pendant les acquisitions, notre approche consistait à enregistrer l'activité **de tous les neurones suffisamment bien isolés**, sans les sélectionner en fonction de leur sensibilité aux paramètres de la tâche. Nos analyses ont porté sur 901 neurones enregistrés dans ces différentes régions cérébrales. Parmi eux, 684 étaient enregistrés dans le PMd, 110 dans le striatum dorsal, 107 dans le GP (GPe pour la grande majorité, moins de 10% d'entre eux ayant été identifiés comme appartenant au GPi). Les données n'ayant été obtenues jusqu'à présent que sur un seul animal et les analyses de ces réponses neuronales étant toujours en cours, je ne décris dans ce document que quelques propriétés observées en lien avec la question du contrôle intégré des décisions et des actions dans la tâche de *foraging*.

---

[5] On parle ici d'exploration pour adopter un vocabulaire propre aux comportements de type *foraging*, mais le singe réalise en réalité un seul mouvement du bras pour passer de l'exploitation d'un *patch* à un autre.

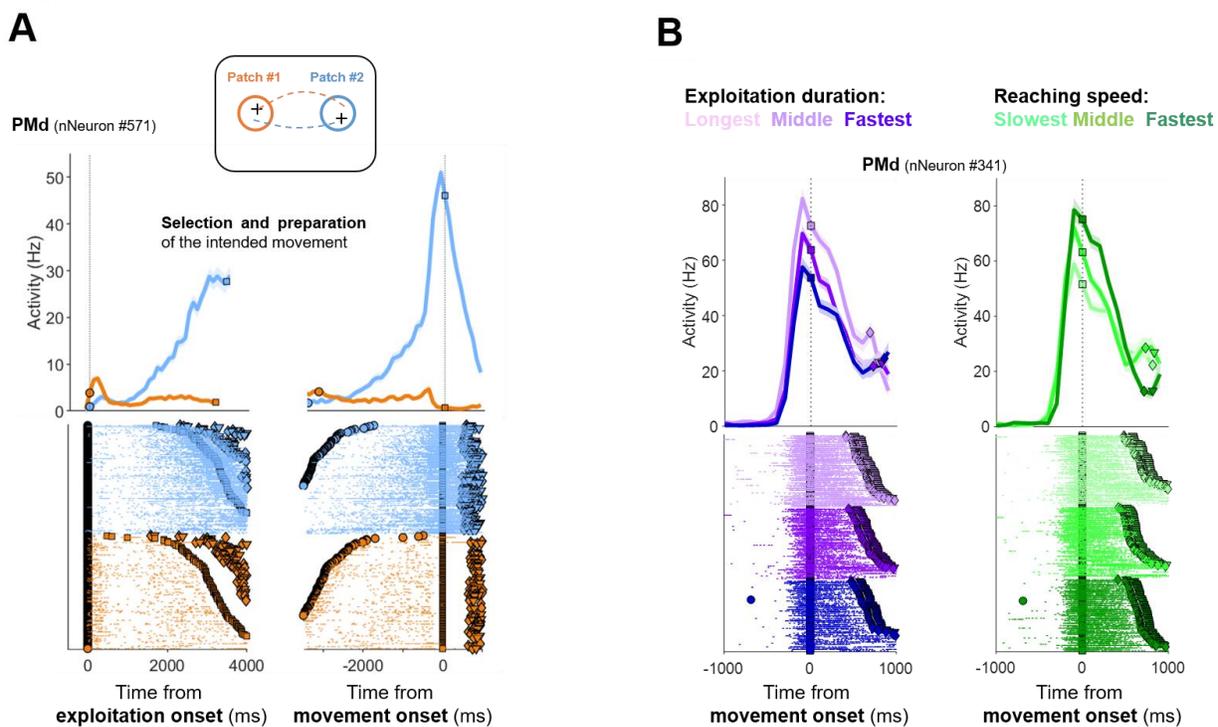

**Figure 16** : *A* : Activité d'un neurone du cortex prémoteur dorsal (PMd) qui augmente son activité dans la tâche de *foraging*. L'activité est illustrée sous forme de courbes de fréquence de décharge moyenne ± l'erreur standard à la moyenne (haut) et de *rasters* (bas). Dans les rasters, chaque ligne correspond à un essai et chaque point d'une ligne correspond à un potentiel d'action. Sur le panneau de gauche, l'activité est alignée sur le début des essais, c'est à dire sur le début de l'exploitation d'un *patch*, avec ces essais triés en fonction de la position du *patch* exploité et de la direction du mouvement exécuté par l'animal (*patch* #1 exploité et mouvement vers le *patch* #2 en bleu ; *patch* #2 exploité et mouvement vers le *patch* #1 en orange). Sur le panneau de droite, la même activité est illustrée mais alignée sur le début du mouvement. *B* : Activité d'un autre neurone de PMd dans la tâche de *foraging*, alignée sur le début du mouvement, avec les essais groupés selon la durée de l'exploitation (en violet, panneau de gauche) ou la vitesse du mouvement (en vert, à droite).

Nous avons par exemple observé que de **très nombreux neurones de PMd augmentaient ou diminuaient leur activité pendant la période d'exploitation** (41% et 22%, respectivement). Pour les cellules qui augmentaient leur activité, un pic de taux de décharge était atteint juste avant l'initiation du mouvement. Un exemple de ce type de profil d'activité est illustré sur la figure 16A. Cette figure montre également une différence d'activité selon le *patch* exploité par l'animal. En effet, dans la plupart des cas (60%), les cellules de PMd avaient une « préférence directionnelle », principalement au moment de l'exécution du mouvement (60%), mais aussi parfois dès le début de l'exploitation (23%). Ceci marque une différence importante avec les neurones enregistrés dans le striatum dorsal et dans les deux sections du globus pallidus qui présentaient **moins de sensibilité à la position du *patch* exploité et à la direction du mouvement d'atteinte exécuté** par l'animal (environ 30% de préférence directionnelle dans ces régions sous-corticales).

Nous avons également entrepris de décrire **le degré de corrélation entre ces activités unitaires et certaines variables comportementales pertinentes pour comprendre les bases neurales de la coordination décision-action**, notamment **la durée des exploitations** (ou la durée de la décision de quitter le *patch* exploité) et **la vitesse de l'exploration** (le mouvement d'atteinte exécuté du *patch* exploité vers le *patch* alternatif). Une simple corrélation de Pearson indique qu'une proportion importante de **neurones de PMd** présente une activité pré-motrice corrélée au pic de vitesse du mouvement (35% des cas), ce qui est somme toute attendu, mais aussi à la durée de l'exploitation (dans 28% des cas). Cette caractéristique concerne à la fois les neurones qui augmentent et qui diminuent leur activité pendant l'exploitation, bien que la proportion de cellules modulées soit globalement plus faible pour cette seconde catégorie. Il est intéressant de noter que la plupart des cellules étaient modulées **par une seule des deux variables comportementales**. Un exemple de ces rares neurones de PMd modulés à la fois par la durée de la décision et par la vitesse des mouvements est illustré dans la figure 16B. Concernant le **striatum dorsal**, ces mêmes analyses indiquent que la proportion maximale de neurones modulés (30 sur 110, soit 27%) concerne **l'activité précédant le mouvement en lien avec la durée de l'exploitation**. Tout comme pour PMd, les cellules sont dans la majorité des cas modulées par une seule des deux variables comportementales. Enfin, dans le **globus pallidus**, on observe que les neurones sont globalement **modulés de la même façon que dans PMd** (la plus grande proportion de neurones est modulée avant le mouvement et en lien avec la vitesse de celui-ci), même si les proportions sont globalement plus faibles dans le GP que dans PMd.

Une analyse de la **dynamique neuronale populationnelle**, analyse en **composantes principales**, a également été effectuée sur l'ensemble des neurones enregistrés pendant la tâche de *foraging*, quelle que soit la région à laquelle appartenaient ces cellules, en alignant l'activité sur l'initiation du mouvement et en groupant les essais en fonction du *patch* exploité par l'animal (et donc de la direction du mouvement effectué pour quitter ce *patch*). Les trois premières composantes principales (*principal component*, PC) **expliquant à elles seules près de 80% de la variance totale partagée par la population** sont interprétables par des éléments du comportement de l'animal. La première composante principale (PC1, expliquant 46% de la variance) reflète **la transition entre la période d'exploitation** (avant la décision de quitter le *patch* actuel) **et l'exécution du mouvement**. La deuxième composante principale (25%) discrimine principalement **la direction du mouvement** exécuté par l'animal et la troisième (11%) semble refléter **le passage du temps**, peut-être l'**urgence**, pendant l'exploitation. Lorsque ces trois premières composantes sont représentées dans le même espace, on remarque que l'état neuronal reste relativement stable entre 3000 ms et 1000 ms avant le début du mouvement. En revanche, à mesure que l'initiation du mouvement approche, on observe **une évolution rapide** de cet état neuronal marquant à la fois la direction du mouvement exécuté dans l'espace capturé par la PC2 et la transition entre l'exploitation et l'action capturé principalement par la PC1 (Figure 17A).

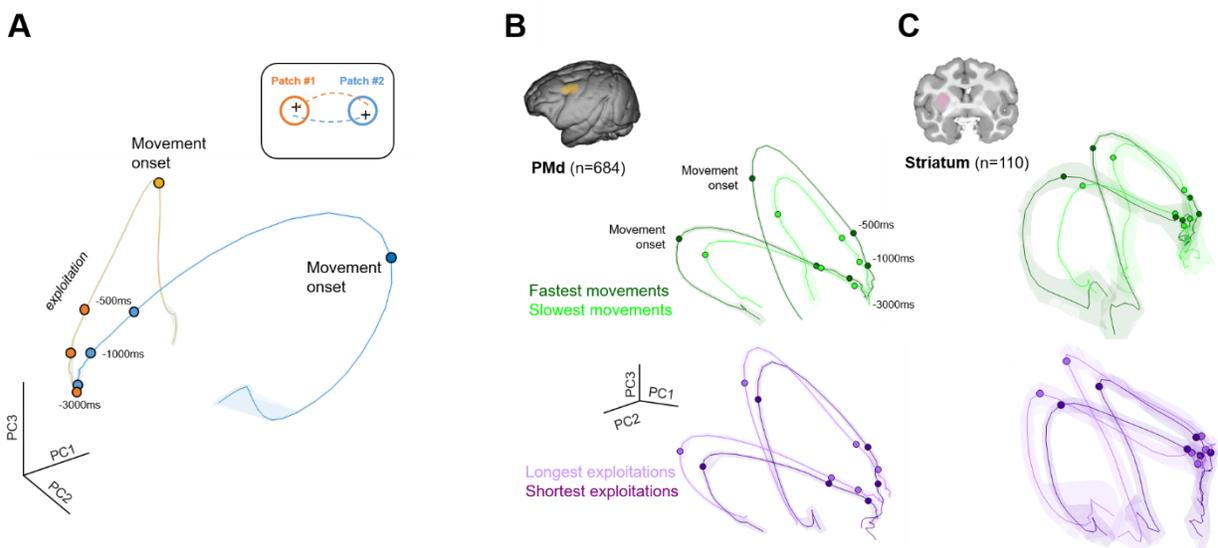

**Figure 17** : Analyse de la dynamique populationnelle pendant la tâche de *foraging*. **A** : Représentation en 3 dimensions des 3 premières composantes principales de l'analyse de dynamique populationnelle (analyse « *state-space* ») alignées sur l'initiation du mouvement, avec les essais groupés selon le *patch* exploité et la direction du mouvement exécuté (bleu : *patch* #1 exploité, mouvement exécuté vers la droite et orange : *patch* #2 exploité, mouvement exécuté vers la gauche). Les intervalles de confiance (non visibles ici) à 95 % sont calculés à l'aide de la méthode statistique de ré-échantillonnage (*bootstrap*). **B** : Trajectoires neuronales alignées sur le début du mouvement dans PMd dans l'espace défini par les 3 premières composantes principales, avec les essais groupés en fonction de la vitesse des mouvements (vert) ou en fonction de la durée de l'exploitation (violet). **C** : Identique à B pour les trajectoires issues de l'activité du striatum dorsal.

Cette analyse permet également de projeter dans l'espace global de la PCA les réponses spécifiques des neurones enregistrés dans les différentes régions étudiées, en regroupant les essais en fonction de critères pertinents pour l'étude du contrôle intégré des décisions et des actions. Par exemple, la figure 17B illustrent les trajectoires neuronales de PMd dans l'espace défini par les trois premières composantes principales décrites ci-dessus, avec les essais groupés **selon la vitesse des mouvements** et selon **la durée des exploitations**. Comme attendu d'une structure souvent associée à la planification et à l'exécution des mouvements, les trajectoires de PMd se séparent nettement dans les deux catégories de vitesse de mouvement, avec une différence principalement visible le long de la première PC. La vitesse du mouvement module donc ces trajectoires dans un même sous-espace, définissant ainsi un plan spécifique pour chaque *patch* exploité. De façon intéressante, **les trajectoires de PMd se différencient de façon similaire lorsque les essais sont groupés en fonction de la durée de l'exploitation**. L'analyse des trajectoires neuronales du striatum dorsal selon les mêmes critères de classification d'essais montre des effets similaires à ceux observés dans PMd (Figure 17C). Ce résultat d'une analyse relativement simple et qu'il reste à approfondir est néanmoins très intéressant car il suggère que **l'état neuronal permettant de déterminer la durée des exploitations est similaire à celui permettant de définir la vigueur des mouvements**.

## 4. Perspectives scientifiques et projets futurs

Les perspectives à court, moyen et long termes que je souhaite évoquer dans le cadre de ce mémoire se répartissent **en deux axes**. À court et moyen termes, je souhaite bien entendu **poursuivre les travaux entrepris depuis mon installation à Lyon en 2018**. Ceci inclut la conclusion du **projet ATIP/Avenir** déjà mentionné à plusieurs reprises ci-dessus, mais aussi **la poursuite d'un projet connexe, financé par l'ANR en 2022**, **initié en 2023** et coordonné par David Robbe (INMED, Marseille). À plus long terme et pour m'orienter toujours plus vers les principes intégratifs qui nous permettront je pense de répondre aux limites évoquées en début de document (section II-1), je souhaite **développer une ligne de recherche dite « naturalistique »**, dans laquelle les singes (et éventuellement les humains) effectueront exactement les même tâches comportementales que celles effectuées jusque-là en condition contrainte, mais dans un environnement leur permettant **d'exprimer les répertoires comportementaux dont ils ont besoin**. Dans les paragraphes ci-dessous, je décris brièvement ces deux axes de perspectives scientifiques.

### a) Poursuite des travaux initiés en 2018

#### i. Comportement des singes dans la tâche de *foraging*

L'abandon (je pense définitif) de la tâche des jetons en 2022 et l'introduction d'une nouvelle tâche comportementale dans le laboratoire présente des **opportunités** mais aussi des **défis** dans l'optique de poursuivre les travaux initiés jusque-là. Parmi ces derniers, évoquons bien-sûr la possibilité de voir **la continuité de ces questions étudiées depuis une dizaine d'années rompue,** due aux spécificités du nouveau paradigme expérimental. Concernant les opportunités, la possibilité de **tester la généralisation des principes** mis en évidence avec une seule et unique tâche représente un intérêt majeur. En effet, la tâche des jetons est une tâche de prise de décision perceptive, où les choix sont guidés par des indices visuels évoluant au cours du temps pendant la délibération, avec un niveau d'urgence induit par l'égrainage des sauts de jetons potentiellement très important. Dès lors, on peut se demander dans quelle mesure les résultats comportementaux mis en évidence dans cette tâche, notamment les modes de coordination décision-action décrits chez les singes et les humains, **ne sont pas spécifiques de ce protocole particulier**. Avec la nouvelle tâche de *foraging*, dans laquelle les choix basés sur la valeur sont guidés par un signal « interne » déterminé par le ratio entre un niveau de récompense qui diminue au cours d'une exploitation et des contraintes motrices contextuelles, sans pression du temps, nous avons la possibilité de tester dans ce protocole radicalement différent **l'aspect non-spécifique** des principes comportementaux observés dans la tâche des jetons.

Pour ce faire, encore faut-il être en mesure de manipuler les mêmes paramètres nous permettant de tester **l'impact des informations guidant le choix** et **du contexte moteur** sur la stratégie de prise de décision et sur la vigueur des mouvements. Comme mentionné ci-dessus, la tâche de *foraging* permet ces manipulations. En effet, concernant les informations guidant le choix, **le rythme auquel sont distribuées les récompenses au cours d'une exploitation[6] varie entre trois niveaux d'un essai à l'autre**, et ce de façon imprévisible par le sujet. Certes, ceci ne représente pas le même niveau d'imprévisibilité que les sauts de jetons dans la tâche perceptive, surtout quand le sujet a effectué de nombreuses sessions et connait parfaitement les rythmes de récompense présentés, mais l'aspect **dynamique de l'information guidant le choix** est néanmoins respecté. Quant à l'impact des coûts moteurs, il peut être testé dans **la tâche de *foraging* exactement de la même manière que dans la tâche des jetons**, en modifiant la taille et la distance des cercles dans lesquels le sujet déplace le levier pour collecter les gouttes de jus de fruit.

Des analyses comportementales dans la tâche de *foraging* ont donc été effectuées, notamment sur les données issues **du singe G** pour lequel les résultats électrophysiologiques sont mentionnés plus haut (section II-3-d-iii). Chez cet animal, **une compensation entre la durée d'exploitation et la vigueur des mouvements** est observée dans la plupart des sessions à l'échelle de l'essai individuel (Figure 18A, gauche). En d'autres termes, plus l'exploitation est longue, plus le mouvement d'atteinte exécuté pour aller exploiter le *patch* suivant est court et rapide. Deux autres singes, Bingo et Diego, sont entrainés à réaliser cette tâche de *foraging* et pour au moins l'un d'entre eux (Bingo), la même stratégie de compensation est observée (Figure 18A, droite). Il est en outre intéressant de souligner que pour cet animal, la relation entre le mode de coordination décision-action (co-régulation ou compensation) et la durée globale des exploitations est très étroite (coefficient de corrélation r = 0.78), à l'image de ce qui a été observé chez les singes dans la tâche des jetons (voir Figure 13C). En d'autres termes, **plus les exploitations du singe B sont longues**, plus il aura tendance **à compenser ses longues exploitations par des mouvements d'exploration courts et rapides**.

Lorsque l'on étudie l'effet **du rythme de récompense** sur la coordination entre décision et action chez le singe G, on observe que, lorsque **ce rythme est faible**, **elle exploite plus longtemps** et à tendance à **effectuer des mouvements légèrement plus vigoureux** (et vice versa pour les rythmes de récompense élevés). Il est à noter que ce résultat concernant la durée d'exploitation n'est pas conforme aux prédictions du *Marginal Value Theorem* (mais ce type de violation de la théorie se retrouve dans d'autres protocoles de *foraging* simplifiés, voir par exemple Sukumar et al., 2024). En revanche, les modulations induites par **les conditions motrices** sont conformes à la théorie. En effet, plus **les contraintes imposées aux**

---

[6] Le rythme de récompense, à ne pas confondre avec le taux de récompense mentionné dans le reste du mémoire.

**mouvements d'atteinte sont importantes** (conditions petits *patchs* ou *patchs* très éloignés l'un de l'autre), **plus la durée d'exploitation est longue** et **plus le mouvement exécuté pour exploiter l'autre source de récompense est rapide**. Ces résultats soutiennent l'hypothèse d'un mécanisme de compensation entre décision et action aussi bien à l'échelle des essais individuel qu'à l'échelle des blocs d'essais. En revanche, pour les singes B et G, les résultats concernant l'impact des rythmes de récompense sur le contrôle intégré des choix et des mouvements ne sont, à ce jour, pas aussi clairs et appellent à davantage d'acquisitions.

À ce stade des analyses néanmoins, les résultats suggèrent que la coordination décision-action, et notamment son mécanisme **de type compensation**, n'est pas **spécifique à la tâche des jetons** et que tout comme dans cette tâche perceptive, **les paramètres permettant de tester les déterminants importants du comportement dirigé vers un but (rythmes de récompense et coûts moteurs), sont pris en compte par les animaux**.

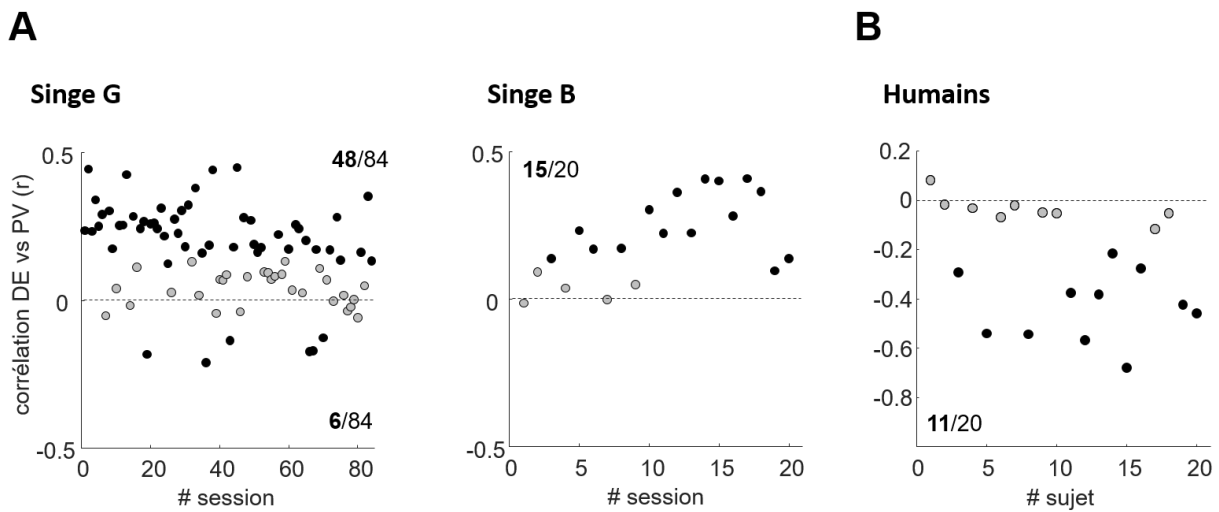

**Figure 18** : Coordination décision-action chez les singes et les humains dans la tâche de *foraging*. *A* : Corrélations de Pearson à l'échelle de l'essai unique entre la durée de l'exploitation (DE) et le pic de vitesse des mouvements d'atteinte (PV) exécutés par le singe G (gauche) et le singe B (droite). Chaque point illustre le résultat de la corrélation essai par essai pour une session, avec les sessions organisées de façon chronologique (axe des abscisses). Une valeur de r positive signifie que plus l'exploitation est longue, plus le mouvement est court et rapide. Les points noirs indiquent les sessions pour lesquelles la corrélation est significative. *B* : Même analyse et mêmes conventions pour une population de 20 sujets humains neurotypiques. Ici, chaque point illustre les données d'un sujet testé.

ii.   Comportement des humains dans la tâche de *foraging*

Toujours dans le but de comparer le comportement des singes macaques avec celui des humains sur les tâches utilisées dans le laboratoire, nous avons mené récemment **une étude pilote dans laquelle 20 sujets humains neurotypiques** (âge : 26 ± 9 ans, 12 femmes, 14 droitier(e)s) ont été testé(e)s sur la même tâche de *foraging* que celle décrite ci-dessus. Il convient cependant de noter que certains paramètres, tels que la taille des stimuli présentés aux sujets ou encore certaines variables temporelles, diffèrent légèrement entre les protocoles utilisés pour les singes et les humains (afin de s'adapter au mieux à chaque espèce étudiée). De plus, l'objectif des sujets différait de celui des animaux, puisque les humains alternaient **des blocs d'essais de 3 minutes** (un essai consistant en une phase d'exploitation durant laquelle des points sont accumulés à un rythme décroissant au cours du temps, suivie d'un mouvement d'atteinte exécuté pour démarrer une nouvelle phase d'exploitation) entrecoupés de courtes pauses, chaque bloc étant caractérisé par un contexte moteur particulier (petits vs grands *patchs* ; petite vs grande distance entre les *patchs*), tandis que les singes changeaient de condition motrice sur la base non d'une durée, mais d'un **nombre d'essais réalisés** (tous les 50 essais).

Parmi les 20 participants testés, une corrélation positive significative entre durée d'exploitation et vigueur de mouvement (corrélation de Pearson essai par essai, $p < 0.05$) est observée chez 11 d'entre eux (Figure 18B). Ceci signifie que les plus longues exploitations sont suivies des mouvements les plus longs (et vice versa pour des exploitations plus courtes), ce qui plaide pour ces sujets en faveur **d'un mode de régulation de type « co-régulation » de la prise de décision et de l'action**. Les singes ayant adopté un mode de régulation **de type « compensation »**, cette différence de stratégie entre les deux espèces pourrait être attribuable au fait que les singes reçoivent des **récompenses immédiates sous forme de gouttes de jus de fruit pendant les exploitations,** tandis que les humains **cumulent des points et tentent implicitement de maximiser ce nombre de points à l'échelle d'une session**. Il est possible que les humains aient été de ce fait moins sensibles à la **dévaluation temporelle de la valeur immédiate de la récompense**, alors que les singes ont compensé de longues exploitations en réalisant des mouvements plus vigoureux pour limiter cette dévaluation temporelle immédiate dans chaque essai. La différence pourrait aussi s'expliquer par **l'objectif demandé aux sujets**, un nombre d'essais pour les singes, une durée pour les humains. Avec une durée de travail fixe comme but à atteindre, les sujets humains n'ont peut-être pas trouvé d'intérêt à adopter un mécanisme de compensation temporel dans leur protocole. Il serait très intéressant de tester dans le futur des sujets humains dans la même tâche de *foraging* mais **en changeant l'objectif des participants afin que celui-ci corresponde à celui donné aux animaux**.

Malgré cette différence, une autre observation pourrait celle-ci indiquer un principe de régulation de la décision et de l'action commun chez ces deux espèces. Comme pour les singes, on s'est en effet également intéressé **au lien entre le mode de coordination décision–action utilisé (co-régulation ou compensation) et la durée d'exploitation moyenne** des 11 participants pour lesquels la corrélation entre durée d'exploitation et vitesse de mouvements à l'échelle des essais est significative. On constate que **les sujets pour lesquels les exploitations sont les plus longues sont ceux qui co-régulent le moins**. Même si cette relation n'est qu'une tendance puisque la corrélation n'atteint pas le seuil de significativité ($r = -0.53$, $p = 0.09$), ce résultat suggère que **le principe déterminant le mode de coordination entre décision et action chez le singe semble dicter également la stratégie des humains** (ici à une échelle d'analyse différente, entre les sujets). Nous prévoyons de poursuivre les acquisitions chez l'humain et avec plus de données, il est possible que nous observions des sujets chez qui les exploitations sont encore plus longues et qui pourraient potentiellement compenser davantage leurs décisions et leurs actions.

iii. Neurophysiologie du réseau cortex sensorimoteur-ganglions de la base chez le singe macaque pendant le *foraging*

La poursuite de l'étude du rôle du réseau cortex sensorimoteur – ganglions de la base dans le contrôle intégré de la prise de décision et de l'action, initiée en octobre 2022 avec les premiers enregistrements électrophysiologiques du laboratoire et dont les résultats préliminaires sont mentionnés dans la section II-3-d-iii, comprend **plusieurs volets**.

Un des premiers objectifs à très court terme sera de **dissocier l'impact de la durée des décisions de celui de la vitesse des mouvements sur la réponse des neurones dans les différentes régions étudiées**. En effet, les analyses préliminaires des activités unitaires et de la dynamique populationnelle chez le singe G ayant effectué une centaine de sessions de la tâche de *foraging* ont mis en évidence un impact important de ces deux variables comportementales sur la réponse neuronale du cortex prémoteur dorsal (PMd), du striatum dorsal et du globus pallidus. Cependant, nous avons également montré que ces deux variables comportementales co-varient fortement dans la tâche de *foraging*, en particulier pour ce même singe G pour lequel, dans la plupart des sessions, des exploitations longues étaient suivies de mouvements d'atteinte rapides. Nous poursuivrons donc l'analyse de ces activités neuronales aux échelles unitaire et populationnelle en utilisant des **modèles de régressions linéaires multiples** permettant de **dissocier la contribution respective de chaque paramètre sur la réponse de ces cellules** (Thura and Cisek, 2017). Ceci devrait nous permettre de répondre à une des questions centrales du projet ATIP/Avenir proposé à l'époque, à savoir déterminer s'il existe **des populations spécifiques de neurones prenant en charge l'urgence décisionnelle et la vigueur des mouvements**, dans le globus pallidus interne notamment.

Un deuxième volet concerne l'étude de l'**impact des déterminants économiques de la tâche de *foraging* (rythmes de récompense et coûts moteurs),** ceux-là même qui influencent le comportement des animaux dans la tâche, dans l'activité des aires composant le réseau cortex sensorimoteur-ganglions de la base. Des analyses très préliminaires de l'activité neuronale du singe G dans la tâche de *foraging* ont déjà montré qu'une proportion importante de cellules des différentes régions testées est modulée par **les coûts moteurs** (entre 30% et 50% selon les régions, voir la Figure 19A-B pour deux exemples dans le striatum et le globus pallidus). Dans le PMd et le striatum dorsal, l'activité neuronale précédant le mouvement est plus fréquemment modulée par le contexte moteur que l'activité observée en début d'exploitation. Dans le globus pallidus, les deux périodes temporelles sont influencées par les coûts moteurs dans des proportions importantes et similaires (~45%) (Figure 19B). Le **rythme de récompense** module également l'activité d'exploitation précoce et l'activité pré-motrice d'une fraction des neurones dans chaque structure, même si la proportion de cellules modulées est significativement inférieure à celle observée lorsque les coûts moteurs sont manipulés, **y compris dans le striatum dorsal**. Indépendamment de la structure considérée, l'activité neuronale précédant le mouvement est globalement autant modulée par le rythme de récompense que l'activité observée en début d'exploitation. À noter que la récompense modulait le plus souvent **le taux** **de décharge** des cellules, mais parfois aussi **l'occurrence des potentiels d'action** pour une même fréquence de décharge, comme l'illustre la figure 19C.

Ces résultats sont d'ores et déjà très intéressants car ils suggèrent **un lien étroit entre l'activité neuronale dans le circuit cortex sensorimoteur - ganglions de la base et l'intégration des informations contextuelles importantes pour le *foraging*.** Cependant, de nombreuses questions subsistent et d'autres analyses sont nécessaires pour comprendre comment ces informations sont traitées dans la boucle cortex-ganglions de la base. Par exemple, il est possible que ces informations soient traitées par **des sous-populations de cellules spécifiques** dans chaque région du réseau. Des analyses préliminaires indiquent en effet un lien entre la sensibilité des cellules du PMd, du striatum dorsal et du globus pallidus, et la **variation du taux de décharge** de ces cellules pendant l'exploitation. En d'autres termes, les cellules ayant un taux de décharge stable pendant la période de délibération sont moins modulées pas les coûts moteurs et les rythmes de récompense que les cellules augmentant ou diminuant leur activité avant le mouvement. L'implication de cette observation reste à établir.

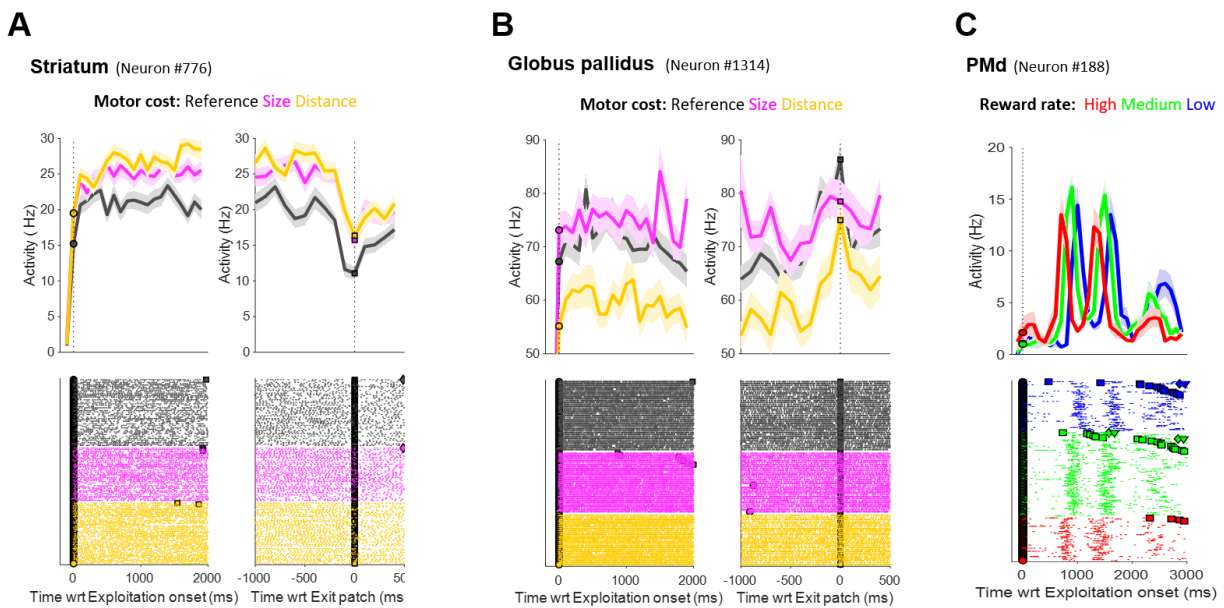

**Figure 19** : Modulation de l'activité des neurones de la boucle cortex-ganglions de la base par les coûts moteurs et les rythmes de récompense dans la tâche de *foraging*. ***A*** : Exemple d'un neurone du striatum dorsal dont l'activité est modulée par le contexte moteur dans lequel se déroule la tâche. L'activité est alignée sur le début de l'exploitation (gauche) et sur le début du mouvement (droite), avec les essais triés en fonction de la condition motrice testée (référence : noir ; petits *patchs* : rose ; *patchs* éloignés l'un de l'autre : orange). ***B*** : Exemple d'un neurone du globus pallidus externe dont l'activité est modulée par le contexte moteur. Mêmes conventions qu'en A. ***C*** : Exemple d'un neurone de PMd dont l'activité est modulée par le rythme de récompense pendant la période d'exploitation. Ici, l'activité est alignée sur le début de l'exploitation et les essais sont groupés en fonction du rythme de récompense dans chaque essai (haut : rouge ; moyen : vert ; bas : bleu).

La question du **type de modulation** des réponses neurales par **le rythme de récompense**, à savoir une variation de taux de décharge ou une modulation de l'occurrence des potentiels d'actions, souvent alignés sur l'arrivée des deux ou trois premières gouttes reçues pour chaque exploitation (Figure 19C), est intrigante et nécessite également plus d'analyses, en combinant l'étude des réponses unitaires avec celle de la dynamique populationnelle. Il est en effet possible que ces modulations de l'occurrence des potentiels d'action en lien avec la délivrance des premières gouttes de jus de fruit uniquement, le plus souvent observées dans le PMd, **servent à estimer le rythme de récompense en cours d'exploitation**, et non à signaler l'arrivée de chaque goutte indépendamment les unes des autres.

Une autre question passionnante concerne **la façon dont les différentes régions communiquent entre elles pour permettre de faire émerger le comportement** observé dans la tâche de *foraging*. Dans le cadre d'un **stage de M2 réalisé par Nathan Lambert** de janvier à juin 2024, nous nous sommes intéressés à la question du niveau de synchronisation d'activité entre le cortex prémoteur dorsal (PMd) et le striatum dorsal, en partant de l'hypothèse selon laquelle le PMd possèderait au moins **deux populations de cellules**, une qui projette sur le striatum pour l'informer du contexte dans lequel se déroule la tâche et une autre population qui reçoit les informations en provenance des ganglions de la base (via le thalamus) pour implémenter le résultat des calculs économiques effectués dans ces noyaux sous-corticaux et influencer la durée de la décision et la vigueur du mouvement (Figure 20A).

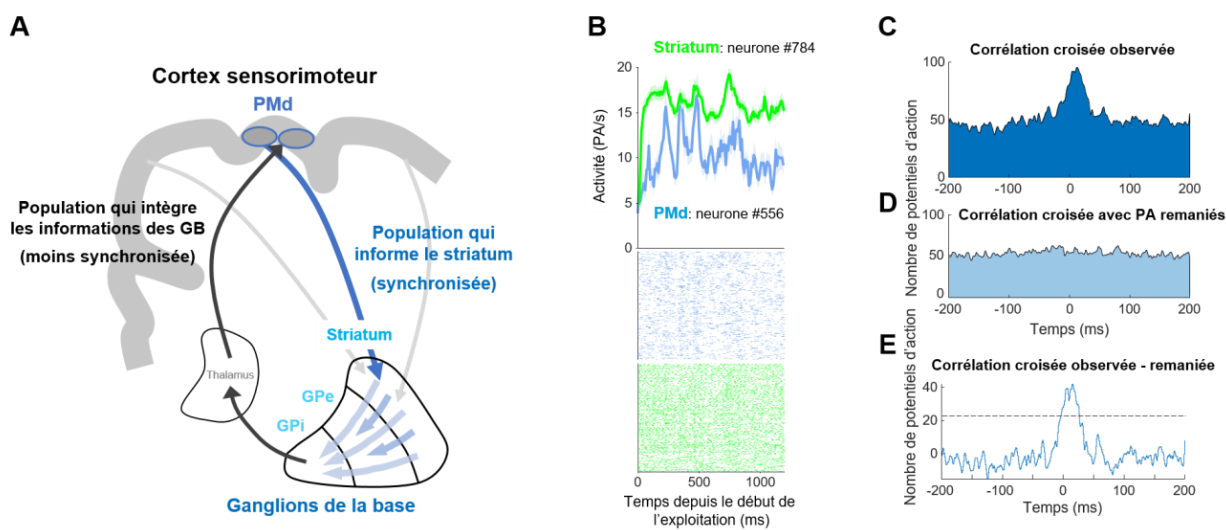

**Figure 20** : Étude de la synchronisation entre les différentes aires de la boucle cortex-ganglions de la base. ***A*** : Hypothèse de la présence de deux populations de neurones dans le cortex prémoteur dorsal (PMd), une qui projette directement sur le striatum dorsal (flèche bleue), une qui reçoit des informations en provenance des ganglions de la base via la thalamus (flèches noires). ***B*** : Activité d'une cellule du striatum dorsal (vert) et d'une cellule de PMd (bleu) alignées sur le début de l'exploitation dans la tâche de *foraging*. ***C*** : Analyse de corrélation croisée de l'activité (potentiels d'action) des deux neurones illustrés en B. ***D*** : Analyse de corrélation croisée entre l'activité du neurone de PMd illustré en B et l'activité remaniée (les potentiels d'actions sont mélangés de façon aléatoire pour chaque essai) du neurone du striatum illustré en B. Cette analyse permet de vérifier que la corrélation entre les activités non remaniées n'est pas due au hasard. ***E***. Différence entre la corrélation réelle calculée en C et la corrélation obtenue avec l'activité striatale remaniée illustrée en D. Cette différence permet de déterminer le niveau de significativité (ligne pointillée noire) de la synchronisation entre les deux neurones.

Les prédictions issues de cette hypothèse sont que l'activité unitaire (les potentiels d'action) d'une fraction des neurones de PMd (ceux qui projettent directement sur le striatum) sera fortement synchronisée avec l'activité des neurones du striatum, et ces cellules de PMd seront également celles qui seront le plus modulées par les informations contextuelles (rythme de récompense, condition motrice) de la tâche de *foraging*. En revanche, les cellules de PMd recevant les informations en provenance du globus pallidus seront moins synchronisées avec les cellules striatales et moins modulées par le contexte de la tâche.

Les premières analyses de corrélation croisée entre potentiels d'action issus des deux régions (Narayanan and Laubach, 2009) confirment la présence de deux populations de neurones de PMd, **une population dont l'activité est fortement synchronisée avec les neurones du striatum** (15% des 687 paires de neurones PMd-striatum testées) (Figure 20B-E) et une autre dont l'activité n'est pas synchronisée de façon significative avec celle du striatum. Contrairement à la seconde partie de notre hypothèse cependant, nous avons observé que la population de neurones de PMd synchronisée avec les cellules du striatum **n'est pas plus modulée par le rythme de récompense ou par les coûts moteurs** que la population de neurones non synchronisée.

D'autres analyses seront nécessaires pour comprendre le rôle fonctionnel joué par ces deux populations. L'étude de la synchronisation entre les différentes aires du réseau sera poursuivie afin de mieux comprendre comment ces régions opèrent pendant le comportement de *foraging*. Ces analyses de synchronisation seront également enrichies avec l'étude des **potentiels de champs locaux** qui, combinée à l'analyse des potentiels d'action, peut nous renseigner non seulement sur **le niveau de communication entre les aires cérébrales, mais aussi sur la direction de cette communication** (Salinas and Sejnowski, 2001; Pesaran, 2010).

Ces analyses seront conduites par **Clara Saleri** dans le cadre de la continuité de ses travaux de thèse.

Enfin, une question importante et qui est à la base du **projet financé par l'ANR en 2022** consiste à approfondir cette idée selon laquelle **les ganglions de la base procèdent aux calculs économiques nécessaires à la détermination de l'utilité du comportement**. Le contexte de ce projet est le suivant : Lors de comportements tels que monter des escaliers pour rendre visite à des amis ou prendre un verre de vin, nous faisons un compromis entre l'effort et le temps nécessaires à la réalisation de ces actions, car les récompenses immédiates ont plus de valeurs que celles qui sont différées dans le temps (Myerson and Green, 1995; Shadmehr et al., 2010). Ainsi, l'**effort** et **le temps** peuvent être considérés comme deux coûts interdépendants qui doivent être optimisés, de manière à **maximiser l'utilité d'une action particulière**. Or, **la façon dont le temps et l'effort interagissent pour affecter à la fois les décisions et les mouvements dirigés vers des récompenses n'a pas été étudiée de manière approfondie**.

L'objectif de ce projet est d'étudier **comment le compromis effort-temps est implémenté dans le cerveau**. Plus précisément, le striatum dorsal et les régions en aval de celui-ci (segments interne et externe du globus pallidus) ont récemment été impliqués dans le contrôle de la vitesse du mouvement (Turner and Desmurget, 2010) et des prises de décision (Redgrave et al., 1999), mais la possibilité que **cette fonction découle d'un rôle de ce circuit sensorimoteur des ganglions de la base dans le compromis effort-temps** n'a pas été testée.

Au cours des 4 prochaines années, nous testerons donc l'hypothèse selon laquelle **le temps et l'effort sont intégrés au sein des ganglions de la base** de sorte que ses noyaux de sortie (globus pallidus interne) **génèrent un signal proportionnel à l'utilité d'une action** donnée et modulent l'activité neuronale dans les régions néocorticales impliquées dans les processus de prise de décision et de contrôle moteur (cortex prémoteur dorsal et cortex moteur primaire notamment, Thura and Cisek, 2017).

Ce projet a été initié par une chercheuse post-doctorante recrutée en octobre 2023, **Claire Poullias**.

b) Vers une étude « naturalistique » des interactions cerveau-environnement

Au cours des vingt dernières années, j'ai eu l'opportunité d'étudier différentes problématiques qui présentaient toutes la particularité de **ne pas considérer le cerveau comme un simple système de traitement de l'information**. À l'inverse, j'ai plutôt travaillé dans un cadre théorique dans lequel le cerveau se compose de **régions fonctionnellement distribuées**, **globalement organisées de manière hiérarchique**, mais possédant néanmoins **un haut niveau de mixité dans les propriétés de leurs réponses neuronales**. Ceci s'est notamment reflété dans l'étude des interactions œil-main dans le système oculomoteur et celle des interactions décision-action dans le système sensorimoteur du bras. De plus, les protocoles expérimentaux et les tâches comportementales utilisés au cours de mes travaux successifs **ont progressivement évolué vers des paradigmes de plus en plus écologiques**. Ce fut le cas avec le passage d'une tâche de coordination œil-main statique très artificielle, à une tâche de *foraging* intégrant des principes comportementaux universels chez les animaux, en passant par une tâche de prise de décision dynamique dans laquelle l'imprévisibilité et les changements de l'environnement étaient testés. Enfin, je suis passé de l'analyse classique de l'activité neuronale en considérant que **la réponse de neurones isolés « représentait » une fonction**, à des analyses populationnelles tenant compte de **l'interconnexion et de la redondance importante des activités neurales**, ainsi que de **leurs natures hautement dynamiques**.

La prochaine évolution du laboratoire (dans un délai de 3 à 5 ans) poursuivra cette volonté de se rapprocher d'une **étude intégrative et écologique du fonctionnement du cerveau**, avec la mise en place d'un dispositif permettant de réaliser des **enregistrements neuronaux sans fil et multicanaux** dans le cerveau des **singes macaques rhésus libres de se déplacer** dans une cage en verre spécialement conçue à cet effet. Les **réponses physiologiques** (rythme cardiaque, température, pression artérielle) seront également collectées et **le comportement** sera capturé par des caméras vidéo et **analysé avec des algorithmes d'intelligence artificielle**, tels que *DeepLabCut* (Mathis et al., 2018).

Grâce à un financement « Équipement De Recherche » obtenu de l'Université Claude Bernard Lyon 1 en 2020, une grande partie de ce matériel a déjà été acquis. Ce dispositif permettra de **s'affranchir d'un haut degré de contraintes** propre à l'approche traditionnelle de l'électrophysiologie chez le singe vigile en comportement.

En effet, cette approche classique se déroule généralement dans un **environnement contrôlé mais non naturel**, dans lequel le niveau de motivation de l'animal est artificiellement standardisé, son répertoire comportemental est minimisé et l'influence de son environnement supprimée. Bien que ceci **facilite l'interprétation** du comportement de l'animal et de l'activité neuronale sous-jacente, cela pose de sérieux problèmes éthiques et limite *in fine* notre compréhension des mécanismes neuronaux qui sous-tendent les fonctions et dysfonctionnements cérébraux. En d'autres termes, **on ignore aujourd'hui si les résultats obtenus en milieu contraint sont valables en conditions naturelles**. En réalité, plusieurs études récentes tendent à montrer que ce n'est pas le cas. Par exemple, il a été démontré récemment que les mouvements spontanés influencent fortement l'activité neuronale dans diverses zones corticales chez la souris et le singe (Musall et al., 2019; Tremblay et al., 2023b), y compris dans les aires visuelles primaires (Parker et al., 2020). Ceci suggère que, dans des conditions naturelles, **le contrôle moteur est si intimement lié aux traitements sensoriels et aux processus cognitifs qu'il n'y a pas de frontières claires entre ces fonctions au sein des circuits cérébraux**.

Bien que ces environnements connaissent un développement croissant ces dernières années, notamment pour les études menées chez le primate non-humain (Testard et al., 2021; Hansmeyer et al., 2023), et que plusieurs laboratoires aient publié des articles décrivant des travaux de neuro-éthologie dans différents scénarios tels que la navigation spatiale, la cognition sociale, le contrôle sensorimoteur ou la recherche de nourriture (*foraging*) (Berger et al., 2020; Mao et al., 2021; Voloh et al., 2023; Shahidi et al., 2024; Testard et al., 2024), cette approche reste associée à de **nombreux défis**, en premier lieu desquels **l'interprétabilité des données collectées en conditions non contrôlées**. En effet, quand un animal est libre d'exprimer la totalité de son répertoire moteur par exemple, les variables comportementales d'intérêt vont souvent, voire systématiquement, **co-varier entre-elles**. Si l'on s'intéresse à de simples mouvements guidés visuellement par exemple, il sera quasiment impossible de déterminer si l'activité neuronale observée est liée à une variable sensorielle qui guide le mouvement, aux commandes descendantes qui produisent le mouvement lui-même, ou à d'autres facteurs « cognitifs » qui varient avec l'un ou les deux processus sensorimoteurs.

Au vu de ce challenge, il serait probablement peu judicieux de procéder **à une transition brutale entre des études réalisées en milieu contraint à d'autres menées en conditions non contraintes**, en espérant que les méthodes analytiques récentes suffisent à décoder le comportement des animaux libres d'exprimer la totalité de leur répertoire comportemental. Une stratégie alternative consiste plutôt à adapter progressivement les protocoles et les dispositifs expérimentaux pour que, pas à pas, chacune de ces évolutions contribue à se rapprocher du comportement naturel des animaux, tout en conservant un contrôle expérimental précis garantissant une interprétabilité satisfaisante des données (Cisek and Green, 2024). Ce procédé devra être associé, du moins dans les étapes initiales de sa mise en place, à **la comparaison systématique** des résultats obtenus en milieux de moins en moins contraints avec ceux obtenus en milieu traditionnel.

L'ambition du laboratoire dans les prochaines années sera donc **de comparer et formaliser le comportement des animaux entre une configuration « contrôlée » et une configuration « naturalistique »**, dans le cadre théorique du *Marginal Value Theorem* et pour l'étude du contrôle intégré de la prise de décision et de l'action lors du *foraging*. De même, **nous comparerons et formaliserons l'activité neuronale dans les deux contextes dans le cadre dit de *Computation Through Dynamics*** (Vyas et al., 2020) qui fournit des outils analytiques pour extraire des réponses neuronales populationnelles les signaux les plus partagés dans une tâche donnée (« *state-space analysis* »). Avec cette approche intégrée de l'étude des interactions cerveau-environnement chez le primate non-humain, **nous visons à fournir une validité écologique aux données collectées dans un environnement contraint**.

## 5. Collaborations

Depuis mon installation à Lyon en 2018 en tant que chercheur indépendant, j'ai à la fois poursuivi et initié plusieurs collaborations.

**Paul Cisek**, Université de Montréal, Montréal, Canada.
Ma collaboration à long terme concerne mon superviseur de formation postdoctorale, Paul Cisek, à l'Université de Montréal. Nous avons travaillé récemment sur des analyses de population de l'activité neuronale collectée dans le PMd, M1, le cortex préfrontal dorsolatéral et les ganglions de la base pour tester l'hypothèse selon laquelle la prise de décision et le contrôle sensorimoteur sont deux modes d'un système dynamique unifié distribué à travers le réseau cortex-ganglions de la base. Cet article a été publié en 2022 dans *Plos Biology* (Thura et al., 2022). Actuellement, je collabore avec Paul et une étudiante en thèse de son laboratoire, Poune Mirzazadeh, sur le développement d'un modèle computationnel permettant d'expliquer les réponses neuronales évoquées ci-dessus.

**David Robbe**, Institut de Neurobiologie de Méditerranée, Marseille, France
Les travaux de David Robbe visent à comprendre comment le striatum contribue au contrôle moteur et à l'apprentissage chez les rongeurs effectuant des tâches de locomotion, en utilisant des méthodes chroniques in vivo, notamment des enregistrements de l'activité striatale (sondes tétrode/silicium, optrodes), des lésions et de l'inactivation pharmacologiques, et des manipulations optogénétiques de l'activité neuronale. David Robbe est le coordinateur du projet financé par l'ANR en 2022 visant à comprendre le rôle des ganglions de la base dans l'intégration de l'effort et du temps pendant le comportement dirigé vers un but, et pour lequel je suis responsable des expérimentations menées chez le primate non-humain.

**Gérard Derosiere** et **Julie Duque**, Université Catholique de Louvain, Bruxelles, Belgique

Au moment de débuter notre collaboration, Gérard Derosière était un chercheur postdoctoral à l'Université Catholique de Louvain, travaillant avec le Dre Julie Duque. Ses travaux portaient sur le rôle du système moteur dans la prise de décision et la planification des actions chez l'humain, utilisant diverses méthodes de stimulation cérébrale basée sur la stimulation magnétique transcranienne (TMS). Gérard a visité le laboratoire de Paul Cisek lors de ma formation postdoctorale et nous avons commencé à travailler ensemble dans le but de mieux comprendre l'impact de l'urgence décisionnelle sur le système moteur chez l'humain. Ces travaux ont conduit à la publication de plusieurs articles dans les années récentes (Derosiere et al., 2019, 2021, 2022). Gérard est maintenant chercheur Inserm dans l'équipe Impact du CRNL, et nous entretenons toujours avec Julie Duque des discussions fréquentes sur nos sujets d'étude respectifs, même si aucun projet collaboratif n'est actuellement en cours d'exécution.

**Fadila Hadj-Bouziane**, Centre de Recherche de Neurosciences de Lyon – Impact, Bron, France

Peu de temps après avoir démarré mon projet ATIP/Avenir à Lyon, j'ai initié une collaboration avec le Dre Fadila Hadj-Bouziane qui travaille sur les substrats neuronaux des mécanismes de perception et d'attention, abordant ce sujet avec diverses approches expérimentales, dont la neuropharmacologie et l'imagerie cérébrale chez le singe. Ensemble, nous avons étudié la relation entre l'urgence de décider et d'agir et deux systèmes majeurs de neuro-modulation, le système dopaminergique (DA) et le système noradrénergique (NA). Ce travail hautement collaboratif a donné lieu à un papier récemment publié dans *Neuropharmacology* (Kaduk et al., 2023).

**Yves Rossetti**, Centre de Recherche en Neurosciences de Lyon – Trajectoire, Bron, France

J'ai également contribué à la publication d'un article décrivant une étude dirigée par le Pr. Yves Rossetti dont le but était de comprendre si l'inhibition motrice liée à la motivation d'agir opère non seulement en amont de la cascade d'actions, pendant la phase de décision et de préparation motrice, mais affecte également le déroulement des mouvements d'atteinte sélectionnés, c'est-à-dire à un niveau de contrôle moteur dit de « bas niveau ». Cet article est paru en 2019 dans *Frontiers in Psychology* (Revol et al., 2019).

## 6. Références bibliographiques

Alexander GE, DeLong MR, Strick PL (1986) Parallel Organization of Functionally Segregated Circuits Linking Basal Ganglia and Cortex. Annual Review of Neuroscience 9:357–381.

Balci F, Simen P, Niyogi R, Saxe A, Hughes JA, Holmes P, Cohen JD (2011) Acquisition of decision making criteria: reward rate ultimately beats accuracy. Atten Percept Psychophys 73:640–657.

Berger M, Agha NS, Gail A (2020) Wireless recording from unrestrained monkeys reveals motor goal encoding beyond immediate reach in frontoparietal cortex. eLife 9:e51322.

Berret B, Castanier C, Bastide S, Deroche T (2018) Vigour of self-paced reaching movement: cost of time and individual traits. Sci Rep 8:10655.

Berret B, Jean F (2016) Why Don't We Move Slower? The Value of Time in the Neural Control of Action. Journal of Neuroscience 36:1056–1070.

Blanchard TC, Hayden BY (2015) Monkeys Are More Patient in a Foraging Task than in a Standard Intertemporal Choice Task. PLOS ONE 10:e0117057.

Boussaoud D, Jouffrais C, Bremmer F (1998) Eye Position Effects on the Neuronal Activity of Dorsal Premotor Cortex in the Macaque Monkey. Journal of Neurophysiology 80:1132–1150.

Britten KH, Shadlen MN, Newsome WT, Movshon JA (1992) The analysis of visual motion: a comparison of neuronal and psychophysical performance. J Neurosci 12:4745–4765.

Bruce CJ, Goldberg ME (1984) Physiology of the frontal eye fields. Trends in Neurosciences 7:436–441.

Bullock D, Grossberg S (1988) Neural dynamics of planned arm movements: emergent invariants and speed-accuracy properties during trajectory formation. Psychol Rev 95:49–90.

Burk D, Ingram JN, Franklin DW, Shadlen MN, Wolpert DM (2014) Motor Effort Alters Changes of Mind in Sensorimotor Decision Making Kiebel S, ed. PLoS ONE 9:e92681.

Buzsáki G, Peyrache A, Kubie J (2014) Emergence of Cognition from Action. Cold Spring Harb Symp Quant Biol 79:41–50.

Calhoun AJ, Hayden BY (2015) The foraging brain. Current Opinion in Behavioral Sciences 5:24–31.

Charnov EL (1976) Optimal foraging, the marginal value theorem. Theor Popul Biol 9:129–136.

Choi JES, Vaswani PA, Shadmehr R (2014) Vigor of Movements and the Cost of Time in Decision Making. Journal of Neuroscience 34:1212–1223.

Churchland MM, Santhanam G, Shenoy KV (2006) Preparatory Activity in Premotor and Motor Cortex Reflects the Speed of the Upcoming Reach. Journal of Neurophysiology 96:3130–3146.

Churchland MM, Shenoy KV (2024) Preparatory activity and the expansive null-space. Nat Rev Neurosci 25:213–236.

Cisek P (1999) Beyond the computer metaphor: Behaviour as interaction. Journal of Consciousness Studies 6:125–142.

Cisek P (2006a) Preparing for Speed. Focus on "Preparatory Activity in Premotor and Motor Cortex Reflects the Speed of the Upcoming Reach." Journal of Neurophysiology 96:2842–2843.

Cisek P (2006b) Integrated neural processes for defining potential actions and deciding between them: a computational model. J Neurosci 26:9761–9770.

Cisek P (2007) Cortical mechanisms of action selection: the affordance competition hypothesis. Philosophical Transactions of the Royal Society B: Biological Sciences 362:1585–1599.

Cisek P, Green AM (2024) Toward a neuroscience of natural behavior. Current Opinion in Neurobiology 86:102859.

Cisek P, Kalaska JF (2010) Neural Mechanisms for Interacting with a World Full of Action Choices. Annual Review of Neuroscience 33:269–298.

Cisek P, Puskas GA, El-Murr S (2009) Decisions in Changing Conditions: The Urgency-Gating Model. Journal of Neuroscience 29:11560–11571.

Constantino SM, Daw ND (2015) Learning the opportunity cost of time in a patch-foraging task. Cogn Affect Behav Neurosci 15:837–853.

Cunningham JP, Yu BM (2014) Dimensionality reduction for large-scale neural recordings. Nat Neurosci 17:1500–1509.

Danielmeier C, Ullsperger M (2011) Post-Error Adjustments. Frontiers in Psychology 2:233.

Derosiere G, Thura D, Cisek P, Duque J (2019) Motor cortex disruption delays motor processes but not deliberation about action choices. Journal of Neurophysiology 122:1566–1577.

Derosiere G, Thura D, Cisek P, Duqué J (2021) Trading accuracy for speed over the course of a decision. Journal of Neurophysiology Available at: https://journals.physiology.org/doi/abs/10.1152/jn.00038.2021 [Accessed July 1, 2021].

Derosiere G, Thura D, Cisek P, Duque J (2022) Hasty sensorimotor decisions rely on an overlap of broad and selective changes in motor activity. PLOS Biology 20:e3001598.

Dotson NM, Hoffman SJ, Goodell B, Gray CM (2017) A Large-Scale Semi-Chronic Microdrive Recording System for Non-Human Primates. Neuron 96:769-782.e2.

Ebitz RB, Hayden BY (2021) The population doctrine in cognitive neuroscience. Neuron:S0896627321005213.

Evans NJ, Hawkins GE (2019) When humans behave like monkeys: Feedback delays and extensive practice increase the efficiency of speeded decisions. Cognition 184:11–18.

Evarts EV (1968) Relation of pyramidal tract activity to force exerted during voluntary movement. J Neurophysiol 31:14–27.

Fedorov A, Beichel R, Kalpathy-Cramer J, Finet J, Fillion-Robin J-C, Pujol S, Bauer C, Jennings D, Fennessy F, Sonka M, Buatti J, Aylward S, Miller JV, Pieper S, Kikinis R (2012) 3D Slicer as an image computing platform for the Quantitative Imaging Network. Magn Reson Imaging 30:1323–1341.

Ferrier D (1875) Experiments on the brain of monkeys.—No. I. Proceedings of the Royal Society of London 23:409–430.

Fetz EE (1992) Are movement parameters recognizably coded in the activity of single neurons? Behavioral and Brain Sciences 15:679–690.

Fine JM, Hayden BY (2022) The whole prefrontal cortex is premotor cortex. Philos Trans R Soc Lond B Biol Sci 377:20200524.

Fodor JA (1983) The Modularity of Mind. The MIT Press. Available at: https://direct.mit.edu/books/book/3985/The-Modularity-of-Mind [Accessed July 2, 2024].

Gallivan JP, Chapman CS, Wolpert DM, Flanagan JR (2018) Decision-making in sensorimotor control. Nat Rev Neurosci 19:519–534.

Georgopoulos AP, Caminiti R, Kalaska JF, Massey JT (1983) Spatial coding of movement: A hypothesis concerning the coding of movement direction by motor cortical populations. Experimental Brain Research 49:327–336.

Georgopoulos AP, Kalaska JF, Caminiti R, Massey JT (1982) On the relations between the direction of two-dimensional arm movements and cell discharge in primate motor cortex. J Neurosci 2:1527–1537.

Gold JI, Shadlen MN (2007) The Neural Basis of Decision Making. Annu Rev Neurosci 30:535–574.

Gomez-Marin A, Ghazanfar AA (2019) The Life of Behavior. Neuron 104:25–36.

Grießbach E, Raßbach P, Herbort O, Cañal-Bruland R (2022) Embodied decisions during walking. J Neurophysiol 128:1207–1223.

Griffiths B, Beierholm UR (2017) Opposing effects of reward and punishment on human vigor. Scientific Reports 7:42287.

Gross CG (2007) The Discovery of Motor Cortex and its Background. Journal of the History of the Neurosciences 16:320–331.

Hanks TD, Summerfield C (2017) Perceptual Decision Making in Rodents, Monkeys, and Humans. Neuron 93:15–31.

Hansmeyer L, Yurt P, Agha N, Trunk A, Berger M, Calapai A, Treue S, Gail A (2023) Home-Enclosure-Based Behavioral and Wireless Neural Recording Setup for Unrestrained Rhesus Macaques. eNeuro 10:ENEURO.0285-22.2022.

Hayden BY, Pearson JM, Platt ML (2011) Neuronal basis of sequential foraging decisions in a patchy environment. Nat Neurosci 14:933–939.

Hebb DO (1949) The organization of behavior; a neuropsychological theory. Oxford, England: Wiley.

Heron CL, Kolling N, Plant O, Kienast A, Janska R, Ang Y-S, Fallon S, Husain M, Apps MAJ (2020) Dopamine Modulates Dynamic Decision-Making during Foraging. J Neurosci 40:5273–5282.

Humphries MD (2020) Strong and weak principles of neural dimension reduction. arXiv:201108088 [q-bio] Available at: http://arxiv.org/abs/2011.08088 [Accessed February 2, 2021].

Hurley S (2001) Perception And Action: Alternative Views. Synthese 129:3–40.

James W (1890) The principles of psychology, Vol I. New York, NY, US: Henry Holt and Co.

Jazayeri M, Afraz A (2017) Navigating the Neural Space in Search of the Neural Code. Neuron 93:1003–1014.

Kaduk K, Henry T, Guitton J, Meunier M, Thura D, Hadj-Bouziane F (2023) Atomoxetine and reward size equally improve task engagement and perceptual decisions but differently affect movement execution. Neuropharmacology 241:109736.

Kalaska JF (2009) From Intention to Action: Motor Cortex and the Control of Reaching Movements. In: Progress in Motor Control (Sternad D, ed), pp 139–178 Advances in Experimental Medicine and Biology. Boston, MA: Springer US. Available at: http://link.springer.com/10.1007/978-0-387-77064-2_8 [Accessed January 29, 2021].

Kalaska JF (2019) Emerging ideas and tools to study the emergent properties of the cortical neural circuits for voluntary motor control in non-human primates. F1000Res 8.

Kalaska JF, Scott SH, Cisek P, Sergio LE (1997) Cortical control of reaching movements. Curr Opin Neurobiol 7:849–859.

Kim JN, Shadlen MN (1999) Neural correlates of a decision in the dorsolateral prefrontal cortex of the macaque. Nat Neurosci 2:176–185.

Leroy É, Koun É, Thura D (2023) Integrated control of non-motor and motor efforts during perceptual decision-making and action execution: a pilot study. Sci Rep 13:9354.

Mante V, Sussillo D, Shenoy KV, Newsome WT (2013) Context-dependent computation by recurrent dynamics in prefrontal cortex. Nature 503:78–84.

Mao D, Avila E, Caziot B, Laurens J, Dickman JD, Angelaki DE (2021) Spatial modulation of hippocampal activity in freely moving macaques. Neuron 109:3521-3534.e6.

Maselli A, Gordon J, Eluchans M, Lancia GL, Thiery T, Moretti R, Cisek P, Pezzulo G (2023) Beyond simple laboratory studies: Developing sophisticated models to study rich behavior. Phys Life Rev 46:220–244.

Mathis A, Mamidanna P, Cury KM, Abe T, Murthy VN, Mathis MW, Bethge M (2018) DeepLabCut: markerless pose estimation of user-defined body parts with deep learning. Nat Neurosci 21:1281–1289.

Milstein DM, Dorris MC (2007) The influence of expected value on saccadic preparation. J Neurosci 27:4810–4818.

Muhammed K, Dalmaijer E, Manohar S, Husain M (2020) Voluntary modulation of saccadic peak velocity associated with individual differences in motivation. Cortex 122:198–212.

Musall S, Kaufman MT, Juavinett AL, Gluf S, Churchland AK (2019) Single-trial neural dynamics are dominated by richly varied movements. Nat Neurosci 22:1677–1686.

Myerson J, Green L (1995) Discounting of delayed rewards: Models of individual choice. J Exp Anal Behav 64:263–276.

Narayanan N, Laubach M (2009) Methods for Studying Functional Interactions Among Neuronal Populations. Methods in molecular biology (Clifton, NJ) 489:135–165.

Padoa-Schioppa C (2011) Neurobiology of Economic Choice: A Good-Based Model. Annu Rev Neurosci 34:333–359.

Palmer J, Huk AC, Shadlen MN (2005) The effect of stimulus strength on the speed and accuracy of a perceptual decision. Journal of Vision 5:1.

Parker PRL, Brown MA, Smear MC, Niell CM (2020) Movement-Related Signals in Sensory Areas: Roles in Natural Behavior. Trends in Neurosciences 43:581–595.

Pasquereau B, Nadjar A, Arkadir D, Bezard E, Goillandeau M, Bioulac B, Gross CE, Boraud T (2007) Shaping of Motor Responses by Incentive Values through the Basal Ganglia. Journal of Neuroscience 27:1176–1183.

Penfield W, Boldrey E (1937) Somatic motor and sensory representation in the cerebral cortex of man as studied by electrical stimulation. Brain: A Journal of Neurology 60:389–443.

Pesaran B (2010) Neural correlations, decisions, and actions. Current Opinion in Neurobiology 20:166–171.

Pezzulo G, Cisek P (2016) Navigating the Affordance Landscape: Feedback Control as a Process Model of Behavior and Cognition. Trends in Cognitive Sciences 20:414–424.

Purcell BA, Kiani R (2016) Neural Mechanisms of Post-error Adjustments of Decision Policy in Parietal Cortex. Neuron 89:658–671.

Pylyshyn Z (1984) Computation and Cognition: Toward a Foundation for Cognitive Science. Cambridge, MA: MIT Press.

Ratcliff R, Smith PL, Brown SD, McKoon G (2016) Diffusion Decision Model: Current Issues and History. Trends in Cognitive Sciences 20:260–281.

Redgrave P, Prescott TJ, Gurney K (1999) The basal ganglia: a vertebrate solution to the selection problem? Neuroscience 89:1009–1023.

Revol P, Collette S, Boulot Z, Foncelle A, Niki C, Thura D, Imai A, Jacquin-Courtois S, Cabanac M, Osiurak F, Rossetti Y (2019) Thirst for Intention? Grasping a Glass Is a Thirst-Controlled Action. Front Psychol 10:1248.

Reynaud AJ, Saleri Lunazzi C, Thura D (2020) Humans sacrifice decision-making for action execution when a demanding control of movement is required. Journal of Neurophysiology 124:497–509.

Roitman JD, Shadlen MN (2002) Response of Neurons in the Lateral Intraparietal Area during a Combined Visual Discrimination Reaction Time Task. J Neurosci 22:9475–9489.

Saleri C, Thura D (2024) Evidence for interacting but decoupled controls of decisions and movements in non-human primates. :2024.01.29.577721 Available at: https://www.biorxiv.org/content/10.1101/2024.01.29.577721v1 [Accessed February 3, 2024].

Saleri Lunazzi C, Reynaud AJ, Thura D (2021) Dissociating the Impact of Movement Time and Energy Costs on Decision-Making and Action Initiation in Humans. Front Hum Neurosci 15:715212.

Saleri Lunazzi C, Thura D, Reynaud AJ (2023) Impact of decision and action outcomes on subsequent decision and action behaviours in humans. European Journal of Neuroscience 57:1098–1113.

Salinas E, Sejnowski TJ (2001) Correlated neuronal activity and the flow of neural information. Nat Rev Neurosci 2:539–550.

Saxena S, Cunningham JP (2019) Towards the neural population doctrine. Current Opinion in Neurobiology 55:103–111.

Schultz W (2006) Behavioral theories and the neurophysiology of reward. Annu Rev Psychol 57:87–115.

Servant M, Logan GD, Gajdos T, Evans NJ (2021) An integrated theory of deciding and acting. Journal of Experimental Psychology: General Available at: http://doi.apa.org/getdoi.cfm?doi=10.1037/xge0001063 [Accessed September 15, 2021].

Shadmehr R, Ahmed AA (2020) Vigor: Neuroeconomics of Movement Control. The MIT Press. Available at: https://doi.org/10.7551/mitpress/12940.001.0001.

Shadmehr R, Orban de Xivry JJ, Xu-Wilson M, Shih T-Y (2010) Temporal Discounting of Reward and the Cost of Time in Motor Control. Journal of Neuroscience 30:10507–10516.

Shahidi N, Franch M, Parajuli A, Schrater P, Wright A, Pitkow X, Dragoi V (2024) Population coding of strategic variables during foraging in freely moving macaques. Nat Neurosci 27:772–781.

Shenoy KV, Sahani M, Churchland MM (2013) Cortical control of arm movements: a dynamical systems perspective. Annu Rev Neurosci 36:337–359.

Straube A, Fuchs AF, Usher S, Robinson FR (1997) Characteristics of Saccadic Gain Adaptation in Rhesus Macaques. Journal of Neurophysiology 77:874–895.

Sukumar S, Shadmehr R, Ahmed AA (2024) Effects of reward and effort history on decision making and movement vigor during foraging. J Neurophysiol 131:638–651.

Summerside EM, Shadmehr R, Ahmed AA (2018) Vigor of reaching movements: reward discounts the cost of effort. Journal of Neurophysiology 119:2347–2357.

Testard C, Tremblay S, Parodi F, DiTullio RW, Acevedo-Ithier A, Gardiner KL, Kording K, Platt ML (2024) Neural signatures of natural behaviour in socializing macaques. Nature 628:381–390.

Testard C, Tremblay S, Platt M (2021) From the field to the lab and back: neuroethology of primate social behavior. Current Opinion in Neurobiology 68:76–83.

Thura D (2007) Influence de la position de la main sur l'exploration oculaire de l'espace péripersonnel. Available at: https://hal.science/tel-02906275/.

Thura D (2016) How to discriminate conclusively among different models of decision making? J Neurophysiol 115:2251–2254.

Thura D (2020) Decision urgency invigorates movement in humans. Behavioural Brain Research 382:112477.

Thura D (2021) Reducing behavioral dimensions to study brain–environment interactions. Behavioral and Brain Sciences 44 Available at: https://www.cambridge.org/core/journals/behavioral-and-brain-sciences/article/abs/reducing-behavioral-dimensions-to-study-brainenvironment-interactions/685BFFF2731C77DF4734B8DE039FF998 [Accessed September 30, 2021].

Thura D, Beauregard-Racine J, Fradet C-W, Cisek P (2012) Decision making by urgency gating: theory and experimental support. J Neurophysiol 108:2912–2930.

Thura D, Boussaoud D, Meunier M (2008a) Hand position affects saccadic reaction times in monkeys and humans. J Neurophysiol 99:2194–2202.

Thura D, Cabana J-F, Feghaly A, Cisek P (2022) Integrated neural dynamics of sensorimotor decisions and actions. PLoS Biol 20:e3001861.

Thura D, Cisek P (2014) Deliberation and commitment in the premotor and primary motor cortex during dynamic decision making. Neuron 81:1401–1416.

Thura D, Cisek P (2016a) On the difference between evidence accumulator models and the urgency gating model. J Neurophysiol 115:622–623.

Thura D, Cisek P (2016b) Modulation of Premotor and Primary Motor Cortical Activity during Volitional Adjustments of Speed-Accuracy Trade-Offs. J Neurosci 36:938–956.

Thura D, Cisek P (2017) The Basal Ganglia Do Not Select Reach Targets but Control the Urgency of Commitment. Neuron 95:1160-1170.e5.

Thura D, Cisek P (2020) Microstimulation of dorsal premotor and primary motor cortex delays the volitional commitment to an action choice. Journal of Neurophysiology 123:927–935.

Thura D, Cos I, Trung J, Cisek P (2014) Context-dependent urgency influences speed-accuracy trade-offs in decision-making and movement execution. J Neurosci 34:16442–16454.

Thura D, Guberman G, Cisek P (2017) Trial-to-trial adjustments of speed-accuracy trade-offs in premotor and primary motor cortex. J Neurophysiol 117:665–683.

Thura D, Hadj-Bouziane F, Meunier M, Boussaoud D (2008b) Hand position modulates saccadic activity in the frontal eye field. Behav Brain Res 186:148–153.

Thura D, Hadj-Bouziane F, Meunier M, Boussaoud D (2011) Hand modulation of visual, preparatory, and saccadic activity in the monkey frontal eye field. Cereb Cortex 21:853–864.

Trapanese C, Meunier H, Masi S (2019) What, where and when: spatial foraging decisions in primates. Biological Reviews 94:483–502.

Tremblay S, Testard C, DiTullio RW, Inchauspé J, Petrides M (2023a) Neural cognitive signals during spontaneous movements in the macaque. Nat Neurosci 26:295–305.

Tremblay S, Testard C, DiTullio RW, Inchauspé J, Petrides M (2023b) Neural cognitive signals during spontaneous movements in the macaque. Nat Neurosci 26:295–305.

Turner RS, Desmurget M (2010) Basal ganglia contributions to motor control: a vigorous tutor. Current Opinion in Neurobiology 20:704–716.

Voloh B, Maisson DJ-N, Cervera RL, Conover I, Zambre M, Hayden B, Zimmermann J (2023) Hierarchical action encoding in prefrontal cortex of freely moving macaques. Cell Reports 42:113091.

Vyas S, Golub MD, Sussillo D, Shenoy KV (2020) Computation Through Neural Population Dynamics. Annu Rev Neurosci 43:249–275.

Yin HH (2014) Action, time and the basal ganglia. Philosophical Transactions of the Royal Society B: Biological Sciences 369:20120473.

Yoo SBM, Hayden BY, Pearson JM (2021) Continuous decisions. Philosophical Transactions of the Royal Society B: Biological Sciences 376:20190664.

Yoon T, Geary RB, Ahmed AA, Shadmehr R (2018) Control of movement vigor and decision making during foraging. Proc Natl Acad Sci USA 115:E10476–E10485.